\documentclass[12pt]{iopart}
\usepackage{graphicx}
\usepackage{xcolor}
\usepackage{subcaption}
\usepackage{url}

\begin{document}

\title[FIW 2D modeling]{Spatiotemporal dynamics of nanosecond pulsed discharge in the form of a fast ionization wave: self-consistent two-dimensional modeling and comparison with experiments under negative and positive polarity.
}


\author{Konstantinos Kourtzanidis}%
\address{Chemical Process \& Energy Resources Institute (CPERI), Centre for Research \& Technology Hellas (CERTH),
6th km Charilaou-Thermi, Thermi, 57001 Thessaloniki, Greece}
\author{Svetlana M. Starikovskaia}%
\address{Laboratory of Plasma Physics (CNRS, Ecole Polytechnique, 
Univ. Paris-Sud, Observatoire de Paris, Sorbonne Universit\'e, 
l’Institut Polytechnique de Paris), Ecole
Polytechnique, route de Saclay, 91128 Palaiseau, France}
\ead{kourtzanidis@certh.gr}

\begin{abstract}

\color{black}
Nanosecond discharges are characterized by a shift in energy branching toward the excitation of electronic levels and dissociation, making them particularly attractive for plasma chemistry. Understanding the spatio-temporal structure of these discharges is especially important. This paper presents a detailed 2D-axisymmetric numerical analysis of a nanosecond discharge propagating in a long tube and in pure nitrogen. The modeling is conducted using a self-consistent plasma fluid solver under the local mean energy approximation (LMEA), including photoionization. The discharge develops at moderate pressures, 1–10~Torr, in the form of a fast ionization wave (FIW). Simulations are performed for both negative and positive polarities of the voltage pulse applied to the high-voltage electrode. The computational results are validated against available experimental data, including FIW velocity within the studied pressure range, electron density, longitudinal electric field, and the radial distribution of N$_2$(C$^3\Pi_u$) emission on a nanosecond timescale.
\color{black}
\end{abstract}

\maketitle

\section{Introduction}

Nanosecond discharges have gained popularity in recent decades, due to their unique properties related to plasma chemistry: the energy branching in these discharges is shifted toward excitation of electronically excited species and dissociation by electron impact. The availability of commercially produced nanosecond high-voltage generators, the development of pulse measurement techniques, "ordinariness" of pico- and femtosecond laser measurements \cite{Goldberg2022,Kulatilaka2023} have rendered nanosecond plasma no longer exotic, which is currently studied for a wide variety of applications, such as plasma-assisted combustion and detonation \cite{Starikovskiy2013, Starikovskaia2014, Ju2015}, plasma-assisted aerodynamics \cite{Leonov2016,Samimy2019}, plasma medicine \cite{Limanowski_Keidar2022} and many more. \color{black} These applications require an understanding of the spatio-temporal structure of discharge development, often in chemically active mixtures (e.g., in the case of combustion or detonation). Such an understanding can be achieved through plasma modeling, validated against a broad set of experimental data.\color{black} 

The nanosecond discharge in long tubes was first reported in 1893 by J.J. Thomson \cite{Thomson1893}. He investigated the propagation of a luminous front in a 15-metre glass tube between two vessels with mercury, which were the  electrodes and the constructive parts of a mercury pump. The voltage at the high-voltage electrode was a pulse with a calculated decay of about a hundred nanoseconds. Measured using a rotating mirror, the propagation speed of the luminescence front was estimated to be "approximately half the speed of light".

Since then, interest in nanosecond discharges has been renewed every 20-30 years, in correlation with the development of experimental techniques and theoretical approaches. Works on nanosecond discharges of the XX~century are reviewed in \cite{Vasilyak1994}. Therein, the authors suggest the term "fast ionization wave" (FIW) for nanosecond discharges developing in long tubes (the length of the tube is much bigger than the diameter of the tube) and propagating with a velocity of a few centimeters per nanosecond. The review includes description of dynamics of the propagation of the FIWs generated by high-voltage pulses from a few kilovolts to a few hundreds of kilovolts and duration of a few tens of nanoseconds. Velocities of the FIW propagation in different gases, from nitrogen, air and helium to quite exotic like SF$_6$, CCl$_4$ or acetone vapors are presented. Non-local effects in the FIWs front, influence of high-energy electrons on the formation and motion of fast ionization waves are discussed. Theoretical models of fast ionization waves at that time mainly described one-dimensional distribution of the electric field and of current in plasma, special attention being attributed to the run-away electrons in the front of nanosecond discharges at low and moderate pressures. 

In the period 1994-2008, a series of papers on the fast ionization waves was published (see for example, \cite{Anikin1998,Anikin2002,Anikin2004,Anikin2008,Zatsepin1998,Pancheshnyi1999,Pancheshnyi2000}), a part of them being summarized in \cite{Starikovskaia2001,Starikovskaia2002}. The papers were devoted to experimental study of the fast ionization waves in so-called kinetic approach when the main interest of study is how to link electrical parameters measured in plasma (the electric field, the electric current, the deposited energy) to the energy branching in the discharge and the efficiency of plasma-chemical processes. The FIWs were triggered in long tubes by pulses of units-tens of kV at moderate pressures, 0.1-100~Torr. 

In particular, it was shown that the peak of high reduced fields \cite{Pancheshnyi1999} in the front of the fast ionization wave, up to several kTd (1~Td$=10^{-17}$~V$\cdot$cm$^2$), does not exceed two to three nanoseconds in duration. Then the field decreases sharply to hundreds of Td, optimal for excitation of electronically excited levels of atoms and molecules \cite{Pancheshnyi1999,Anikin2002}. In the pressure range from units of Torrs to tens of Torr, optimal for the development of fast ionisation waves, the peak value of the electric field $E$ increases less than twofold; the reduced electric field $E/N$ decreases almost by an order of magnitude. It was shown, by means of numerical calculations, that at the FIW front, the electron energy distribution function, EEDF, is close to the distribution formed at relaxation of the electron beam in the gas media  with predominance of high energy electrons responsible for the homogeneous organisation of the gas \cite{Starikovskaia2001}. At the same time, the main production of electrons and major energy deposition occur behind the front of the FIW, 5-10 nanoseconds after the peak. 

The progress in numerical modeling dureing the two last decades opened possibilities for more detailed description of nanosecond discharges. Topical review \cite{Naidis2018} considers sub-nanosecond breakdown in atmospheric pressure gases. The authors of the review claim that, even if numerical simulation of fast pulsed discharges does not reproduce all the complexity of the  physical processes involved in the discharge development (i) it provides a much wider set of discharge characteristics comparing to experimental measurements; (ii) it allows easier than in experiment, sensitivity analysis revealing the main factors responsible for the observed phenomena. Finally, it is a synergy of experimental measurements and numerical modeling which provides better understanding of the physical event under study.

As described in the review \cite{Vasilyak1994}, the first theoretical/numerical investigations of fast ionization waves date back to the late 1970s and early 1980s Soviet era. These studies were based on one-dimensional fluid (hydrodynamic) or long-wavelength approximations and self-similar solutions were obtained for the FIW propagation, which inherently assumes constant properties in the plasma front. In the same review paper, a more general one-dimensional approach based on the telegraph and total current equations is also described which has been used to describe and comprehend several features of the FIW development. Since then, and besides some very interesting theoretical/analytical studies on the solutions of streamer-like planar fronts ~\cite{ebert1996streamer} and FIW propagation in long shielded tubes~\cite{sinkevich2000propagation}, several one-dimensional models (both kinetic and fluid based) have been used to describe FIWs. Acknowledging the large computational cost of a kinetic description of plasmas based on particles~\cite{verboncoeur2005particle, birdsall2018plasma} (either using a Monte-Carlo-Collision or Particle In Cell method), which require the solution of the Boltzmann equation to describe the behavior of single particle species, and and its dramatic increase in 2D systems despite recent advancements \cite{rodriguez2020implementation}, most of FIW numerical studies incorporate the fluid (macroscopic) approximation, where macroscopic quantities such as density, mean velocity and mean energy are used to describe the particle species behaviour and which correspond to velocity moments of the Boltzmann equation (and relevant closure approximations). The corresponding computational cost is largely reduced compared to kinetic approaches, while the solution of the system of fluid equations (momentum, continuity, energy) can leverage efficient algorithms and discretization schemes on cartesian or unstructured meshes, from the very active Computational Fluid Dynamics (CFD) community. Further assumptions lead to more simplified (and thus computationally efficient) models: neglecting inertia terms in the momentum equation (focusing on non-magnetized plasma), assuming isothermal conditions and collisionality, we derive the drift-diffusion model (see Sec.\ref{sec:numerical} for details), where each species evolution is governed by a single continuity equation in a convection-diffusion-reaction form self-consistently coupled with the electric field through the Poisson equation. Transport and reaction rates depend non-linearly on the reduced electric field under the Local Field Approximation (LFA - 1st order moment model) or the mean electron energy under the Local Mean Energy Approximation (LMEA - 2nd order moment model). Under the LMEA, non-local effects owed to a considerable variation of the electric field over the electron energy relaxation length are accounted for and the model requires the inclusion of an equation for the electron mean energy.     

One of the first two-dimensional numerical studies of regular ionization waves in long tubes was performed by Brok et al. \cite{brok2003model} in 2003. The authors studied the development of anode directed, DC-excited ionization waves sustained by thermionic emission in moderate pressure (few Torr) Argon and demonstrated the capabilities of fluid modeling to qualitatively capture the breakdown and propagation mechanism, while emphasized the importance of dielectric surface charging. The influence of surface charges and memory effects in such rather slow ionization waves have recently been investigated~\cite{viegas2022quantification} both experimentally and numerically. Fast ionization waves have been studied self-consistently in dry air and nitrogen ~\cite{takashima2012development, takashima2011experimental} but also in Helium~\cite{huang2018effect} and in millimeter and capillary tubes in low pressure N$_{2}$ discharges~\cite{klochko2014investigation, zhu2020scaling, chen2021modeling, timshina2020numerical}, 
and very recently in short gaps in air and CO$_{2}$~\cite{wong2024computational}. Effects of field emission and relative influence of runaway electrons have been studied with 1D PIC simulations for atmospheric pressure nitrogen ~\cite{levko2016influence, levko2023kinetics}. Although this list of numerical works related to FIWs is not exhaustive, it showcases that in the current state of numerical models and High Performance Computing infrastructure, fluid modeling is the typical method for two-dimensional simulations of FIWs. Nevertheless, the computational cost of fluid models is strongly linked on the highly discrepant nature of spatial and temporal timescales of FIWs as well as the dimensions of the simulated discharge region. Accurate, robust and efficient time integration schemes are necessary to account for the extremely different temporal timescales related to electrons and ions drift, assuring that stability (majorly linked to timestep constraints linked to the dielectric relaxation time) is guaranteed. In addition, the development of FIWs in long tubes (several tens of centimeter long as the ones studied in this work), require appropriate meshing in the whole interelectrode gap which renders such 2D simulations extremely time demanding and only possible when parallel computing techniques are efficiently implemented. The influence of photoionization in gases that are known to efficiently produce photons that can ionize neutral molecules (such as oxygen-nitrogen mixtures) adds up to the numerical challenges.    

In this work, we present a detailed 2D axisymmetric self-consistent numerical study of nanosecond moderate pressure discharge in nitrogen propagating in a long tube as a fast ionization wave (FIW). The main goal of this paper is the comparison of self-consistent numerical simulations with the experimental results on FIW propagation obtained within the research group of one of the co-authors and reviewed in \cite{Starikovskaia2001}. We have deliberately chosen the results which never were a subject of detailed numerical analysis. In particular, the dependence of the FIW velocity and the change in the radial structure of the FIW with pressure depending on the discharge polarity have never been the subject of detailed numerical modeling. Thus, the numerical results are  compared with experimental measurements of the FIW propagation velocity, the reduced electric field, the electron density and the N$_2$(C$^3\Pi_u$) number density distribution for negative and positive polarities of the applied high voltage pulse. Emphasis is given on both the temporal and spatial development of the discharge, aiming to reveal and understand the particularities of each pulse polarity at different gas pressure inside the 1 - 10 torr range.

The article is structured as follows: in section \ref{sec:experimental}, we provide a brief overview of the experimental setup and findings. In section \ref{sec:numerical}, we present the physical and numerical models used in this study. In section \ref{sec:case}, we present the computational case and numerical parameters. In section~\ref{sec:results} we present numerical results on electric current, FIW velocity, spatiotemporal distribution of plasma parameters and produced species, comparing them with experimental findings and discussing the results. Last, in section \ref{sec:conclusion}, we present  the conclusions and propose future directions. 

\section{A brief information about experimental methods and observed results}\label{sec:experimental}

\begin{figure}[h]
\centering
\includegraphics[width=\linewidth]{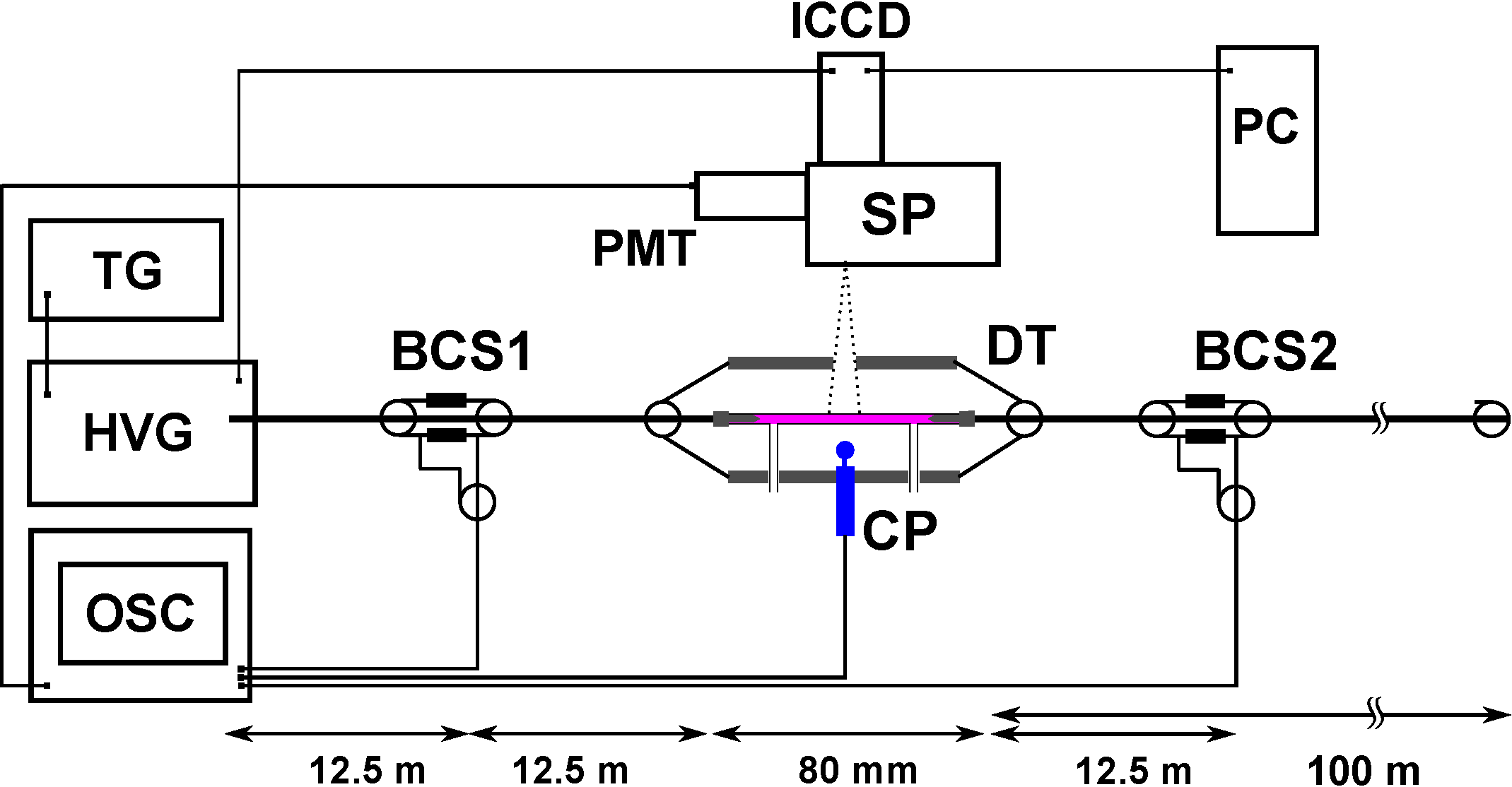}
  \caption{Typical scheme of the experimental setup with the main experimental equipment \protect{\cite{Lepikhin2018}}. The general scheme of the experimental setup with the main experimental equipment. TG is triggering generator, HVG is high voltage generator, OSC is oscilloscope, BCS1 and BCS2 are back current shunts, CP is capacitive probe, DT is discharge tube, PMT is photomultiplier tube, ICCD is intensiﬁed charge-coupled device (camera), SP is spectrometer, PC is personal computer.}
    \label{Exp_setup}
 \end{figure}

A typical experimental scheme is presented in Fig.~\ref{Exp_setup}. In the considered experiments, the high-voltage generator (HVG) is always connected to the discharge tube (DT) by a long high-voltage coaxial cable. For the most of experiments with FIWs at moderate (1-50~Torr) pressures, glass or quartz discharge tubes of different lengths (20 to 60 cm) and diameters (1 to 5 cm) are  used. Under these conditions, the application of a voltage pulse of 10 to 50~kV amplitude, a few nanosecond rise time and a few tens of nanoseconds FWHM to the high-voltage electrode, initiates an ionization wave which starts always from the high voltage electrode and propagates along the tube.  Typically, experiments incorporate a grounded screen linked to the shielding of the high-voltage cable. However, even in the absence of the screen, the ionization wave can propagate as the current closes over the surrounding space. The end load of the discharge tube can be different, often set to zero or infinity; or the low voltage electrode and the grounded screen can be connected to another coaxial cable. Typically, the nanosecond discharge is highly repetitive, ans it is possible to accumulate any signal at low frequencies (5-30~Hz) having a subnanosecond synchronization of the discharge and diagnostics. 

Spatial structure of the propagating FIW is studied by sub-nanosecond ICCD imaging (0.5-2~ns typical ICCD gates). To control the voltage pulse arriving the the high voltage electrode, to measure the current through plasma and the deposited energy, calibrated custom made back current shunts (BCSs) are used. They are soldered in the grounded shielding of the high voltage cable between the high-voltage generator and the discharge cell so that the incident and reflected from the electrical mismatched load pulses are separated in time. We note here that the voltage on the high-voltage electrode at the moment of the discharge start is doubled comparing to the voltage in the cable as a result of  the constructive interference between the incident pulse and the pulse reflected from the discharge tube due to mismatch of the electrical load: 10~kV in the middle of the long cable provides 20~kV at the high-voltage electrode. A description of the back current shunt technique in application to nanosecond discharges can be found in \cite{Pavan2024}.

The velocity of the fast ionization wave and the longitudinal electric field are measured with the help of calibrated custom made capacitive probes (CPs). The electron density is calculated knowing the current, the field and the drift velocity. Time-resolved behavior of electronically excited species is studied using optical emission spectroscopy with the help of a spectrometer linked to the ICCD camera or to the photomultiplier tube (PMT). All electrical signals are measured by nanosecond oscilloscopes.  

A general picture of the development of the nanosecond discharge under mentioned conditions can be summarized as follows: for both negative and positive polarities of the high--voltage pulse, the discharge starts from the high--voltage electrode and propagates with a velocity of a few centimeters per nanosecond \cite{Vasilyak1994,Anikin2002}. There is a well-defined maximum of the FIW velocity upon pressure. For example, for the voltage amplitude about 10~kV in the cable and the tube diameter of a few centimeters, the maximum velocity is observed around 4~Torr. The position of the maximum shifts to higher pressures as the voltage amplitude increases.

The reduced electric field in the front of the fast ionization wave is high, a few kTd \cite{Pancheshnyi1999}. The duration of this high peak is a few nanoseconds which corresponds, for a given velocity of propagation, to centimeters or even tens of centimeters in length. After the high peak of the electric field, the fields optimal for excitation of electronic levels of molecules and dissociation, a few hundreds of Td, are usually installed. These fields keep constant or only slightly decreasing up to the end of the high voltage pulse. The first high peak of the electric field during the FIW propagation is responsible for the main initial ionization of the gas, resulting in diffuse plasma.  In the constant field after the discharge front the electron density can continue to increase, albeit at a significantly reduced rate.

At the dimensions of discharge tubes and the parameters of the high-voltage pulses mentioned above, the specific energy deposited in gas (SED) is typically small, on the order of $10^{-4}-10^{-3}$~eV/molecule. Typical electron densities are $n_e \sim 10^{12}-10^{13}$~cm$^{-3}$, providing the ionization degree around $10^{-5}-10^{-4}$. This means that the recombination time, 1-10~microseconds, is much longer than the duration of the discharge, and the electron density stays constant in time in the discharge and near afterglow \cite{Pancheshnyi1999}. 

\begin{figure}[h]
\centering
\includegraphics[width=0.75\linewidth]{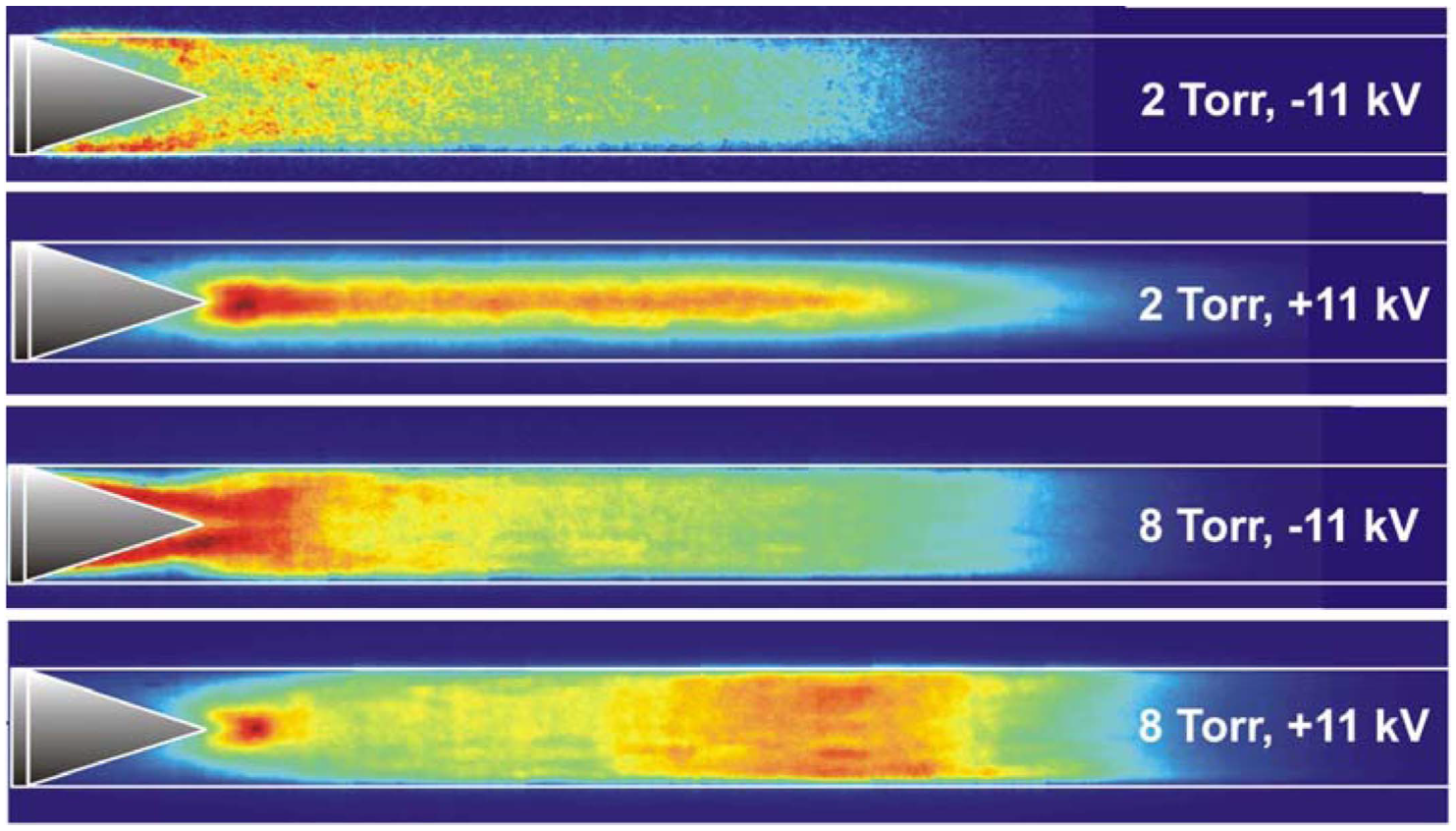}
  \caption{Development of the fast ionization wave in air for different pressures and polarities of the high-voltage pulse. The diameter of the quartz tube is 17.5~mm; the tube length is 50~cm. Emission of the 2+ system of molecular nitrogen. ICCD gate is 1~ns; spectral response is 300–800~nm. Conical high-voltage electrode is on the left-hand side \protect{\cite{Anikin2008}}.}
    \label{2D_ICCD}
 \end{figure}

Although the negative and positive polarity FIWs are similar in physical parameters, they differ essentially in their spatial structure. This was observed for air and for nitrogen, as example of ICCD imaging for air is provided by Fig.~\ref{2D_ICCD} \cite{Anikin2008}. The images present raw experimental data, without the inverse Abel transform. The measurements are made for the second positive system of molecular nitrogen, N$_2$(C$^3\Pi_u$) $\rightarrow$ N$_2$(B$^3\Pi_g$) transition. At these pressures, the life time of N$_2$(C$^3\Pi_u$) in air at 1-10~Torr is long, $40-20$~ns \cite{Pancheshnyi2000}, still the pattern of the discharge is clearly seen. At negative polarity, the emission is stronger near the walls, while at positive polarity, the main emission, especially at low pressures, comes from the near--axis region. 

\color{black} 
 In nitrogen, FIWs have a similar space structure, often with more pronounced  difference between the axis and the near-wall reagion. This can be seen from ICCD images in rectangular cross-section double-pulse setup~ \cite{takashima2012development} or from N$_2$(C$^3\Pi_u$) radial profiles in nanosecond second capillary discharge~\cite{lepikhin2018fast}.

\color{black}

The observed difference in the spatial patterns of the FIW for each polarity is not well understood till date. Thus, in the following sections we leverage high-fidelity numerical modeling to explain this difference and explain features of the FIW for both polarities at different gas pressure. 

\section{Numerical modeling}\label{sec:numerical}
We employ a fluid (continuum) description of the plasma which is the core of the physical model used in the computational software COPAIER~\cite{kourtzanidis2020self, dufour2015numerical} - a multi-species and multi-temperature plasma fluid solver allowing for self-consistent description of the plasma spatial and temporal evolution (the main author is a co-developer of the solver).  Each of the species, $s$, considered in the pre-defined plasma-gas chemistry is governed by a continuity equation :
  \begin{equation}
 \label{eq:continuity}
{{\partial{n_{s}}} \over {\partial{t}}}  +\vec{\nabla} \cdot \vec{\Gamma}_{s} =S_{s},
\end{equation}

\noindent
where $S_{s}$ is the net rate of production of species $s$ due to chemical reactions and it reads: 
 \begin{equation}
 \label{eq:chemistry}
S_{s}=\sum_{r}c_{s,r}R_{r} + S_{ph},
\end{equation}

\noindent
where $c_{s,r}$ is the net number of particles of species $s$ created or lost in one reaction of type $r$ (so it can be positive or negative). Note that $S_{ph}$ is the photo-ionization source term based on the three-group Helmholtz approximation model \cite{bourdon2007efficient} (see below). 

The species number flux $\vec{\Gamma}=n_{s}\vec{v_{s}}$ is given by the momentum balance equation which is approximated by the drift-diffusion equation :

  \begin{equation}
 \label{eq:driftdiffusion}
\vec{\Gamma_{s}}=sgn(q_{s})n_{s}\mu_{s}\vec{E} - \vec{\nabla}(n_{s}D_{s}) - n_{s}\vec{u},
\end{equation}

\noindent
where $q$ is the species charge, $\mu_{s}$ and $D_{s}$ are the species $k$ mobility and diffusion coefficient respectively, $u$ is the mean mass fluid convection velocity. The electric field is defined as $\vec{E}=-\nabla \Phi$, with $\Phi$ being the electrostatic potential defined by the Poisson's equation 

  \begin{equation}
 \label{eq:poisson}
\nabla \cdot (\epsilon \nabla \Phi) = -\rho=-\sum_{s}q_{s}n_{s},
\end{equation}

\noindent
where $\epsilon$ is the dielectric permittivity. 

In Eq.~(\ref{eq:driftdiffusion}) the diffusion coefficient is related to the mobility term by Einstein's relation, $D_{s}=\mu_{s}k_{B}T_{s}/q$. The rest of transport (mobility $\mu$) and reaction rate coefficients ($k$) are tabulated with an electric energy dependence under the Local Mean Energy Approximation (LMEA).
Under the LMEA, transport and rate coefficients depend on the mean electron energy instead of the electric field (Local Field Approximation - LFA). 

To justify our choice of the LMEA over the LFA, we calculate the electron energy relaxation length which can be approximated using\cite{piskin2019challenges}: 

  \begin{equation}
\lambda_{e} = \sqrt{D_{e}\tau_{e}},
\end{equation}

\noindent
where $\tau_{e}^{-1}={{2m_{e}}\over{m_{n}}}\nu + \nu^{*}$, $m_{n}$ is the neutral species mass, $\nu$ is the elastic collision frequency and $\nu^*$ is the inelastic collision frequency. In our conditions, we estimate $\lambda_{e}$ in the 5-10~mm range. The discharge gap, $L$, is 60 cm and most importantly, the FIW front (as we'll see below) presents strong gradients of the electric field in length scales that are millimetric,   which indicates that non-local effects are important and the validity~\cite{grubert2009local} of the Local Field Approximation (LFA) and local field equilibrium is violated.   

The mean electron temperature can be obtained by the electron energy conservation given as : 
 
  \begin{equation}
 \label{eq:energy}
{{\partial{n_{\epsilon}}} \over {\partial{t}}}  +\vec{\nabla} \cdot \vec{\Gamma}_{\epsilon} =S_{\epsilon},
\end{equation}

\noindent
where $n_{\epsilon}=n_{e}\epsilon$ is the electron energy density and $\epsilon$ is the electron mean energy which by assuming that is a results mainly from random motion relates to the electron mean temperature by $k_{B}T_{e}=2/3\epsilon$.

The electron energy flux reads :

  \begin{equation}
 \label{eq:energyflux}
 \vec{\Gamma}_{\epsilon} ={{5}\over{3}}\epsilon \vec{\Gamma}_{e} + \vec{q}
\end{equation}

The heat flux, $\vec{q}$ is assumed proportional to the gradient of the electron mean energy :
  \begin{equation}
 \label{eq:heatflux}
 \vec{q} =-{{5}\over{3}} n_{e} D_{e} \nabla \epsilon
\end{equation}

The source term, $S_{\epsilon}$ includes heating by the electric field (electron Joule heating) and energy losses in inelastic and elastic collisions as:
 \begin{equation}
 \label{eq:sourceenergy}
S_{\epsilon}=-e\vec{\Gamma}_{e} \cdot \vec{E} - \sum_{i=1}^{N_{react}}\Delta E_{i}r_{i} -  {{3} \over {2}} k_{b} n_{e} {{2m_{e}} \over {m_{g}}} (T_{e} - T_{g}) \nu_{m}
\end{equation}

The first term in the right hand side corresponds to the electron Joule heating term. The second term is the sum for each electron impact reaction, $i$, of the inelastic electron energy loss, where $\Delta E_{i}$ is the energy threshold (thermal energy loss) in reaction $i$ and $r_{i}$ is the rate of progress. The summation is performed over the total number of gas-phase reactions, $N_{react}$.The last term, corresponds to the energy loss due to elastic collisions which can also be approximated with an effective rate coefficient and integrated directly to the inelastic energy loss term. In this last term, $\nu_{m}$ is the electron momentum-transfer frequency, $T_{g}$ is the gas density, $m_{g}$ is the heavy species particle mass,  $m_{e}$ is the electron particle mass.

The density of the background species is given by the ideal gas law :
  \begin{equation}
 \label{eq:ideal}
p=\sum n_{s} k_{B} T_{s} 
\end{equation}

Surface charge accumulation on dielectric surfaces is taken into account by assuming that particles stick to the dielectric surface and no diffusion occurs. Thus the surface charge is given by: 
\begin{equation}
\sigma=\int \vec{j}\cdot \vec{n} dt,
\end{equation}
 
\noindent
where the plasma current density reads : 
 \begin{equation}
\vec{j}=\sum_{s}q_{s}\vec{\Gamma}_{s}
\end{equation}

Lastly, we include a photoionization model involving the solution of three Helmholtz type equations~\cite{bourdon2007efficient} in the form: 

 \begin{equation}
 \label{eq:photo}
\nabla^{2}S_{ph, j}- (\lambda_{j}p)^{2} =  - A_{j}p^{2}I_{ion}
\end{equation}

\noindent
for $j=1,3$,  where $S_{ph, j}$ is the photoionization source term in $\mathrm{m^3/s}$ for equations $j=1,3$, the gas mixture pressure is denoted as $p$  and $I_{ion}$ is the ionization rate. The total photoionization source term is then simply given by: 

 \begin{equation}
 \label{eq:photo_total}
S_{ph} = \sum_{j=1}^{3}S_{ph, j}.
\end{equation}

The six coefficients, $\lambda_{j}$ and $A_{j}$ are exponentially fitting parameters for the photon propagator function which is given as: 

 \begin{equation}
 \label{eq:photo_propagator}
{\Phi_{0} \over {p}} (pr) = (pr) \sum_{j=1}^{3}A_{j}e^{\lambda_{j}pr}
\end{equation}

The function $\mathrm{{\Phi_{0} \over {p}} (pr)}$ has been calculated for pure nitrogen in low pressure using the PHOTOPiC software~\cite{yifei2020photo,zhu2020photopic}. The fitting coefficients have been calculated by minimizing an ordinary least square cost function and are presented in Table~\ref{table:photo}. Although photoionization occurs more efficiently when small amounts of oxygen are present in the gas as UV photons mainly produced from radiative decay of nitrogen excited species ionize oxygen molecules, the mechanism is present in pure nitrogen~\cite{penney1970photoionization} also (although poorly understood and studied), and our numerical results showed that its inclusion is important for a correct representation of the FIW propagation speed. 

 \begin{table*}
\caption{ Fitting coefficients for the photoionization three-group Helmholtz model}
\label{table:photo}
\lineup
\begin{center}
\item[]\begin{tabular}{@{}*{3}{l}}
\br                        
$\0\0\textbf{j}$&$\mathbf{A_{j}}$&$\mathbf{\lambda_{j}}$\cr
\br
\mr
\0\01 & $\mathrm{4.6657\times 10^{-5}}$ & $\mathrm{0.51102}$ \cr
\mr
\0\02 & $\mathrm{3.8146\times 10^{-4}}$ & $\mathrm{0.951}$ \cr
\mr
\0\03 & $\mathrm{7.4994\times 10^{-4}}$ & $\mathrm{4.2468}$ \cr
\br
\0\0$\mathrm{A_{j}}$ in  $\mathrm{cm^{-2}Torr^{-2}}$ & \cr
\0\0$\mathrm{\lambda_{j}}$ in  $\mathrm{cm^{-1}Torr^{-1}}$ \cr
\end{tabular}
\end{center}
\end{table*}

A detailed description of several numerical schemes available in the COPAIER solver can be found in~\cite{dufour2015numerical, kourtzanidis2020self, kourtzanidis2023full}.  Here we summarize the main features used in this specific study. A finite volume approach is used to solve the drift-diffusion equations. The Scharfetter-Gummel flux scheme~\cite{scharfetter1969large} is used here to discretize the convection-diffusion operator of species continuity equations. Temporal integration for the species system of continuity equations  is performed via an implicit time-marching algorithm based on the Gauss-Seidel technique~\cite{dung2024implicit}. 

 In Eqs~(\ref{eq:energy}) and (\ref{eq:sourceenergy}), the evaluation of the energy loss source term explicitly can lead to non-physical oscillations that can damage the stability of the whole system. In order to circumvent this issue, an implicit treatment of the joule heating source terms has been adopted based on~\cite{hagelaar2000speeding}. The energy loss source term is treated fully implicitly as well as all electron-impact reaction source terms involving neutral species as reactants. The Finite Element Method has been used to solve the Poisson and Helmholtz equations, with a P1-Finite
Element formulation. A semi-implicit formulation for the Poisson equation, similar to the one proposed in~\cite{villa2013asymptotic} has been implemented, allowing to circumvent the dielectric relaxation timestep constraints. The numerical solver is parallelized based on the MPI paradigm and a non-overlapping domain decomposition method.

\subsection{Plasma chemistry model}\label{sec:chemistry}
We choose to use a reduced $\mathrm{N_{2}}$-plasma chemistry, where nine (9) modeled species are included, consisting of electrons, positive  $\mathrm{N_{2}^{+}}$ and $\mathrm{N_{4}^{+}}$  ions,  neutral $\mathrm{N_{2}}$ molecules as well as the following excited states:  N$_2$(A$^3\Sigma^+_u$),  N$_2$(B$^3\Pi_g$),  N$_2$(C$^3\Pi_u$),  $\mathrm{N_{2}(a'^1\Sigma_u)}$.
Sixteen (16) chemical reactions (in addition to photoionization) are taken into account which and listed in Table~\ref{tabfour}. Transport coefficients and reaction rates for R1-R6 and R6-R7 have been calculated using the two-term Boltzmann solver, BOLSIG+~\cite{hagelaar2005solving} including electron-electron, electron-ion and superelastic collisions and under the exponential growth (PT) model. We assume a gas composition of 100\% $\mathrm{N_{2}}$. The ionization degree is assumed $10^{-5}$, the electron density $10^{18}$ $\mathrm{m^{-3}}$ and the gas temperature 300 K. Cross-sections for $\mathrm{N_{2}}$ used as inputs in BOLSIG+ have been taken from the Phelps database~\cite{PhelpsLXCAT} and includes 26 electron-neutral scattering cross sections describing dissociative attachment, effective momentum transfer, ionization, rotational, electronic and vibrational excitation. The direct electron-impact ionization (15.6 eV) cross section dataset has been extended to high electron energies based on~\cite{murphy_2025_etc6n-fnp04, cinar2014relativistic}. To include associated electron impact energy losses, all excited (rotationally, vibrationally in different levels and electronically) species which are not tracked have been lumped to a single species, denoted as $\mathrm{N_{2}(lump)}$ which is also not tracked (assumed as $\mathrm{N_{2}}$). The rate coefficient for R15 is then calculated as follows: we equate the sum of contributions of all these untracked reactions to the inelastic power loss term, with the loss of the fictitious lumped reaction (R15). Assuming that the energy threshlold for R15 is 1 eV, we get the following formula for the rate coefficient: 

\begin{equation}
k_{lump}=\sum_{i=1}^{n}( \Delta E_{i}k_{i}),
\end{equation}

\noindent
where $\Delta E_{i}$ and $k_{i}$ are the energy threshold and rate coefficient for reaction $i$, and $n$ are the total number of excitation reactions (not tracked). N atoms are not tracked (assumed as $\mathrm{0.5~N_{2}}$ in R8) as well as vibrationally excited $\mathrm{N_{2}}$ states (assumed in ground state of $\mathrm{N_{2}}$). This reduced chemistry set is of course rather simplified but provides a good balance between accuracy and required computational time, allowing us to capture the main features of the FIW dynamics, while stabilizing the numerical algorithm under the LEA scheme, by incorporating the main electron energy loss mechanisms.

 Reduced ion mobilities for $\mathrm{N_{2}^{+}}$ and $\mathrm{N_{4}^{+}}$ (in $\mathrm{N_{2}}$) have been taken from LxCAT swarm data (Phelps and Viehland databases~\cite{PhelpsLXCAT, viehland1995relating}). Constant (reduced) diffusion coefficient of $\mathrm{1.98 \times  10^{-5}}$ $\mathrm{m^{2}/s}$ was assumed for all neutral species. We neglect gas flow dynamics as their timescales are much larger than the simulation time-scales. The gas temperature ($T_{g}$) is assumed constant at 300 K. This is a good approximation on the nanosecond FIW propagation timescales, as (i) we mainly consider the FIW front propagation, at this time scale the heating can be neglected; (ii) the specific energy delivered to plasma is low. 
 

Electrons and positive $\mathrm{N_{2}^{+}}$ and $\mathrm{N_{4}^{+}}$  ions are initialized with a number density of $2\times 10^{14}$ $\mathrm{m^{-3}}$  and $10^{14}$ $\mathrm{m^{-3}}$ respectively, rest of neutral species with a density of $10^{3}$ $\mathrm{m^{-3}}$. This roughly corresponds to levels of residual charges from the repetitive nanosecond pulses of 10 Hz.  

A floor density of  $10^{9}$ $\mathrm{m^{-3}}$ is used for all charged species to emulate the effects of cosmic rays induced, background ionization. These values also correspond to the floor density for each species respectively. The secondary electron emission (SEE) coefficient for the electrode boundary is set to $10^{-3}$ while for the dielectric rods and barrier to 1$\times$$10^{-2}$. The influence of these parameters on the discharge evolution falls outside the scope of this work.

 \begin{table*}
\caption{ Chemical reaction model used in the simulations.}
\label{tabfour}
\lineup
\begin{center}
\item[]\begin{tabular}{@{}*{3}{l}}
\br                        
$\0\0\textbf{Reaction}$&\textbf{Rate coefficient}&\textbf{Comment - Source}\cr
\mr
\0\0R1: e +  $\mathrm{N_{2}}$ $\rightarrow$ $\mathrm{N_{2}^{+}}$ + 2e &f(E) & $\mathrm{\Delta E_{i}=15.6}$ eV, Ref.~\cite{PhelpsLXCAT}$ ^{a}$\cr
\mr
\0\0R2: e +  $\mathrm{N_{2}}$ $\rightarrow$ $\mathrm{N_{2}^{+}}$ + 2e &f(E) & $\mathrm{\Delta E_{i}=18.8}$ eV \\
\0\0 & & N$_2^+$(B$^2\Sigma^+_u$) excitation, Ref.~\cite{PhelpsLXCAT}$ ^{a}$\cr
\mr
\0\0R3: e +  $\mathrm{N_{2}}$ $\rightarrow$ $\mathrm{N_{2}(A)}$ + e &f(E) & v0-4, Ref.~\cite{PhelpsLXCAT}$ ^{a}$\cr
\mr
\0\0R4: e +  $\mathrm{N_{2}}$ $\rightarrow$ $\mathrm{N_{2}(B)}$ + e &f(E) & Ref.~\cite{PhelpsLXCAT}$ ^{a}$\cr
\mr
\0\0R5: e +  $\mathrm{N_{2}}$ $\rightarrow$ $\mathrm{N_{2}(C)}$ + e &f(E) & Ref.~\cite{PhelpsLXCAT}$ ^{a}$\cr
\mr
\0\0R6: e +  $\mathrm{N_{2}}$ $\rightarrow$ $\mathrm{N_{2}(a')}$ + e &f(E) & Ref.~\cite{PhelpsLXCAT}$ ^{a}$\cr
\mr
\0\0R7: e +  $\mathrm{N_{4}^{+}}$ $\rightarrow$ $\mathrm{N_{2}}$ +  $\mathrm{N_{2}}$  & $\mathrm{2\times 10^{-6}(300/T_{e})^{0.5}}$ & Ref.~\cite{kossyi1992kinetic}$ ^{b}$\cr
\mr
\0\0R8: e +  $\mathrm{N_{2}^{+}}$ $\rightarrow$ $\mathrm{N}$ +  $\mathrm{N(2D)}$  &$\mathrm{2\times 10^{-7}(300/T_{e})^{0.5}}$& N atoms not tracked \\ 
\0\0 & & (assumed  $\mathrm{0.5~N_{2}}$) , Ref.~\cite{kossyi1992kinetic}$ ^{a}$\cr
\mr
\0\0R9: $\mathrm{N_{2}(C)}$ $\rightarrow$ $\mathrm{N_{2}(B)}$ +  $\mathrm{hv}$  & $\mathrm{2.45\times 10^{7}}$ & Ref.~\cite{capitelli2013plasma}$ ^{d}$\cr
\mr
\0\0R10: $\mathrm{N_{2}(C)}$ + $\mathrm{N_{2}}$ $\rightarrow$ $\mathrm{N_{2}(a')}$ +  $\mathrm{N_{2}(v)}$  & $\mathrm{1\times 10^{-11}}$ &  Vib. state not tracked \\ 
\0\0 & & (assumed  $\mathrm{N_{2}}$) , Ref.~\cite{kossyi1992kinetic}$ ^{b}$\cr
\mr
\0\0R11: $\mathrm{N_{2}(B)}$ + $\mathrm{N_{2}}$ $\rightarrow$ $\mathrm{N_{2}(A)}$ +  $\mathrm{N_{2}(v)}$  &  $\mathrm{1\times 10^{-11}}$  &Vib. state not tracked \\ 
\0\0 & & (assumed  $\mathrm{N_{2}}$), Ref.~\cite{kossyi1992kinetic}$ ^{b}$\cr
\mr
\0\0R12: $\mathrm{N_{2}(a')}$ + $\mathrm{N_{2}}$ $\rightarrow$ $\mathrm{N_{2}(v)}$ +  $\mathrm{N_{2}(v)}$  & $\mathrm{2\times 10^{-13}}$  &Vib. state not tracked \\ 
\0\0 & & (assumed  $\mathrm{N_{2}}$), Ref.~\cite{kossyi1992kinetic}$ ^{b}$\cr
\mr
\0\0R13: $\mathrm{N_{2}(A)}$ + $\mathrm{N_{2}(A)}$ $\rightarrow$ $\mathrm{N_{2}(C)}$ +  $\mathrm{N_{2}(v)}$  & $\mathrm{1.6\times 10^{-10}}$ &Vib. state not tracked \\ 
\0\0 & & (assumed  $\mathrm{N_{2}}$), Ref.~\cite{kossyi1992kinetic}$ ^{b}$\cr
\mr
\0\0R14: $\mathrm{N_{2}(A)}$ + $\mathrm{N_{2}(A)}$ $\rightarrow$ $\mathrm{N_{2}(B)}$ +  $\mathrm{N_{2}(v)}$  &  $\mathrm{7.7\times 10^{-11}}$ &Vib. state not tracked \\ 
\0\0 & & (assumed  $\mathrm{N_{2}}$), Ref.~\cite{kossyi1992kinetic}$ ^{b}$\cr
\mr
\0\0R15: e +  $\mathrm{N_{2}}$ $\rightarrow$ $\mathrm{N_{2}}$ + e &f(E) & Lumped state - see text, \\
\0\0 & &  Ref.~\cite{PhelpsLXCAT}$ ^{a}$\cr
\mr
\0\0R16:  $\mathrm{N_{2}^{+}}$ +  $\mathrm{2N_{2}}$ $\rightarrow$  $\mathrm{N_{4}^{+}}$ +   $\mathrm{N_{2}}$ &$\mathrm{5\times 10^{-29}}$ & Ref.~\cite{kossyi1992kinetic}$ ^{c}$\cr
\br
\0\0$ ^{a}$  Bolsig+.& \cr
\0\0$ ^{b}$  Units of $\mathrm{cm^{3}/s}$. $\mathrm{T_{e}}$ in K. \cr
\0\0$ ^{c}$  Units of $\mathrm{cm^{6}/s}$.  \cr
\0\0$ ^{d}$  Units of $\mathrm{1/s}$. \cr

\end{tabular}
\end{center}
\end{table*}

\section{The case under study, computational and initial parameters}\label{sec:case}

\begin{figure}[h]
\centering
\includegraphics[width=1\linewidth]{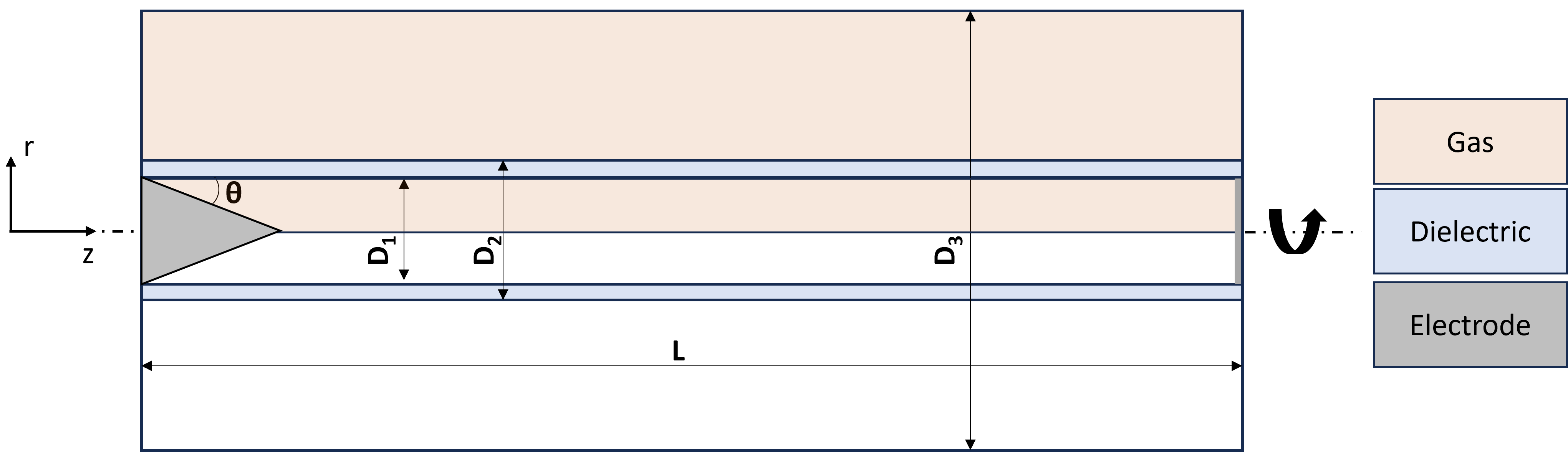}
  \caption{Illustrative graph of the discharge tube. The axisymmetric computational (CPU) domain corresponds to the upper half. Dimensions not in scale.}
    \label{fig:domain}
 \end{figure}

The computationally domain mimics the experimental setup: a 60~cm long quartz tube is used with inner diameter, 17.5~mm, outer diameter, 21.5~mm, with metallic electrodes at the ends of the tube. The high-voltage (HV) electrode has a  25~mm long conical
shape with a cone angle of 30 degree. The low-voltage (grounded) electrode is plate-shaped. Thus, while the inter-electrode gap, $L$, equals the tube's length (60 cm), the axial distance between the HV electrode's apex and the ground electrode is approx. 56.7 cm. The tube is surrounded by a cylindrical, grounded, metallic shield of inner
diameter 60~mm. Fig.~\ref{fig:domain} presents an illustration of the computational domain and relevant dimensions. 

In the experiments, a pulse of 13.5~kV amplitude, 25~ns duration at half width and 6~ns rise time was supplied from a high-voltage generator with a 40~Hz repetition frequency. In the simulations, we apply a similar waveform (one pulse) but with amplitude of 27 kV to emulate the constructive interference between the forward and backward propagating pulses.  In Table~\ref{tabtwo}, we summarize the geometrical and operational conditions for the simulations (see also Fig.~\ref{fig:domain}).
 
\begin{table}
\caption{Geometrical and operational conditions for the simulations.} 
\label{tabtwo}
\begin{indented}
\lineup
\item[]\begin{tabular}{@{}*{2}{l}}
\br                              
$\0\0\textbf{HV electrode half-angle}$&$\theta$ = $15$  degrees\cr
\mr
$\0\0\textbf{Discharge tube internal diameter}$&$D_{1}$ = $17.6$ mm\cr
\mr
$\0\0\textbf{Discharge tube external diameter}$&$D_{2}$ = $21.6$ mm\cr
\mr
$\0\0\textbf{Shielding cage diameter}$&$D_{3}$ = $60$ mm\cr
\mr
$\0\0\textbf{Inter-electrode gap}$&$L$ = $60$ cm\cr
\mr
$\0\0\textbf{Relative permitivity of dielectric}$& $3.2$\cr
\br
\end{tabular}
\end{indented}
\end{table}

\begin{figure}[h]
\centering
\includegraphics[width=0.75\linewidth]{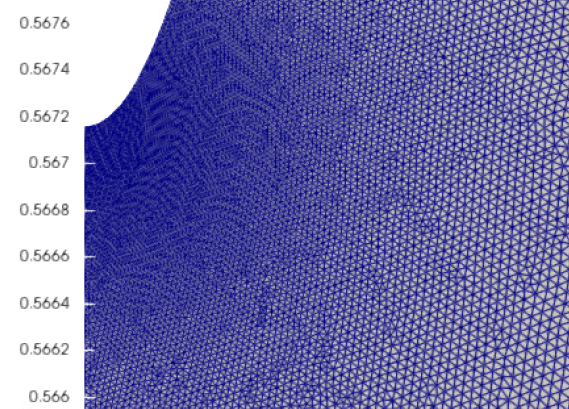}
  \caption{Unstructured mesh refinement near the HV electrode tip. Similar refined regions are used near the dielectric surface and both electrode surfaces.}
    \label{fig:mesh}
 \end{figure}

The computational mesh is an unstructured one, created in the Gmsh~\cite{geuzaine2020three} software, properly refined near the electrodes and near the dielectric surface. It consists of 589.445 triangular cells. The minimum cell size is approximately 10~$\mu$m at the ``boundary layer'' zone near the electrodes and adjacent to the dielectric surface in order to resolve all plasma-related aspects such as sheaths and cathode layer formation. The computational mesh at the region around the HV electrode tip is shown in Fig.~\ref{fig:mesh}. The domain has been decomposed in 32 sub-domains and the calculations were run on a local HPC workstation (Dell PowerEdge R7515) over 32 computational units (AMD EPYC 7352), leveraging MPI-parallelism.

 \section{Results and discussion}\label{sec:results}

\begin{figure}[h]
\centering
 \begin{subfigure}{.5\textwidth}
 \centering
\includegraphics[width=1\linewidth]{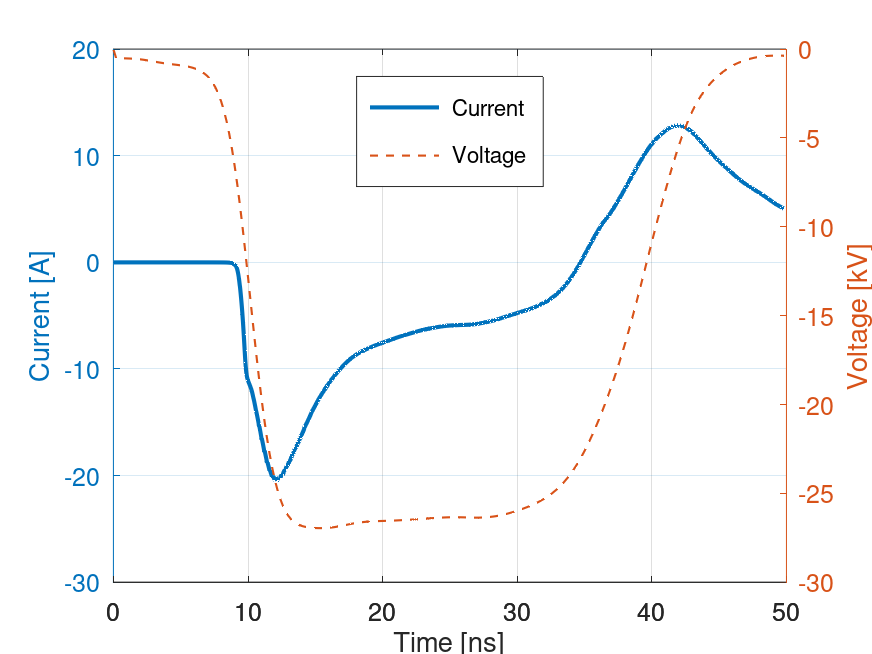}
\caption{ Negative Polarity  }
 \end{subfigure}
  \begin{subfigure}{.5\textwidth}
  \centering
\includegraphics[width=1\linewidth]{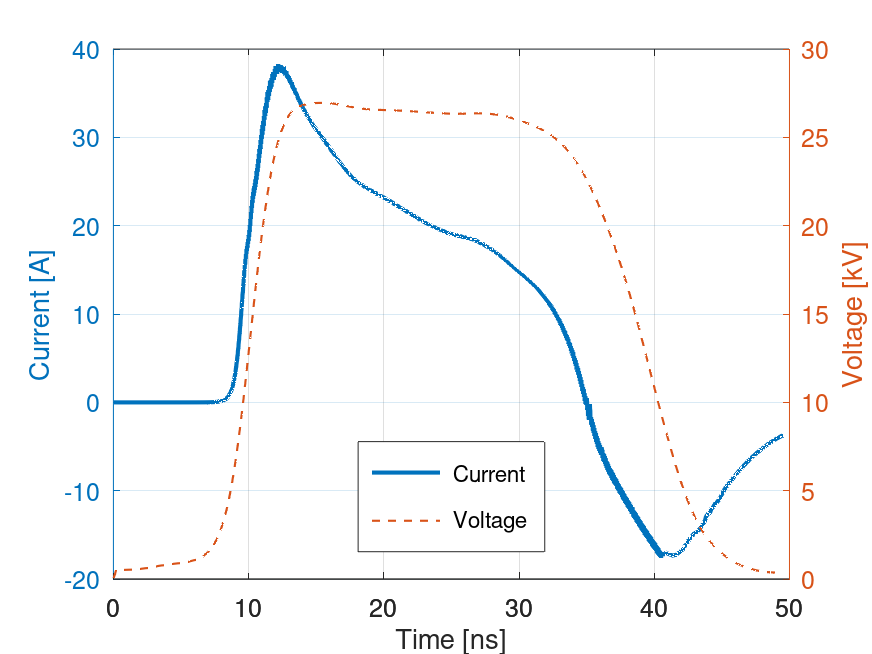}
\caption{ Positive Polarity }
 \end{subfigure}
  \caption{Experimental voltage waveforms (orange dashed curves), used as applied voltage in the numerical solver; calculated current through the discharge cell, (solid blue curves). Measurements are made by capacitive probe; the velocity is calculated by the 0.5~level from the amplitude of the signal. Nitrogen, pressure 10 Torr.}
    \label{fig:IV_-+}
 \end{figure}

Fig.~\ref{fig:IV_-+} presents an example of the  synchronized voltage and calculated current waveforms for nitrogen, 10 Torr pressure. The voltage waveform is the initial data of the problem, while the current waveform is calculated during the simulations. Different shapes of the current are observed for negative and positive polarity: the current behaviour at negative polarity is similar to the surface discharge, when two peaks of different polarity are observed at rising front and falling edge of the high voltage pulse; the current at positive polarity is high after the front. Typical values for current for both polarities are tens of A. 

\subsection{FIW velocity}
 

\begin{figure}[h]
\centering
\includegraphics[width=\linewidth]{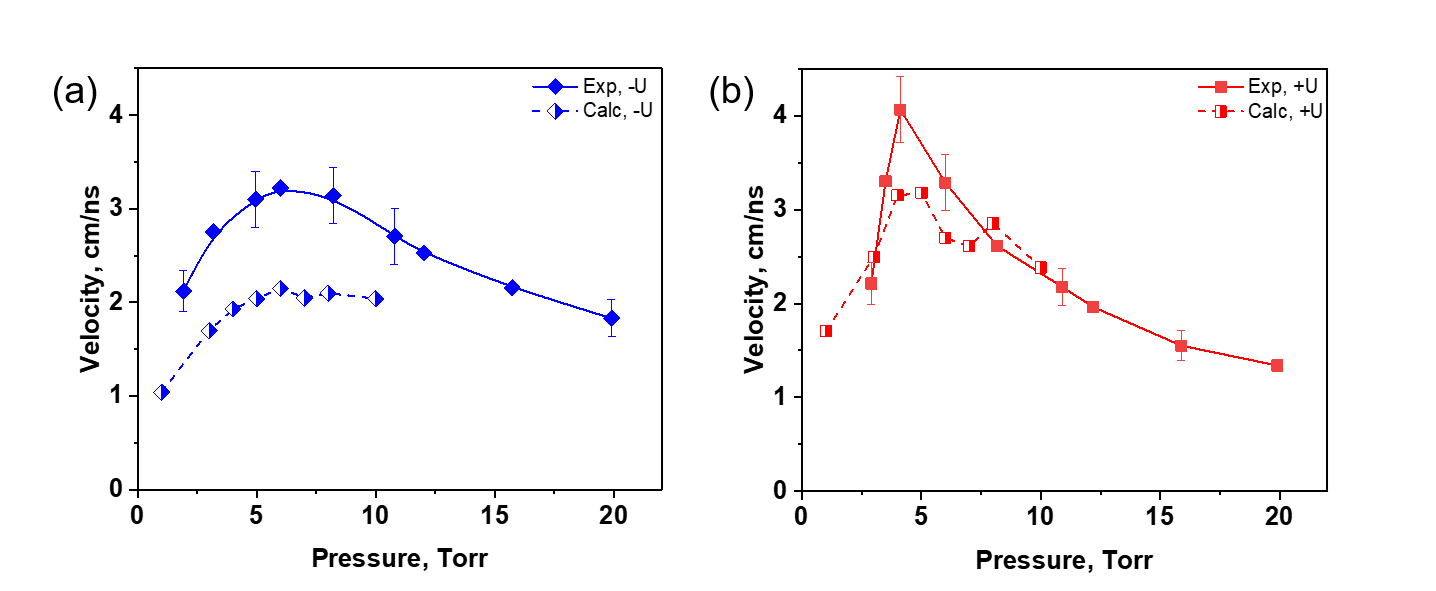}
  \caption{Experimental \protect{\cite{AnikinPhD}} (filled symbols) and calculated (half-filled symbols) velocities of the fast ionization wave (FIW) propagation. Nitrogen, (a) negative polarity; (b) positive polarity.}
    \label{Minus_plus_velocity}
 \end{figure}

Velocity, measured as a gradient of the electric charge along the discharge tube \cite{AnikinPhD} and maximum velocity calculated from the propagating front and the position of maximum total charge (in absolute value), are presented together for negative and positive polarities in Fig.~\ref{Minus_plus_velocity}. Increase of the FIW velocity up to a few Torr and then decrease is confirmed by results of the modeling. A reasonable agreement between the calculated and the measured absolute values of the velocity for negative polarity and a quantitative agreement for positive polarity are clearly seen. Typical values of the FIW propagation speed along the tube are $2-4$~cm per nanosecond. 

We note here that (i) at negative polarity, a double structure of the ionization wave, with a lower amplitude (``a precursor'') has been observed \cite{Anikin2002}; (ii) the calculated values have been extracted only from the saved output data from the simulations and not at every time step (to reduce the generated data and due to storage limitations) and as such, the maximum velocities reported here might deviate slightly from the reality. This fact also can explain the small jump observed at 8~Torr under positive polarity. 

In any case, the observed trends and orders of magnitude are quite satisfactory when compared with the experimental findings. We also need to emphasize that the calculated velocities represent the maximum velocity of the FIW: under most of the conditions studied here, the calculated FIW velocity quickly increases to the maximum value and then slowly decays during its propagation, thus the velocity is non-constant.

The optimum pressure for development of a nanosecond discharge for given amplitude of the high-voltage pulse, nitrogen pressure and the discharge tube diameter is about 5~Torr. The trends observed in the numerical results (which are in good agreement with the experimental data) suggest a rather complex dependence of the propagation speed on the plasma structure and the discharge main parameters (electric field, plasma density etc). As we will see in the following sections, under different polarities and different background pressures, the plasma development and properties differ significantly which renders a straight-forward explanation of the FIW speed behavior difficult to obtain. Nevertheless, we can extract some interesting information based on the theoretical work~\cite{sinkevich2000propagation}. Therein, the authors provide analytical self-similar solutions of the \color{black} (averaged over the cross-section) \color{black} problem of high-speed ionizing wave (FIW) propagation in long shielded tubes. A rather simple equation is then provided to describe the FIW speed: 

\begin{equation}
\label{eq:analytical_eq}
v_{FIW}={{\phi_{f}} \over {E_{0}}} {{\nu(T_{e})} \over {e}}
\label{Sinkevich_speed}
\end{equation}

\begin{figure}[h]
\centering
\includegraphics[width=0.7\linewidth]{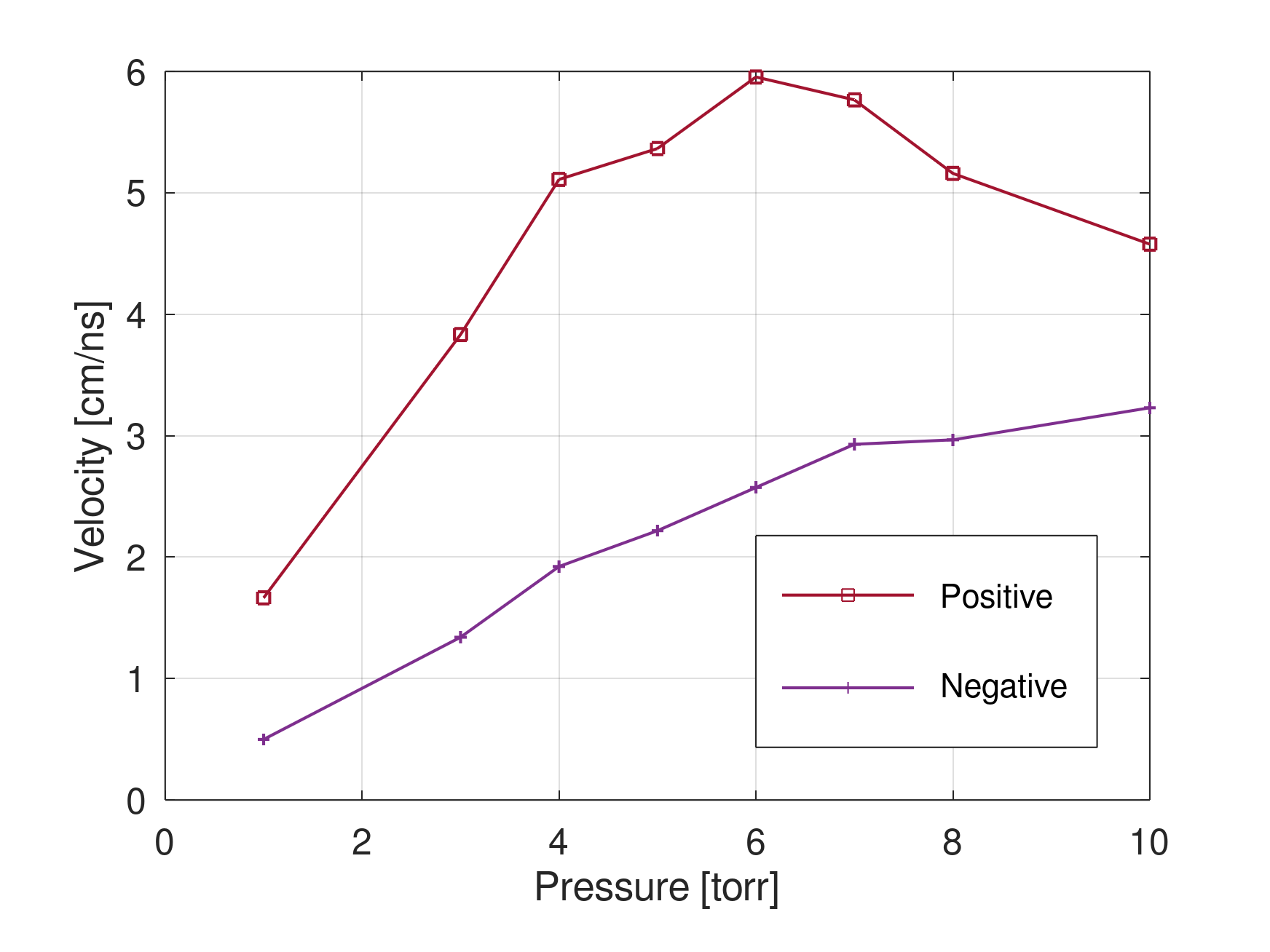}
  \caption{Analytical velocities (Eq. \ref{eq:analytical_eq}) of the fast ionization wave (FIW) for positive and negative polarity versus pressure.}
    \label{fig:analytical}
 \end{figure}

\noindent
In the above formula, ${\phi_{f}}$, $E_{0}$ and $\nu(T_{e})$ are the electrical potential, the maximum of the electric field and the ionization frequency at the FIW front, respectively, while $e$ is the base of the natural logarithm. Note that we have used the electron temperature dependence of the ionization frequency, $\nu(T_{e})$, due to our LEA modeling approach. All these parameters are direct outputs of our simulations and can be used  to plot the analytical dependence of the FIW speed over pressure for both polarities. This is shown in Fig.~\ref{fig:analytical}, which demonstrates a good qualitative agreement with the calculated values from the simulations. 

\color{black}
Our calculations (see following sections for details) show that, for a negative polarity FIW, the potential ${\phi_{f}}$ and the maximum electric field at the front, $E_{0}$, increase monotonically with pressure by nearly an order of magnitude, while their ratio remains constant. Consequently, the speed of the FIW front is approximately proportional to the ionization frequency $\nu(T_{e})$. Will note that, despite their increase, the potential and electric field remain approximately half of the corresponding values observed for positive polarity. This behavior reflects the fact that, under a negative polarity pulse, electrons propagate ahead of the FIW front region. With this pre-ionization, the FIW can propagate even in a relatively low field.

For positive polarity FIW, the electrons more backward, to the bulk of plasma. Both the potential ${\phi_{f}}$ and maximum electric field at the front $E_{0}$ remain almost constant over the studied pressure range. Electric field being constant, the reduced electric field $E/N$ drops by an order of magnitude between 1 and 10 Torr. In the calculations, a sharp drop of $T_e$ is observed when pressure increases from 1 to 10~Torr. As a result, at low pressures, very high $T_e$ and low collision frequencies the ionization efficiency, the electron density and the FIW speed are low, the $E/N$ values are higher than the most efficient ones for the ionization. At high pressures, on the right decaying branch of $v_{FIW}$, the $E/N$ values are lower than the most efficient ones for the ionization, and the maximum value of $v_{FIW}$ is reached in between, corresponding to maximum ionization. 

For a positive polarity FIW, electrons move backward, to the plasma bulk. Both the potential ${\phi_{f}}$ and the maximum electric field at the front, $E_{0}$, remain nearly constant over the studied pressure range. With the electric field being constant, the reduced electric field $E/N$ decreases by an order of magnitude between 1 and 10 Torr. The calculations reveal a sharp drop in $T_e$ with pressure increase. Consequently, at low pressures, ionization efficiency, electron density, and FIW speed remain low, and $E/N$ exceeds the optimal values for ionization. At high pressures, on the right decaying branch of $v_{FIW}$, $E/N$ falls below the optimal one for ionization, with the maximum value of $v_{FIW}$ occurring at an intermediate pressure.

So for both polarities of the high-voltage pulse, the main parameter influencing the FIW propagation speed according to formula~(\ref{Sinkevich_speed}) is the ionization frequency $\nu(T_{e})$ at the FIW front. Discrepancies in the absolute values of the FIW speed between the analytical formula and results of calculations (see Fig.~\ref{Minus_plus_velocity}) can be attributed to the choice of values at the FIW front and limitations of the self-similar analytical approach. 

The electron density as a function of pressure exhibits a maximum at approximately the same pressure as the FIW front speed. On the right decaying branch of $v_{FIW}$, the conditions are optimal for dissociation through the excitation of molecular electronic levels, with the dissociation degree reaching its maximum at a higher pressure than the FIW speed maximum. Experimental data on electron density measured via microwave interferometry \cite{Aleksandrov2007} and O$_3$ production measured via UV absorption \cite{Starikovskaia1995} strongly support this reasoning regarding the dependence of discharge parameters on pressure and the relative positions of the maxima.
\color{black}

To trace the most characteristic interval of the FIW propagation and retrieve information on  discharge aspects that influence the FIW speed for both polarities, simulation results are presented below for 1, 5 and 10~Torr.



  \begin{figure}[h]
\centering
 \begin{subfigure}{.51\textwidth}
 \centering
\includegraphics[width=1\linewidth]{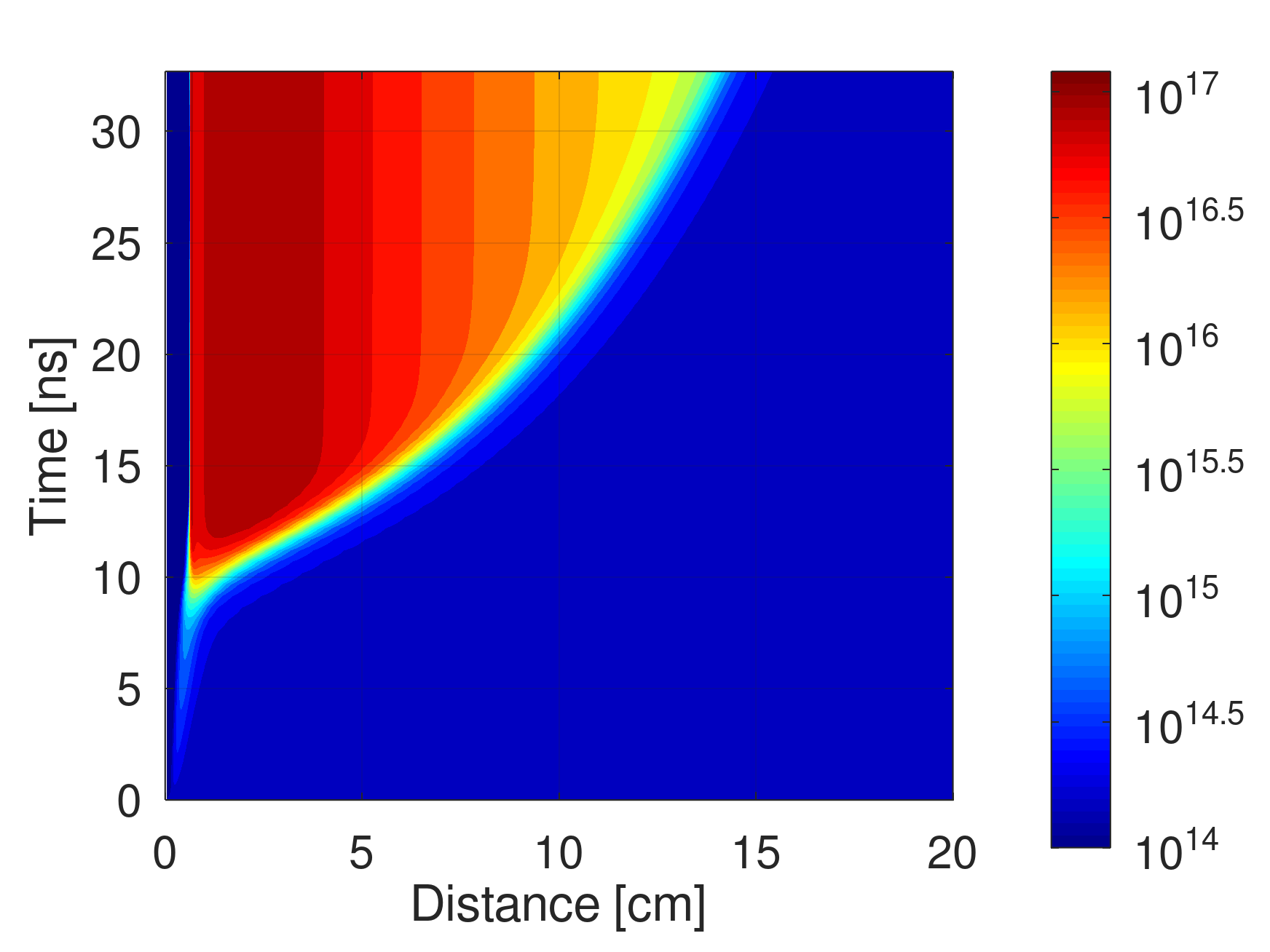}
\caption{ Electron density[$\mathrm{m^{-3}}$]   }
 \end{subfigure}
  \begin{subfigure}{.51\textwidth}
  \centering
\includegraphics[width=1\linewidth]{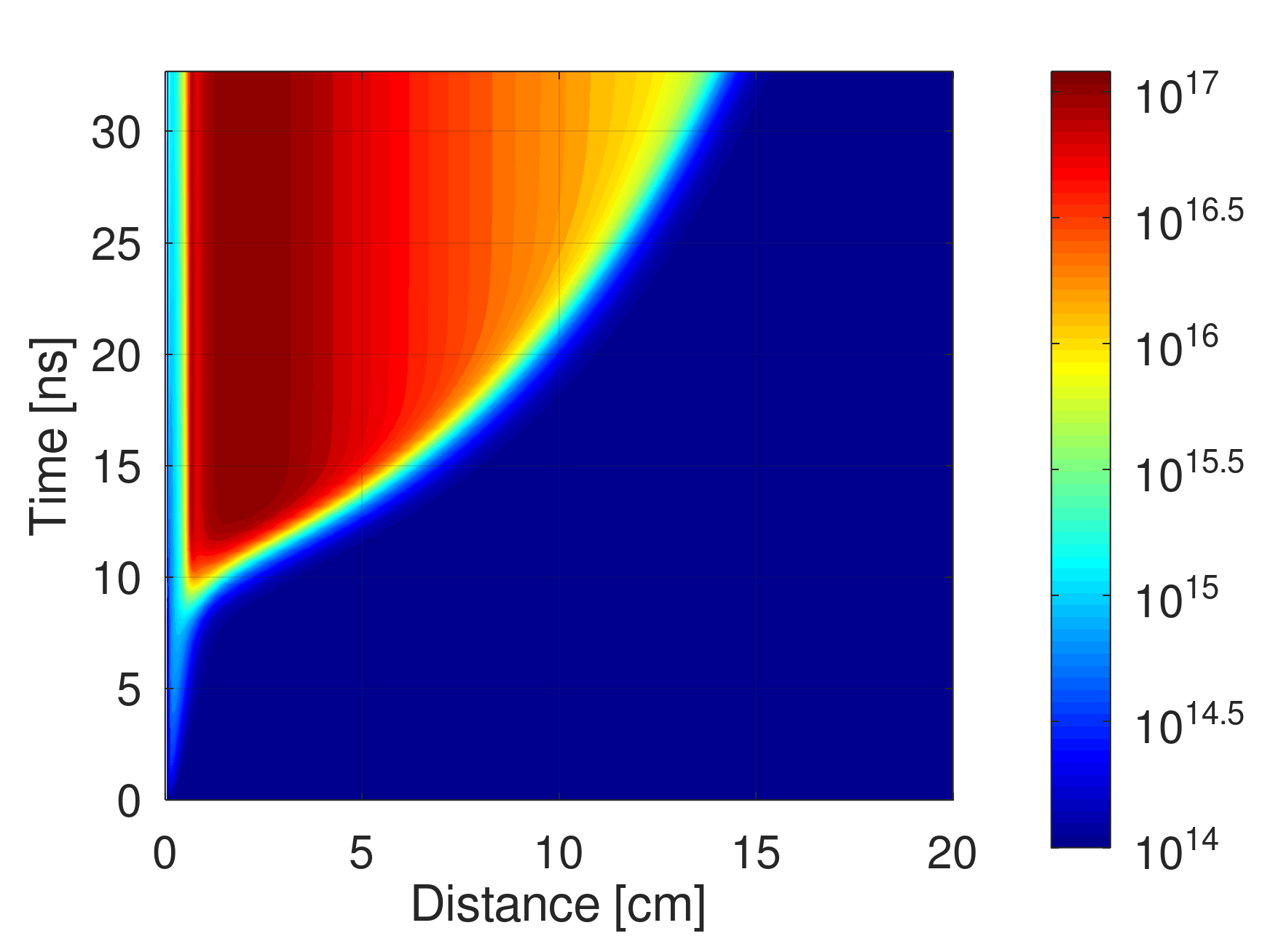}
\caption{ N2+ density[$\mathrm{m^{-3}}$] }
 \end{subfigure}
   \begin{subfigure}{.51\textwidth}
  \centering
\includegraphics[width=1\linewidth]{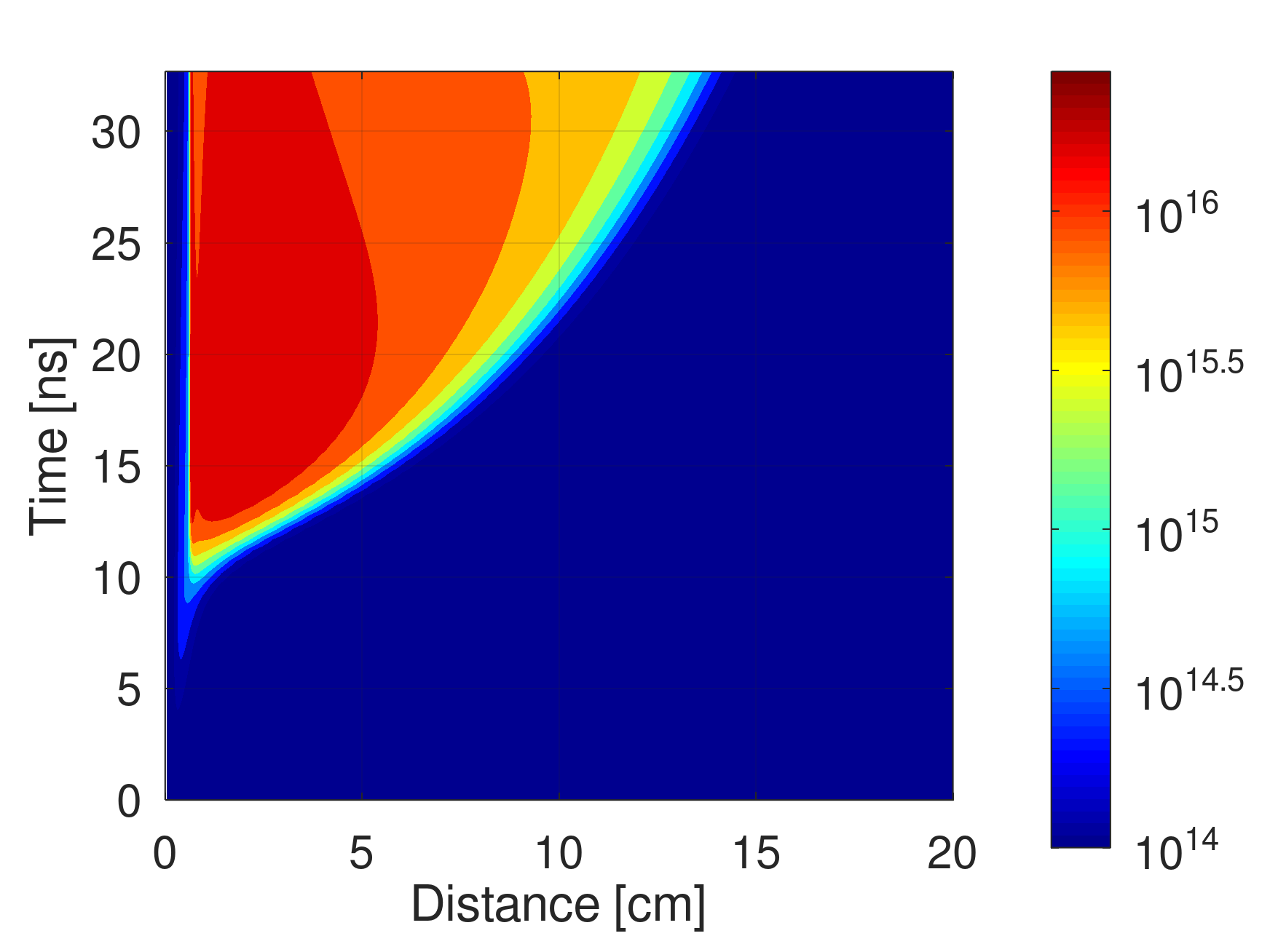}
\caption{ N2C density[$\mathrm{m^{-3}}$]  }
 \end{subfigure}%
  \begin{subfigure}{.51\textwidth}
  \centering
\includegraphics[width=1\linewidth]{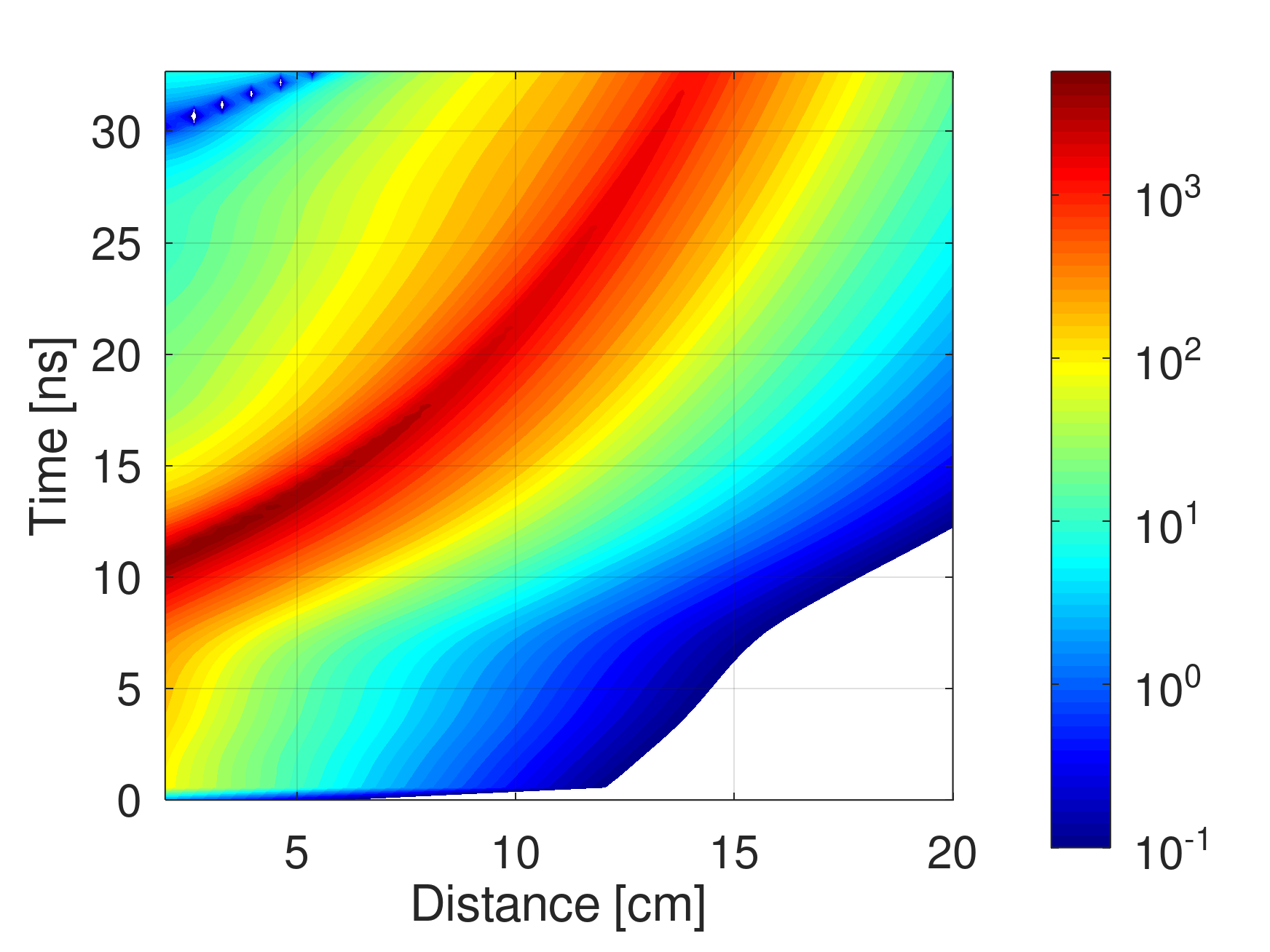}
\caption{ $\mathrm{|E_z|/N_{g}}$ [Td]  }
 \end{subfigure}
   \begin{subfigure}{.51\textwidth}
  \centering
\includegraphics[width=1\linewidth]{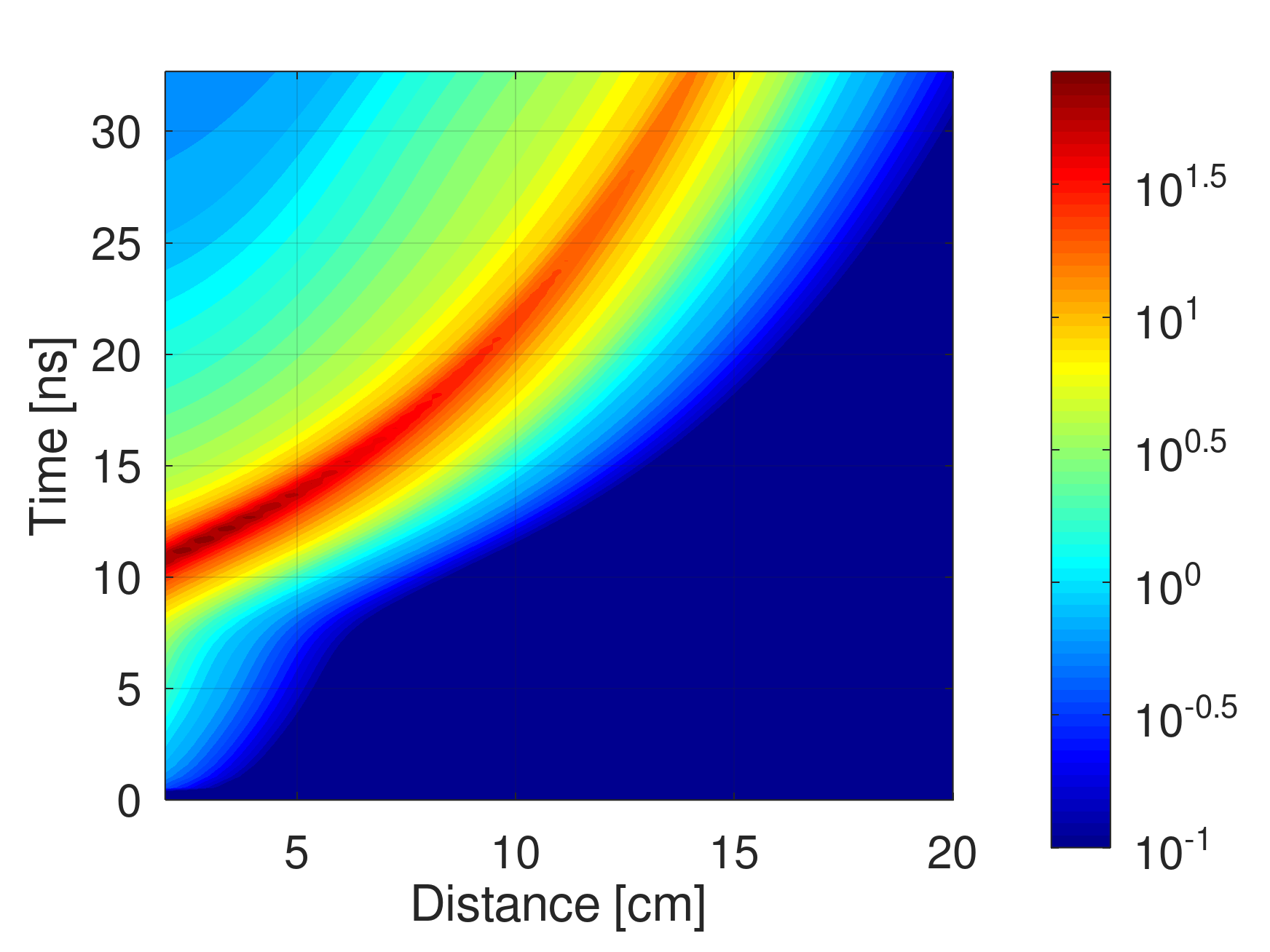}
\caption{ $\mathrm{T_{e}}$ [eV]  }
 \end{subfigure}
  \caption{Negative polarity at 1 Torr: Axial distance vs time contour plot of electron,  N$_2^+$,  N$_2$(C$^3\Pi_u$) species density [$\mathrm{m^{-3}}$], axial component of reduced electric field magnitude [Td] and electron temperature [eV] at r$=$0.  All plots in log-scale. }
    \label{fig:1torr_-_xt}
 \end{figure}

     \begin{figure}[h]
\centering
 \begin{subfigure}{.51\textwidth}
 \centering
\includegraphics[width=1\linewidth]{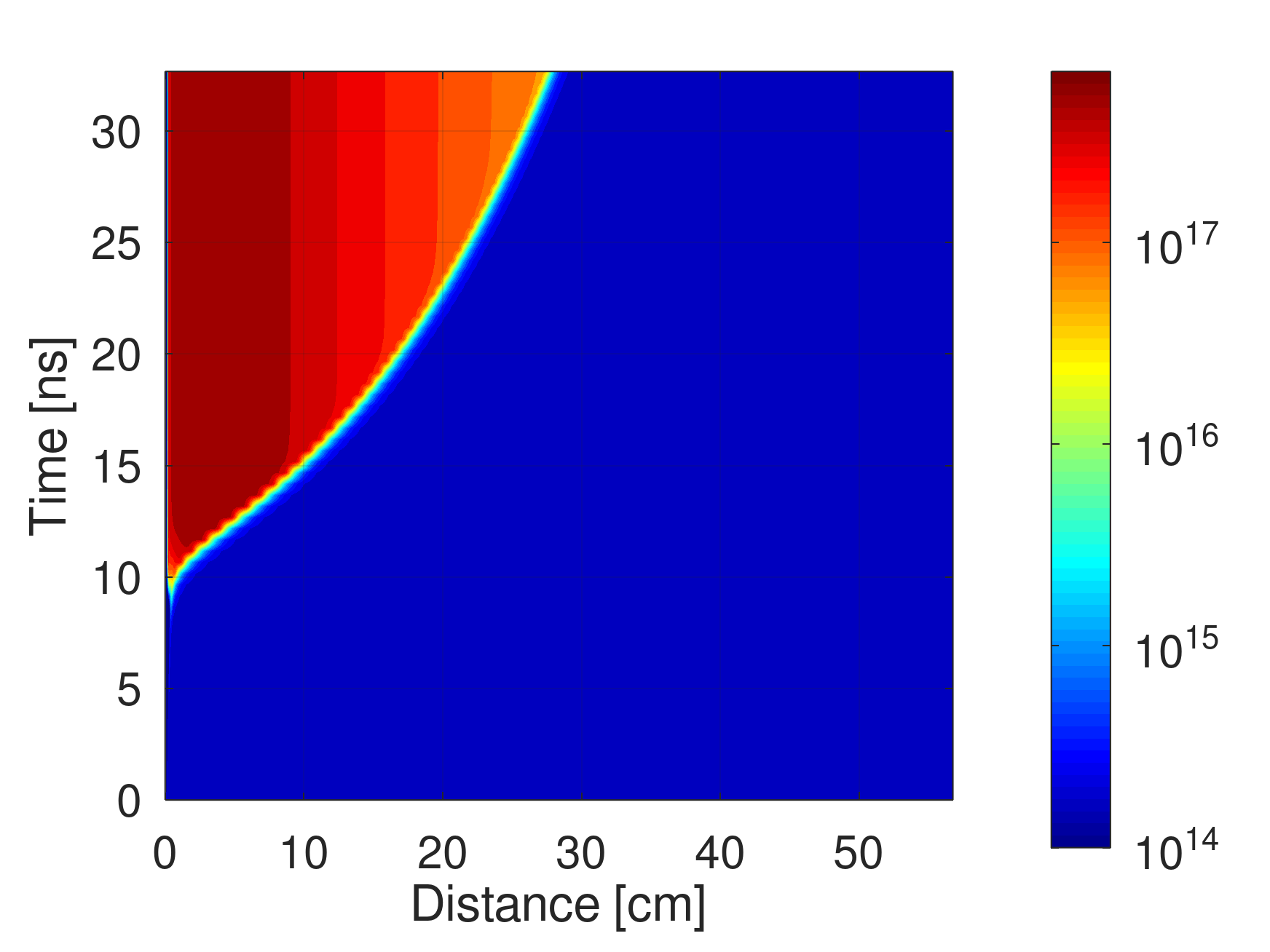}
\caption{ Electron density[$\mathrm{m^{-3}}$]   }
 \end{subfigure}
  \begin{subfigure}{.51\textwidth}
  \centering
\includegraphics[width=1\linewidth]{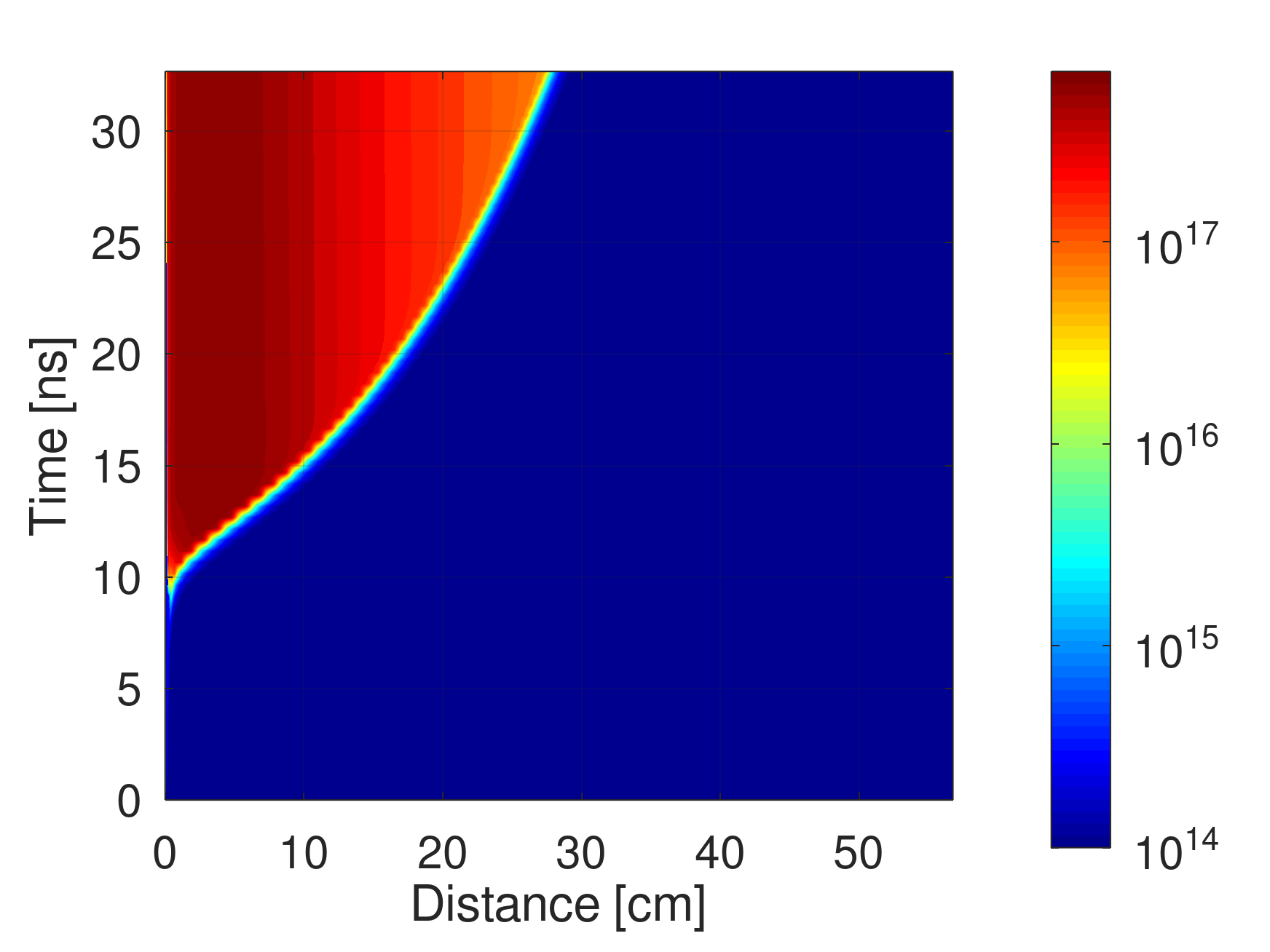}
\caption{ N2+ density[$\mathrm{m^{-3}}$] }
 \end{subfigure}
   \begin{subfigure}{.51\textwidth}
  \centering
\includegraphics[width=1\linewidth]{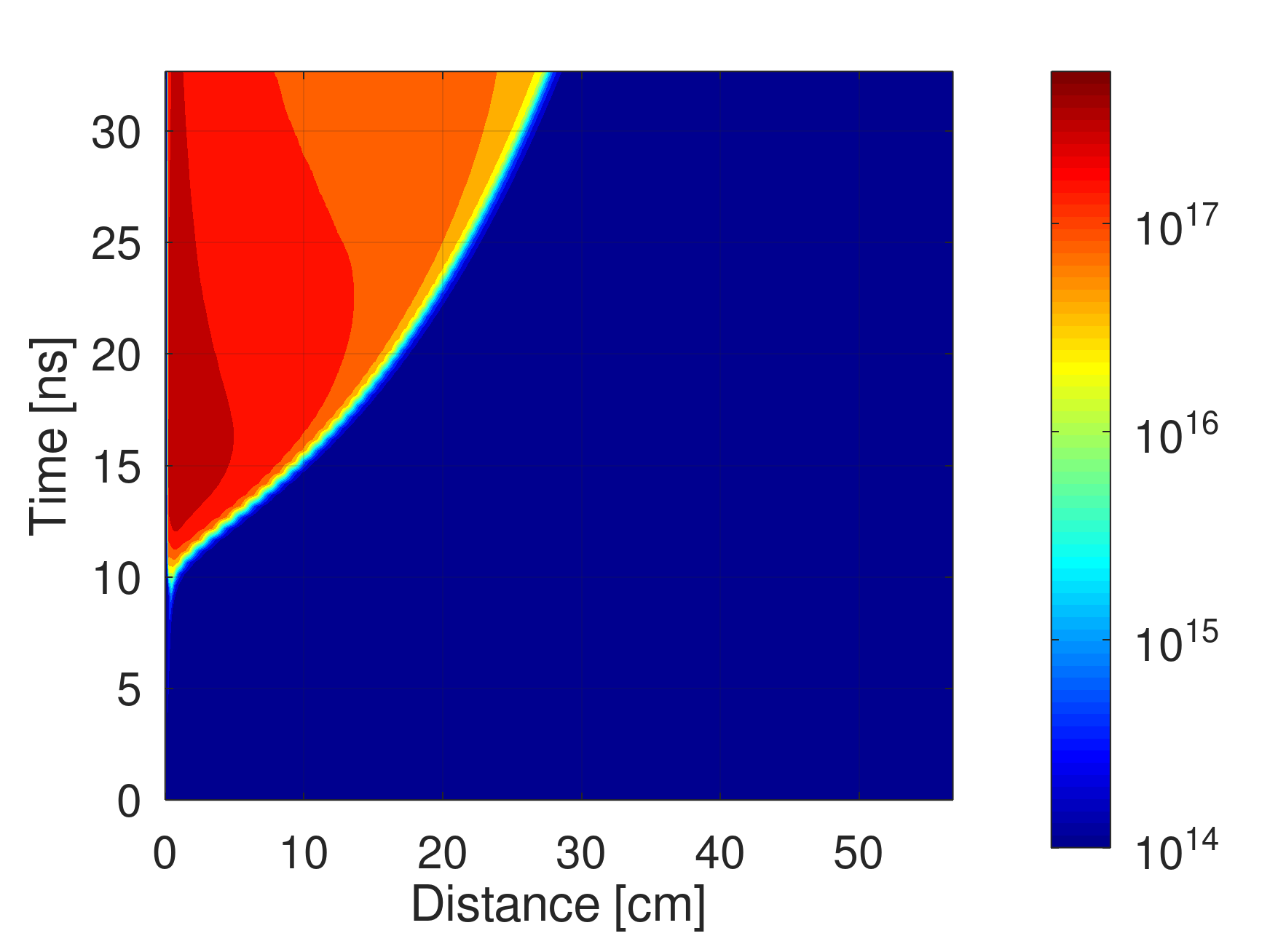}
\caption{ N2C density[$\mathrm{m^{-3}}$]  }
 \end{subfigure}%
  \begin{subfigure}{.51\textwidth}
  \centering
\includegraphics[width=1\linewidth]{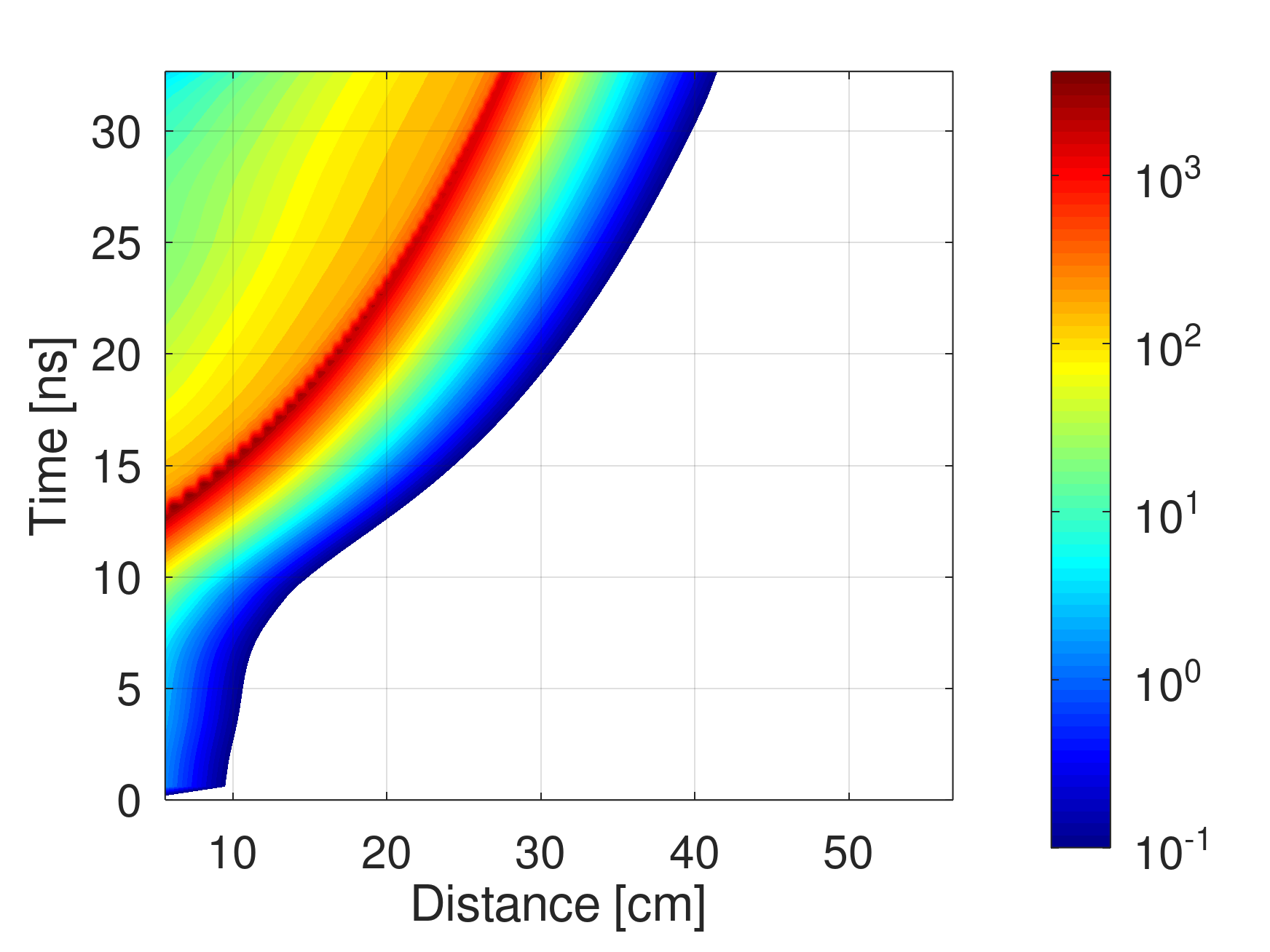}
\caption{ $\mathrm{|E_z|/N_{g}}$ [Td]  }
 \end{subfigure}
   \begin{subfigure}{.51\textwidth}
  \centering
\includegraphics[width=1\linewidth]{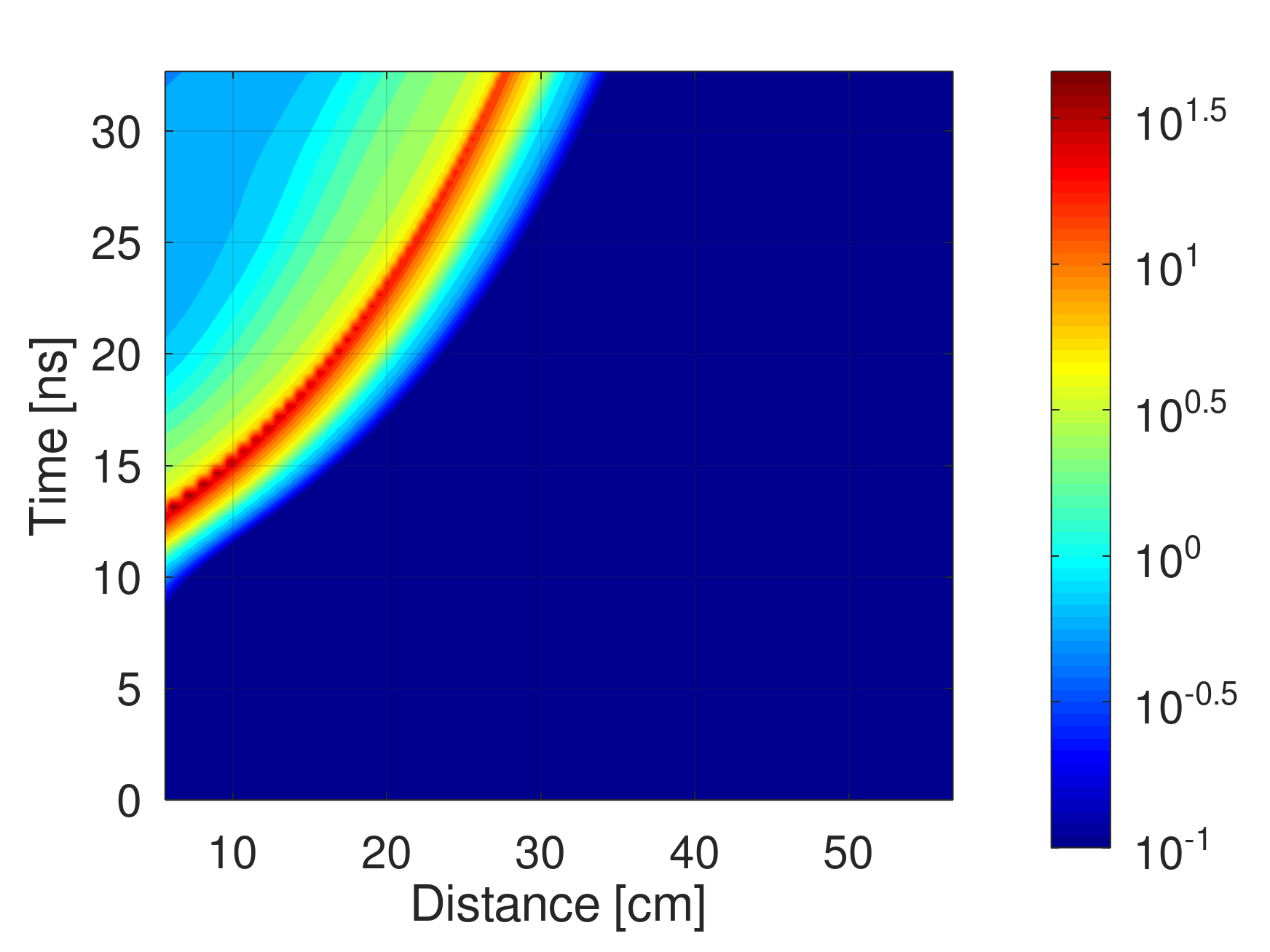}
\caption{ $\mathrm{T_{e}}$ [eV]  }
 \end{subfigure}
  \caption{Negative polarity at 5 Torr: Axial distance vs time contour plot of electron,  N$_2^+$,  N$_2$(C$^3\Pi_u$) species density [$\mathrm{m^{-3}}$], axial component of reduced electric field magnitude [Td] and electron temperature [eV] at r$=$0.  All plots in log-scale. }
    \label{fig:5torr_-_xt}
 \end{figure}

     \begin{figure}[h]
\centering
 \begin{subfigure}{.51\textwidth}
 \centering
\includegraphics[width=1\linewidth]{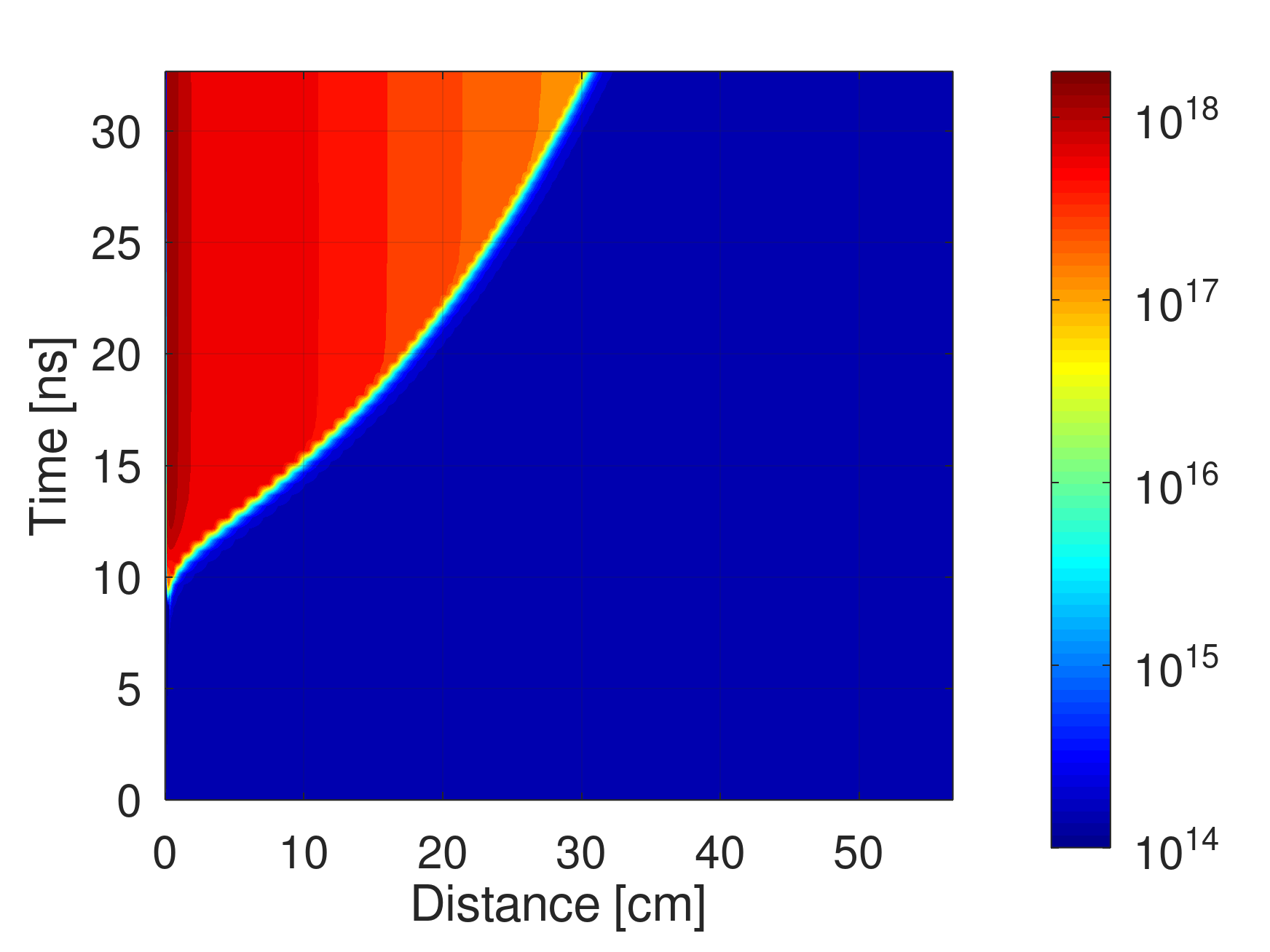}
\caption{ Electron density[$\mathrm{m^{-3}}$]   }
 \end{subfigure}
  \begin{subfigure}{.51\textwidth}
  \centering
\includegraphics[width=1\linewidth]{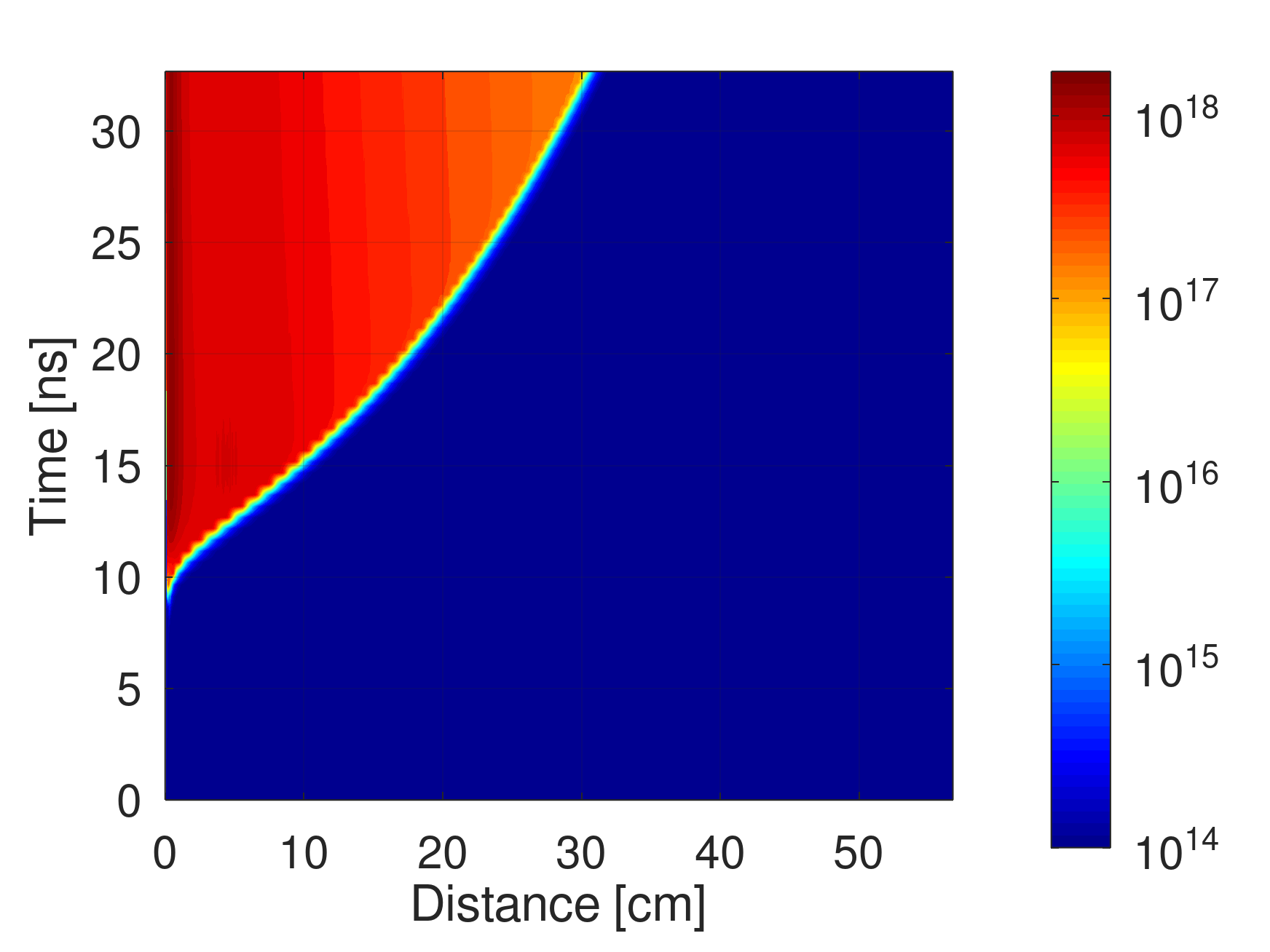}
\caption{ N2+ density[$\mathrm{m^{-3}}$] }
 \end{subfigure}
   \begin{subfigure}{.51\textwidth}
  \centering
\includegraphics[width=1\linewidth]{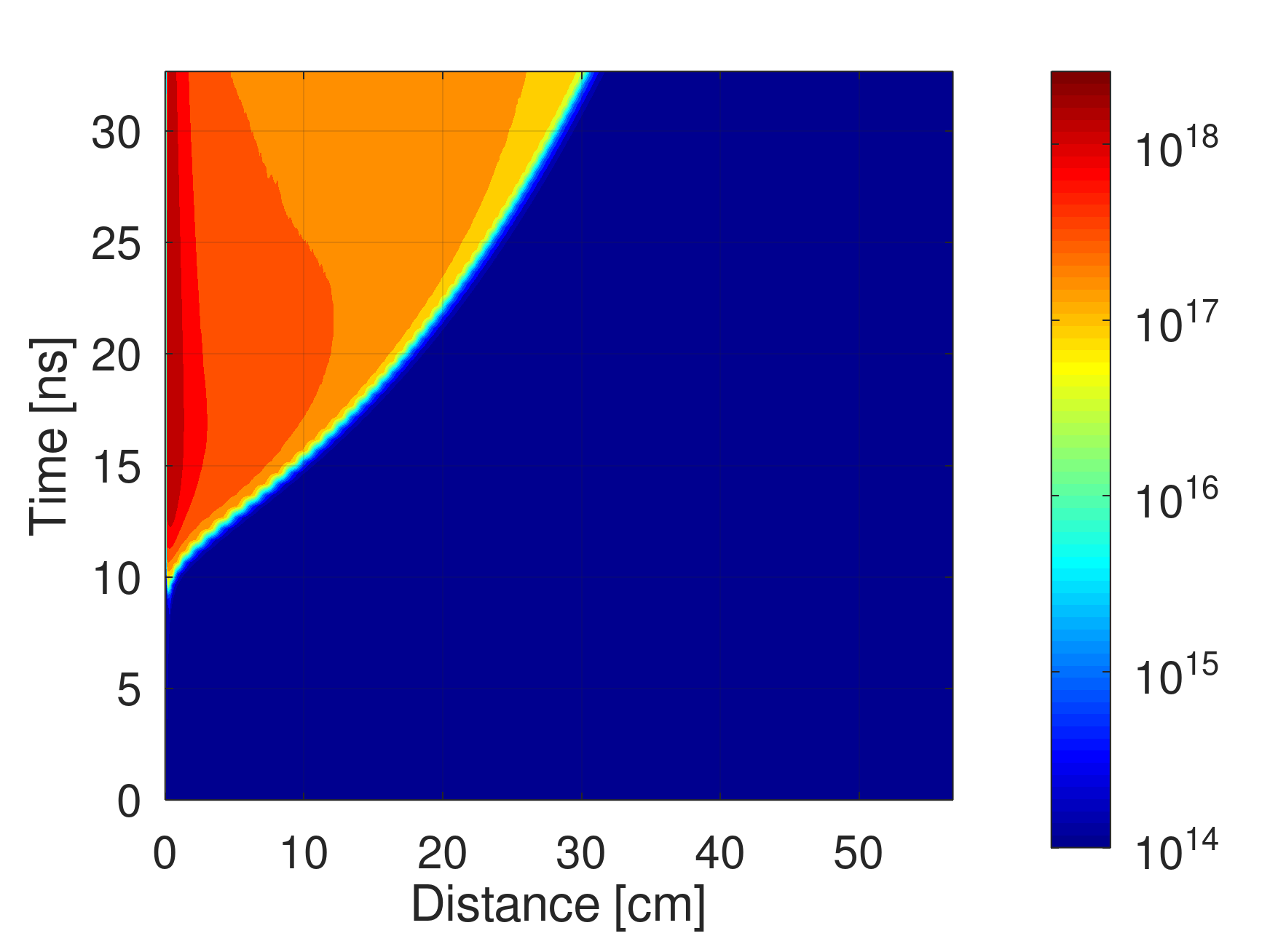}
\caption{ N2C density[$\mathrm{m^{-3}}$]  }
 \end{subfigure}%
  \begin{subfigure}{.51\textwidth}
  \centering
\includegraphics[width=1\linewidth]{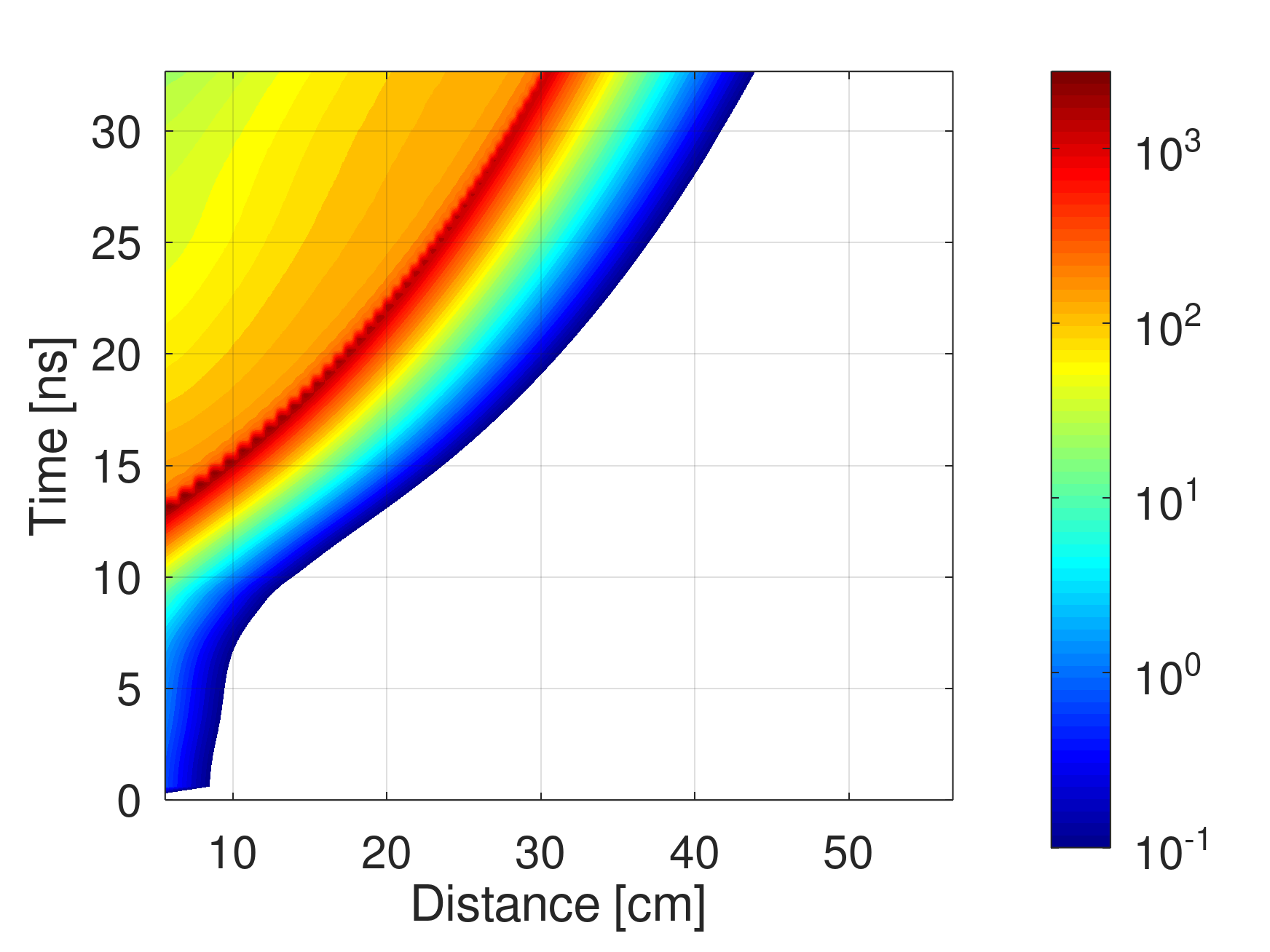}
\caption{ $\mathrm{|E_z|/N_{g}}$ [Td]  }
 \end{subfigure}
   \begin{subfigure}{.51\textwidth}
  \centering
\includegraphics[width=1\linewidth]{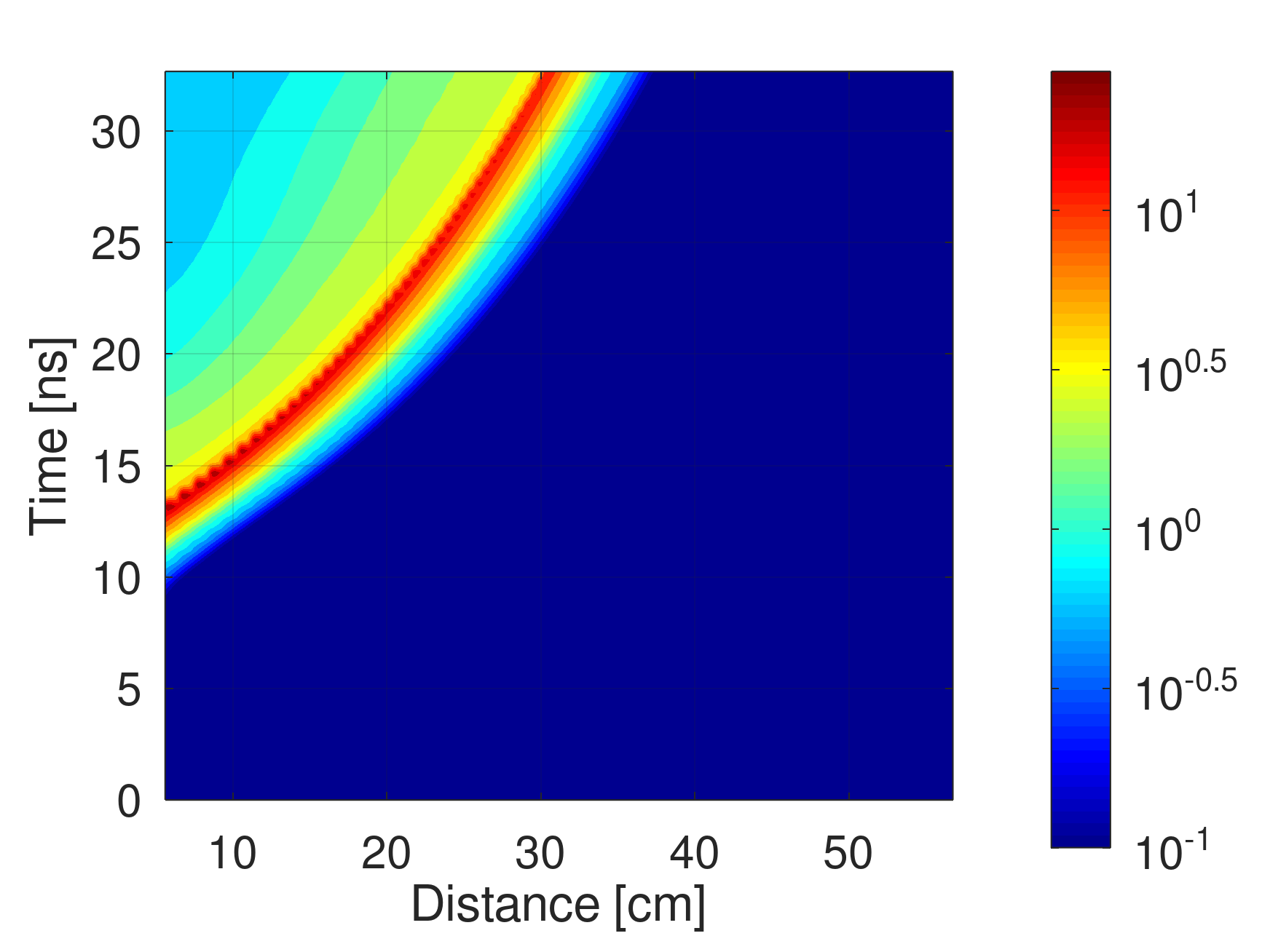}
\caption{ $\mathrm{T_{e}}$ [eV]  }
 \end{subfigure}
  \caption{Negative polarity at 10 Torr: Axial distance vs time contour plot of electron,  N$_2^+$,  N$_2$(C$^3\Pi_u$) species density [$\mathrm{m^{-3}}$], axial component of reduced electric field magnitude [Td] and electron temperature [eV] at r$=$0.  All plots in log-scale. }
    \label{fig:10torr_-_xt}
 \end{figure}


\subsection{Spatiotemporal distribution of electric field, electron temperature, charged species densities and density of \protect{\rm{N$_2$(C$^3\Pi_u$)}}.}

Calculated spatiotemporal distribution, represented as $x-t$ diagrams, of the electron and ion densities,  density of N$_2$(C$^3\Pi_u$), the longitudinal component of the reduced electric field $|E_z|/N_g$ and the electron temperature $T_e$ for negative and positive polarities of the high-voltage pulse are presented and discussed in this section. All values are taken on the axis of the discharge, at $r=0$. The point $\left(Z;t\right)=\left(0;0\right)$ corresponds to the high-voltage electrode, $Z=60$~cm corresponds to the low-voltage electrode, and $t>30$~ns corresponds to the falling edge of the high-voltage pulse. Note that for visualization purposes, the graphs of electron temperature and reduced electric field are excluding the sheath zones near the electrodes (where very high values are observed). Similarly, minimum values of $E/N$, $T_{e}$ and species density limited to 0.1 Td, 0.1 eV and $10^{14}$~m$^{-3}$ respectively, also for visualization purposes, while maximum values for species densities are limited to $10^{20}$~m$^{-3}$ (only in the cases of positive polarity such limitation acts in the near-electrode regions).

\subsubsection{Negative polarity}

Figures~\ref{fig:1torr_-_xt},~\ref{fig:5torr_-_xt},~\ref{fig:10torr_-_xt} show the spatiotemporal behavior for negative polarity discharge at gas pressure $P=1, 5$ and 10~Torr respectively. The discharge onset and formation of high electron density is clearly seen in the interval 5-10~ns, depending on pressure. The discharge velocity is higher at the beginning, decelerating as the FIW propagates along the tube. This is a distinctive feature of the negative polarity discharge, which is more pronounced for low pressures: for example, for $P=5$ torr, the discharge velocity gradually decreases from 2 cm/ns (at 13.64 ns) to 0.68 cm/ns (at 32.1 ns) as the FIW develops in the gap.

The front of the fast ionization wave, or a region with a high electric field, lasts a few nanoseconds, in agreement with experimental observations. The duration of this high peak of E-field is a function of pressure, decreasing from 5-10~ns at 1 Torr to 1-2~ns at 10~Torr. At 1 Torr, a plasma sheath with thickness of approx. 0.6 cm is clearly visible near the left electrode (at distance z$=$0). The electron density produces mainly in the FIW front. The front represents a gradient of both electron and $\mathrm{N_{2}^{+}}$ ions, with a length of approximately 4~cm. Behind the FIW front, the quasi-neutral plasma body sustains electron and $\mathrm{N_{2}^{+}}$ densities in the order of $10^{11}$ - $10^{12}$~cm$^{-3}$. Electron-ion recombination for such a period of time can be disregarded and the decrease of the level excitation rate is caused by a decrease of the electron energy. Hence, the electron density monotonically builds up during the impulse and becomes stationary. Behaviour of density of N$_2^+$ ions repeats the behaviour of electrons; the densities of other ions are much lower.  

In the plasma bulk, increasing electron density, and still high despite the decreasing electric fields, lead to efficient excitation of N$_2$(C$^3\Pi_u$)-state, producing a major part of optical emission of the nanosecond moderate pressure discharges in nitrogen. The delay between the FIW front and N$_2$(C$^3\Pi_u$) excitation is clearly seen. The density of N$_2$(C$^3\Pi_u$)-state inside the plasma bulk is in the order of $10^{11}$~cm$^{-3}$. The picture corresponds completely to experimental observations of paper \cite{Pancheshnyi1999} describing population of nitrogen molecule electron states and structure of the fast ionization wave. 

The calculated reduced electric field and electron temperature inside the plasma bulk are in the order of $10 - 100$~Td and $1-5$~eV while the FIW front sustains quite high fields in the order of 1000~Td and higher, with electron temperatures in the $10-30$~eV range. Such high values of $E/N$ are in agreement with several modeling results~\cite{chen2021modeling,klochko2014investigation, zhu2020scaling}.         
 

    \begin{figure}[t!]
\centering
 \begin{subfigure}{.51\textwidth}
 \centering
\includegraphics[width=1\linewidth]{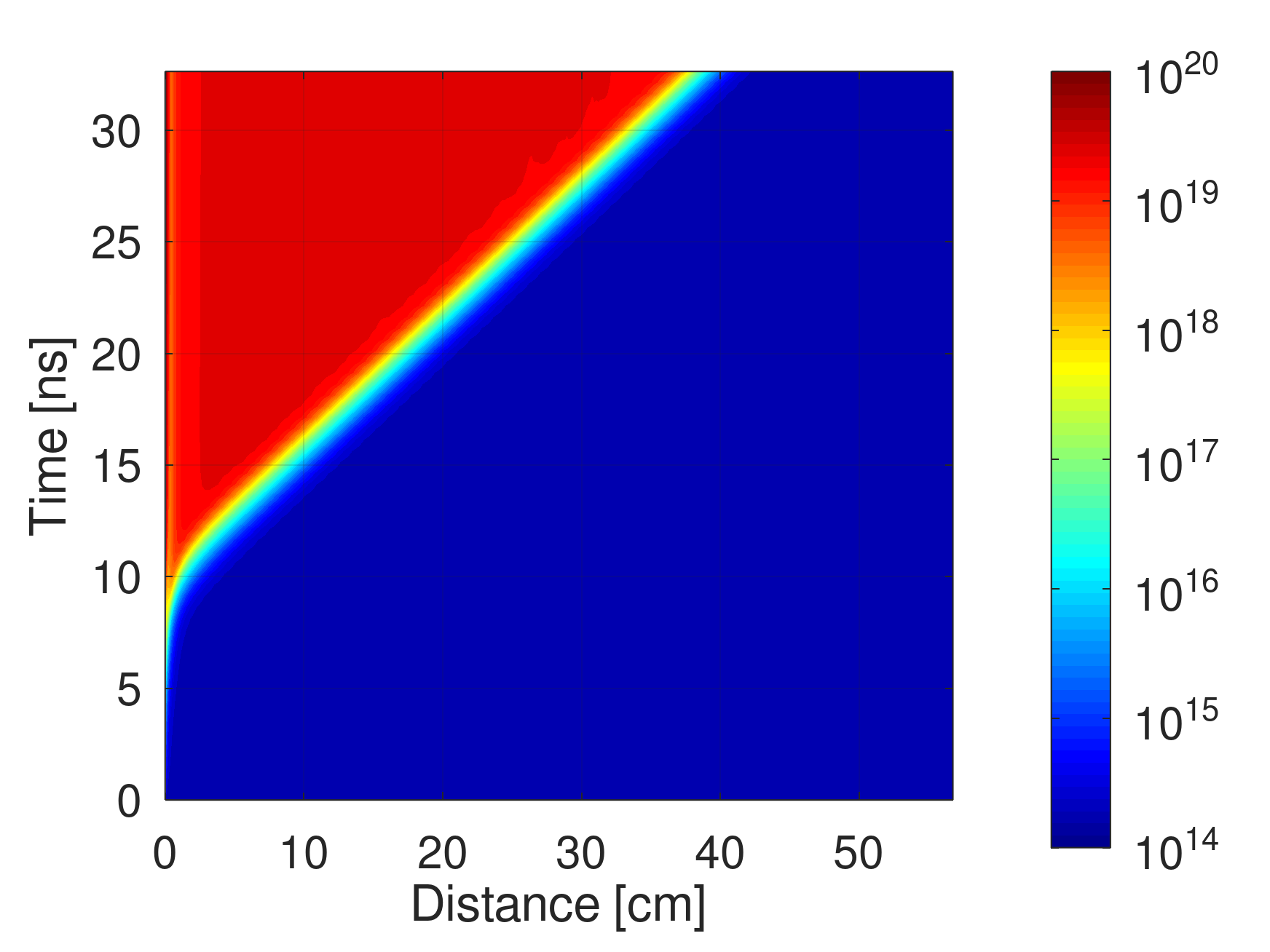}
\caption{ Electron density[$\mathrm{m^{-3}}$]   }
 \end{subfigure}
  \begin{subfigure}{.51\textwidth}
  \centering
\includegraphics[width=1\linewidth]{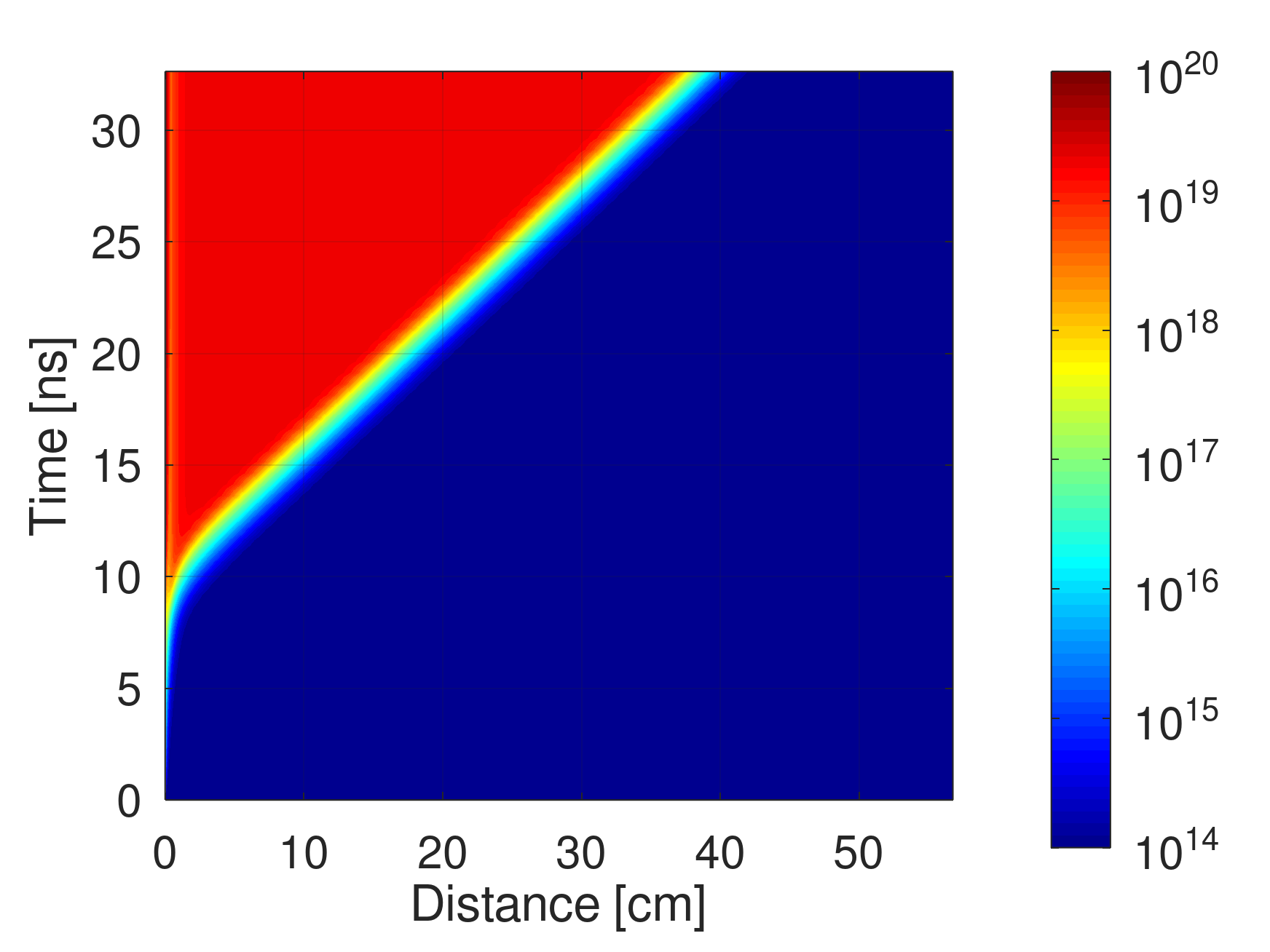}
\caption{ N2+ density[$\mathrm{m^{-3}}$] }
 \end{subfigure}
   \begin{subfigure}{.51\textwidth}
  \centering
\includegraphics[width=1\linewidth]{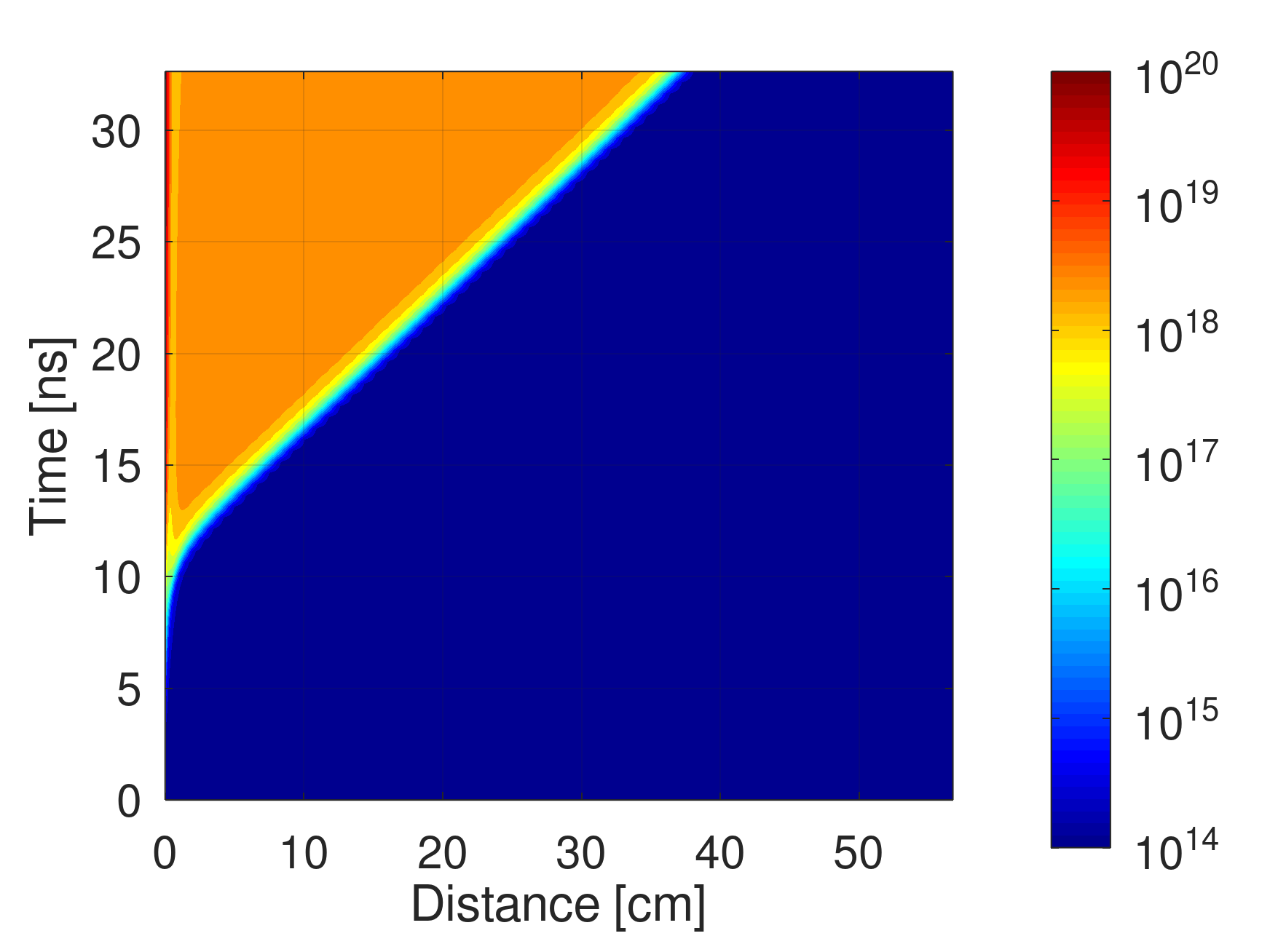}
\caption{ N2C density[$\mathrm{m^{-3}}$]  }
 \end{subfigure}%
  \begin{subfigure}{.51\textwidth}
  \centering
\includegraphics[width=1\linewidth]{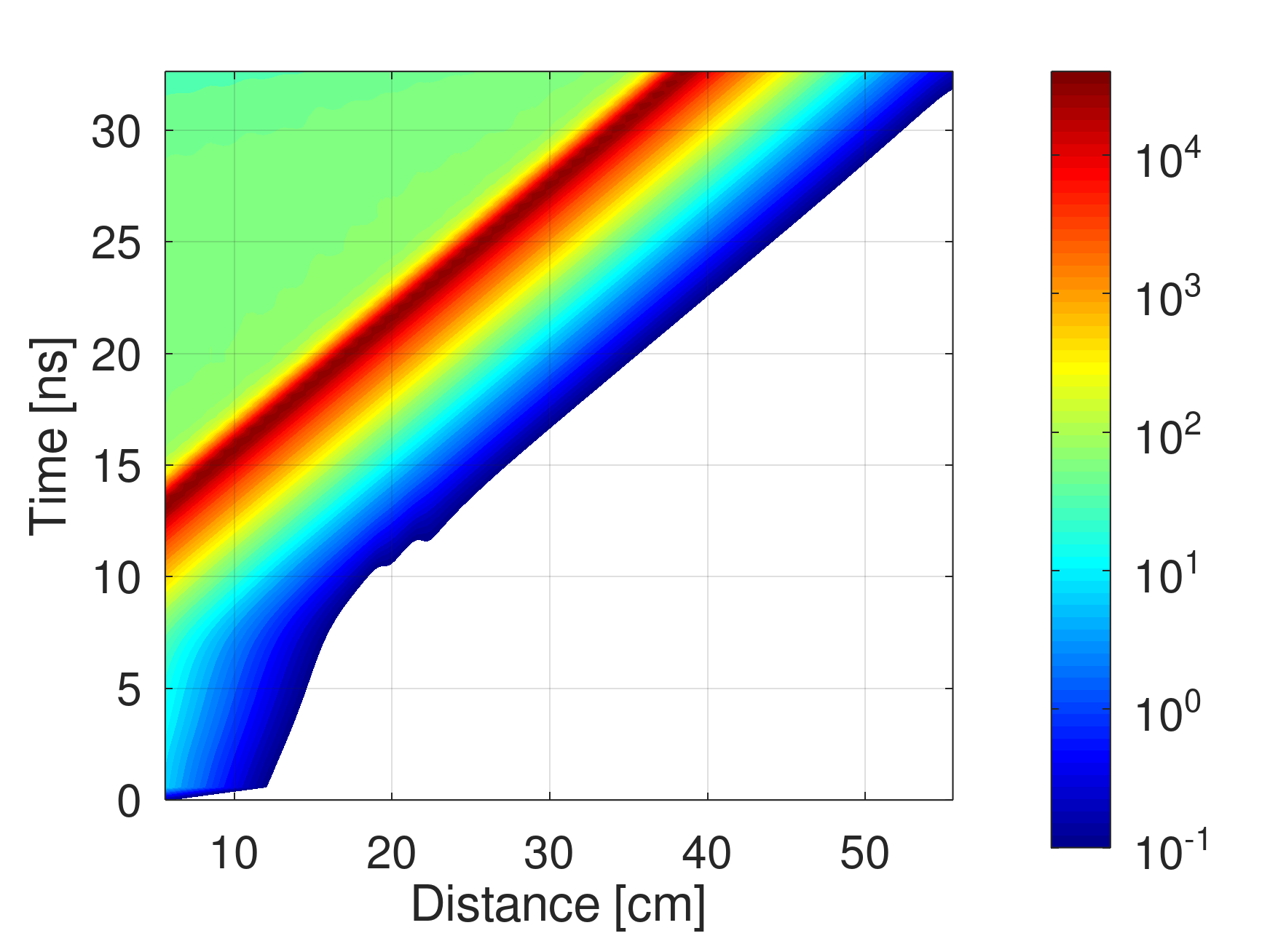}
\caption{ $\mathrm{|E_z|/N_{g}}$ [Td]  }
 \end{subfigure}
    \begin{subfigure}{.51\textwidth}
  \centering
\includegraphics[width=1\linewidth]{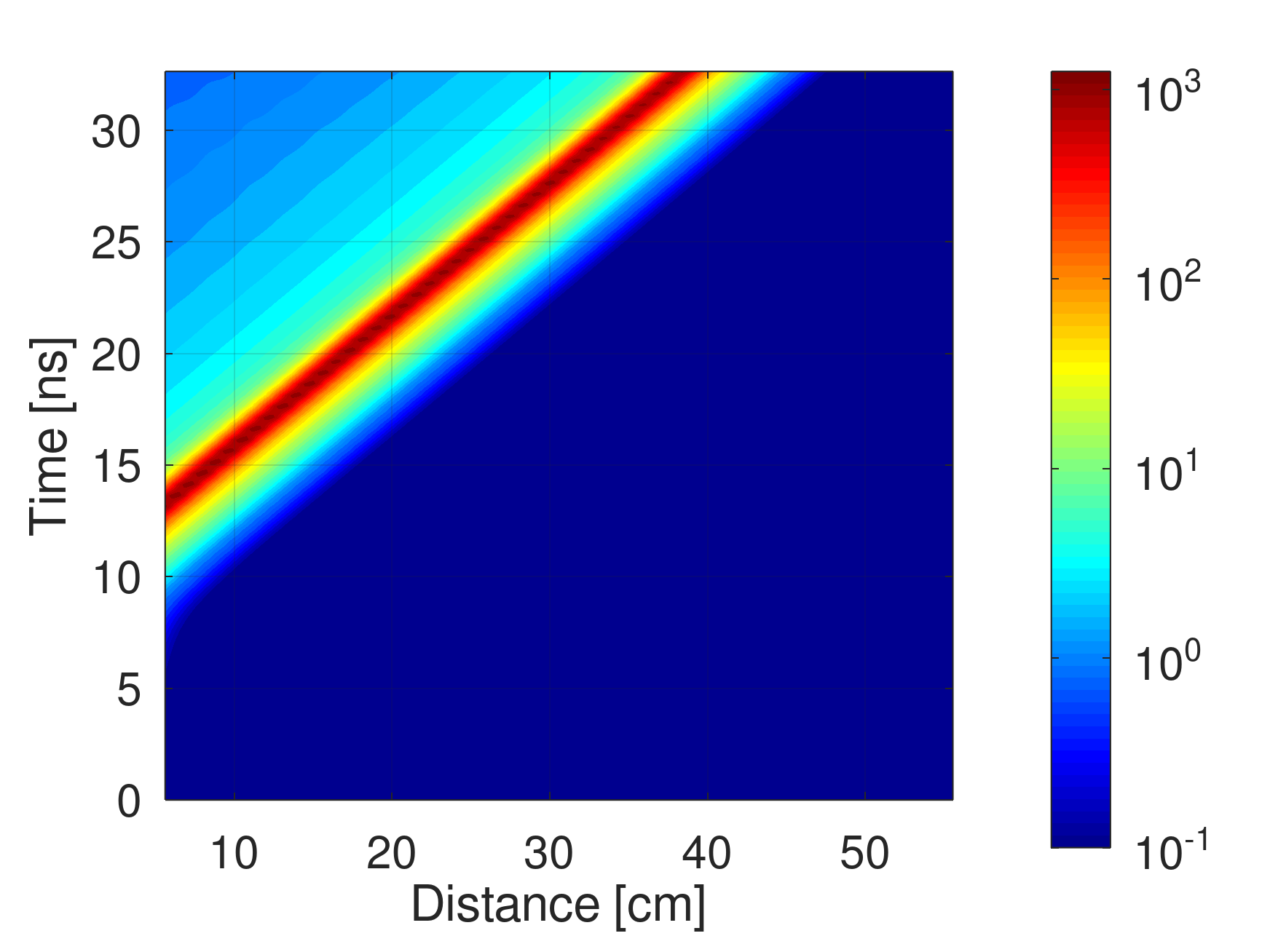}
\caption{ $\mathrm{T_{e}}$ [eV]  }
 \end{subfigure}
  \caption{Positive polarity at 1 Torr: Axial distance-time contour plot of electron,  N$_2^+$,  N$_2$(C$^3\Pi_u$) species density [$\mathrm{m^{-3}}$], axial component of reduced electric field magnitude [Td] and electron temperature [eV] at r$=$0. All plots in log-scale. }
    \label{fig:1torr_+_xt}
 \end{figure}
 
      \begin{figure}[h]
\centering
 \begin{subfigure}{.51\textwidth}
 \centering
\includegraphics[width=1\linewidth]{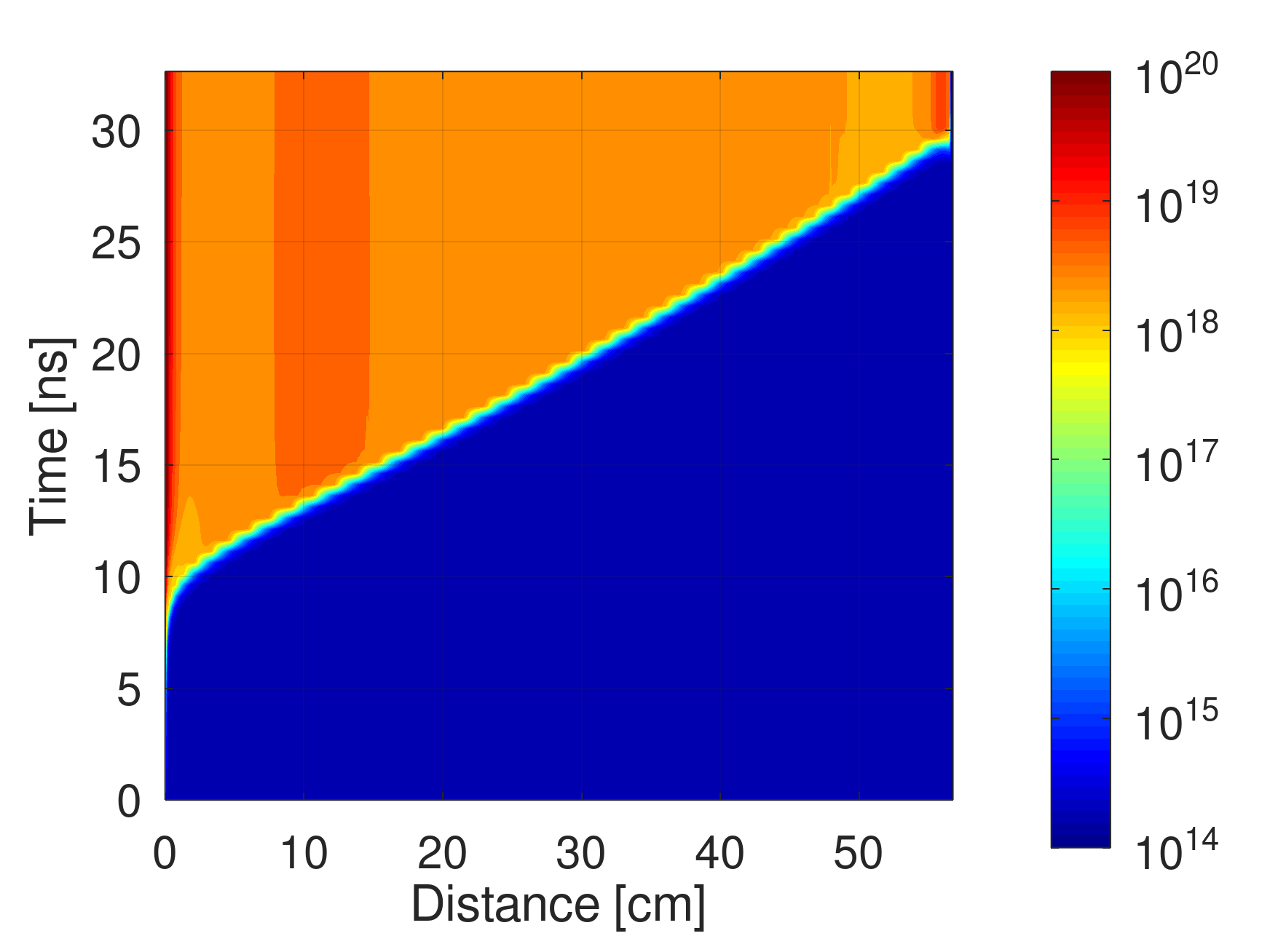}
\caption{ Electron density[$\mathrm{m^{-3}}$]   }
 \end{subfigure}
  \begin{subfigure}{.51\textwidth}
  \centering
\includegraphics[width=1\linewidth]{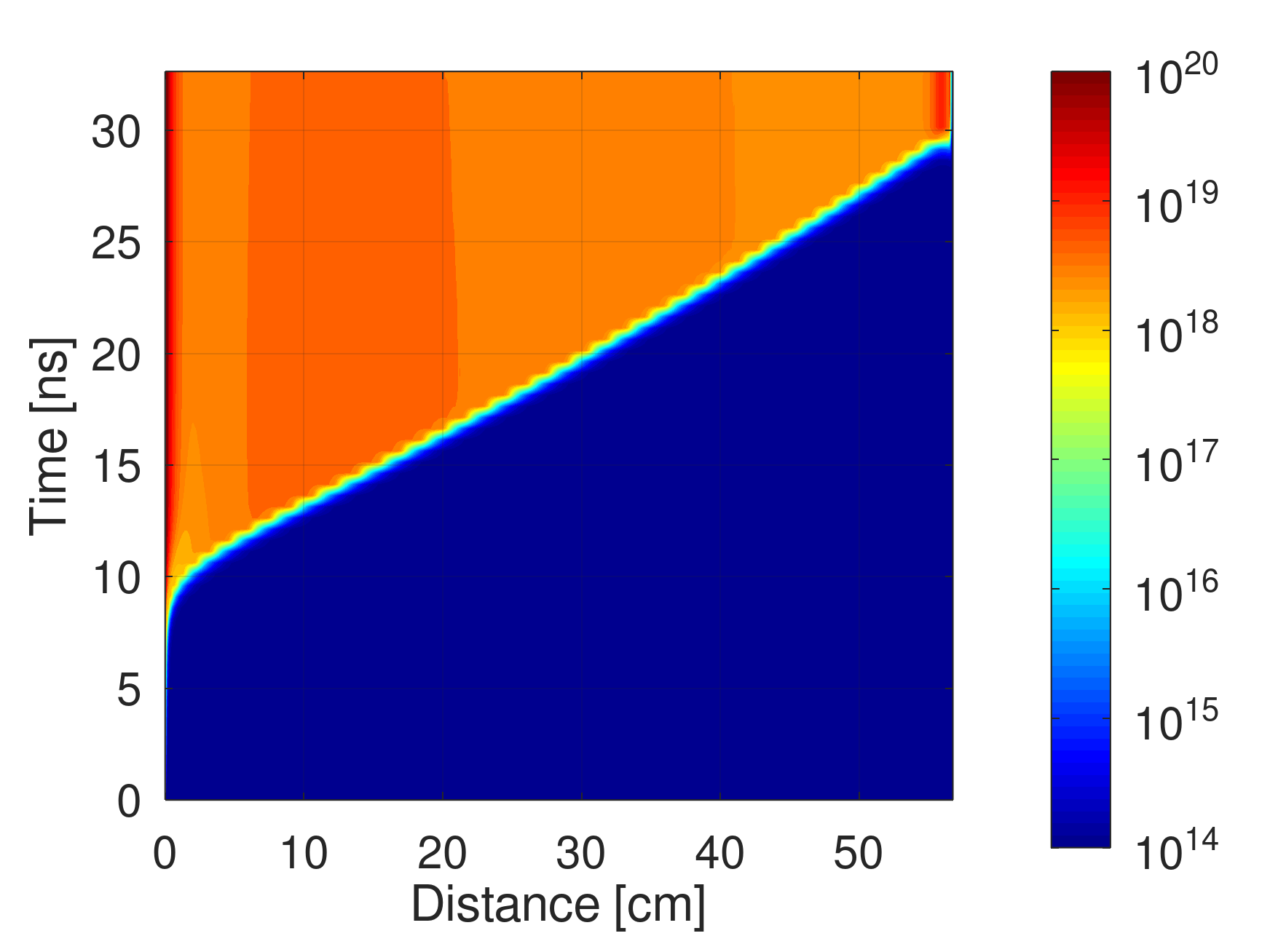}
\caption{ N2+ density[$\mathrm{m^{-3}}$] }
 \end{subfigure}
   \begin{subfigure}{.51\textwidth}
  \centering
\includegraphics[width=1\linewidth]{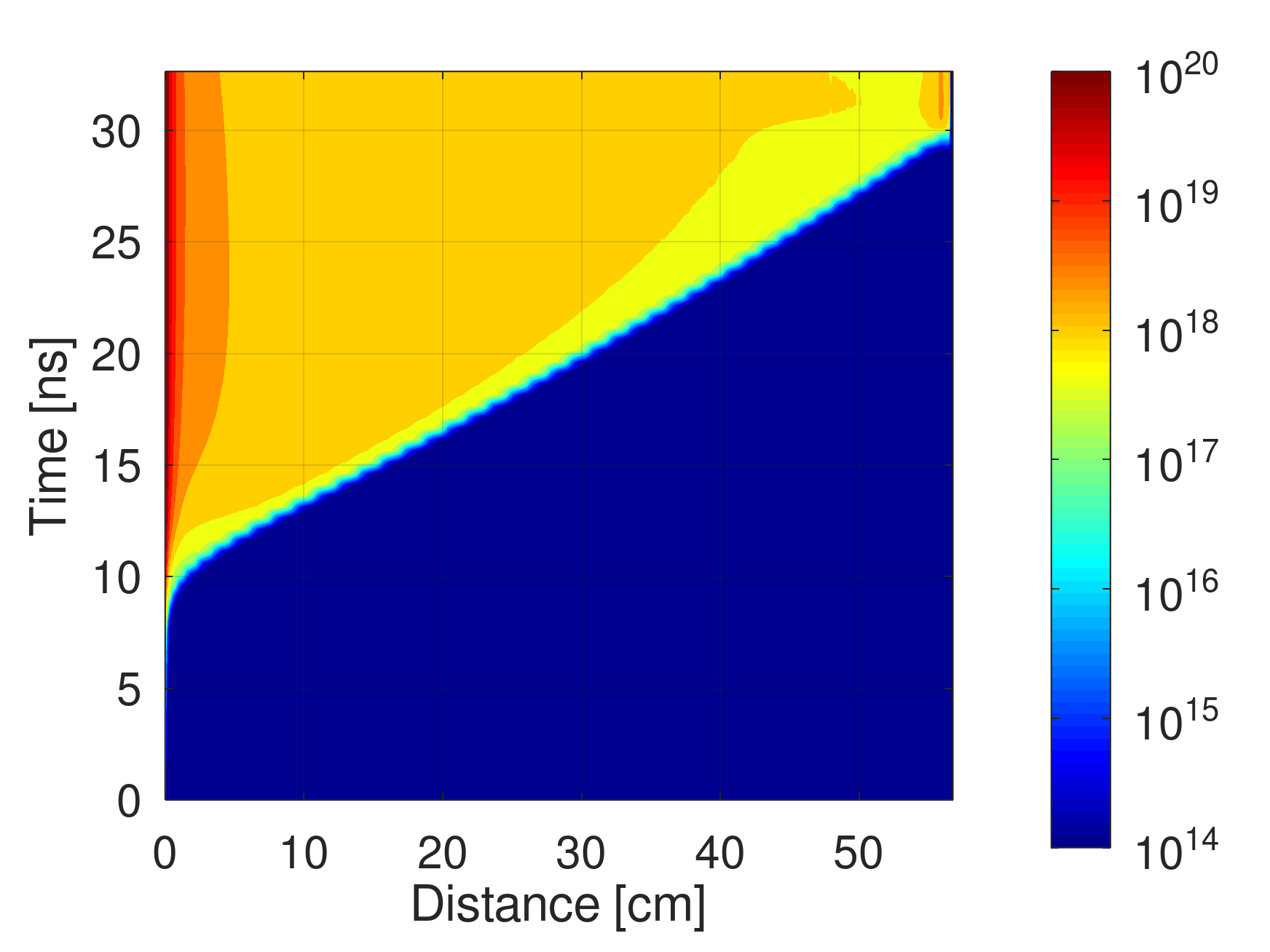}
\caption{ N2C density[$\mathrm{m^{-3}}$]  }
 \end{subfigure}%
  \begin{subfigure}{.51\textwidth}
  \centering
\includegraphics[width=1\linewidth]{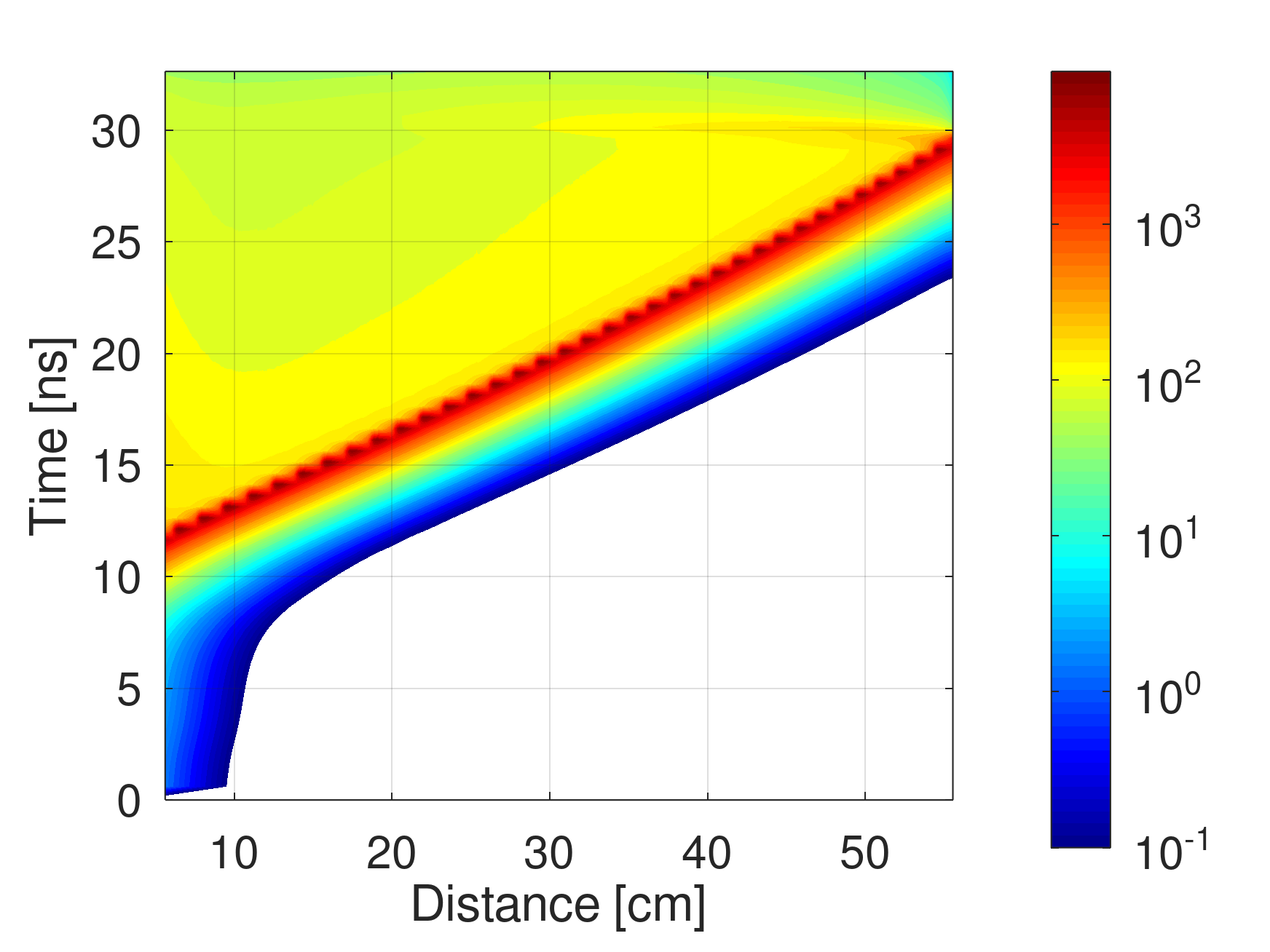}
\caption{ $\mathrm{|E_z|/N_{g}}$ [Td]  }
 \end{subfigure}
    \begin{subfigure}{.51\textwidth}
  \centering
\includegraphics[width=1\linewidth]{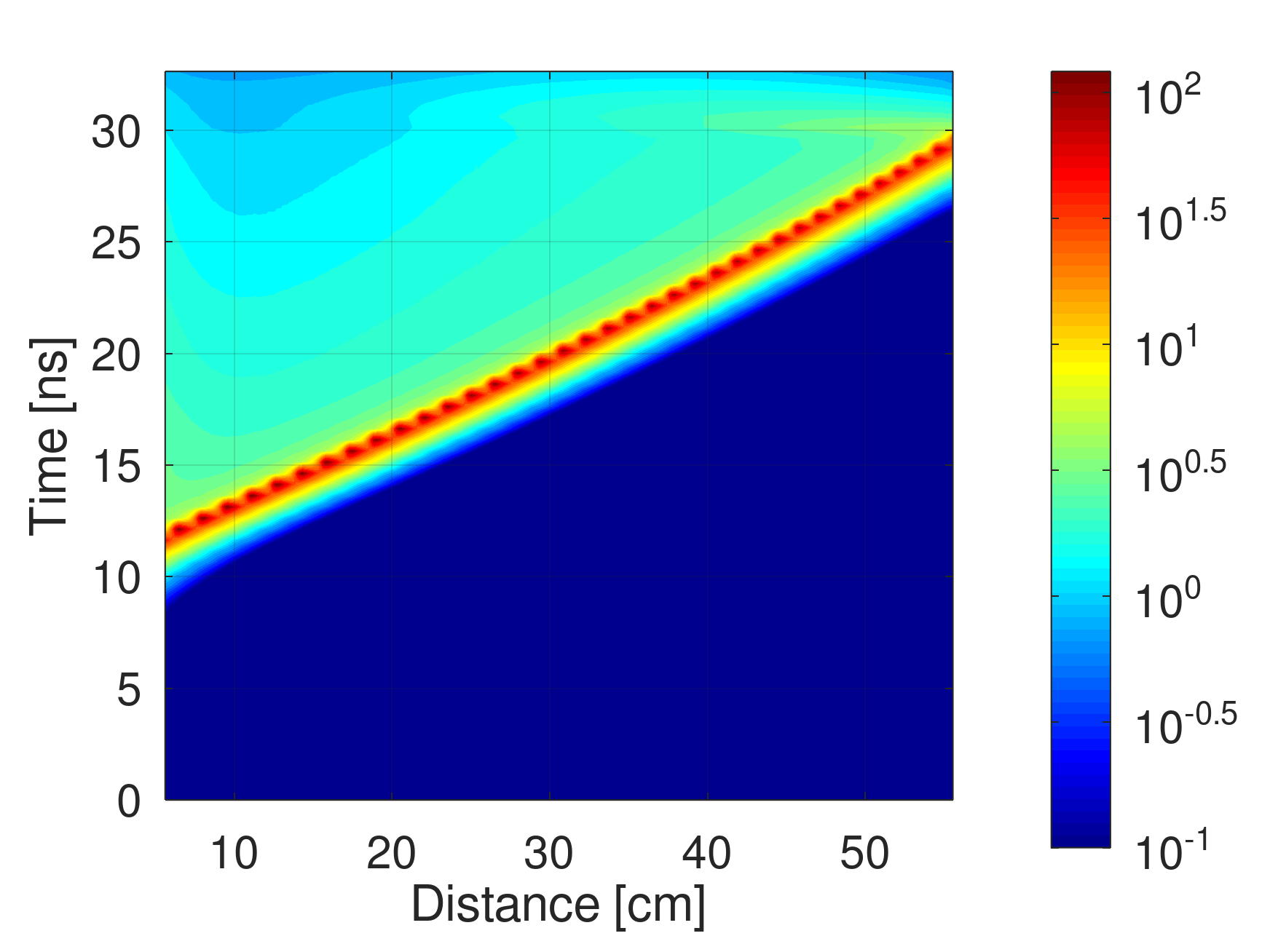}
\caption{ $\mathrm{T_{e}}$ [eV]  }
 \end{subfigure}
  \caption{Positive polarity at 5 Torr: Axial distance vs time contour plot of electron,  N$_2^+$,  N$_2$(C$^3\Pi_u$) species density [$\mathrm{m^{-3}}$], axial component of reduced electric field magnitude [Td] and electron temperature [eV] at r$=$0. All plots in log-scale. }
   \label{fig:5torr_+_xt}
 \end{figure}
 
     \begin{figure}[h]
\centering
 \begin{subfigure}{.5\textwidth}
 \centering
\includegraphics[width=1\linewidth]{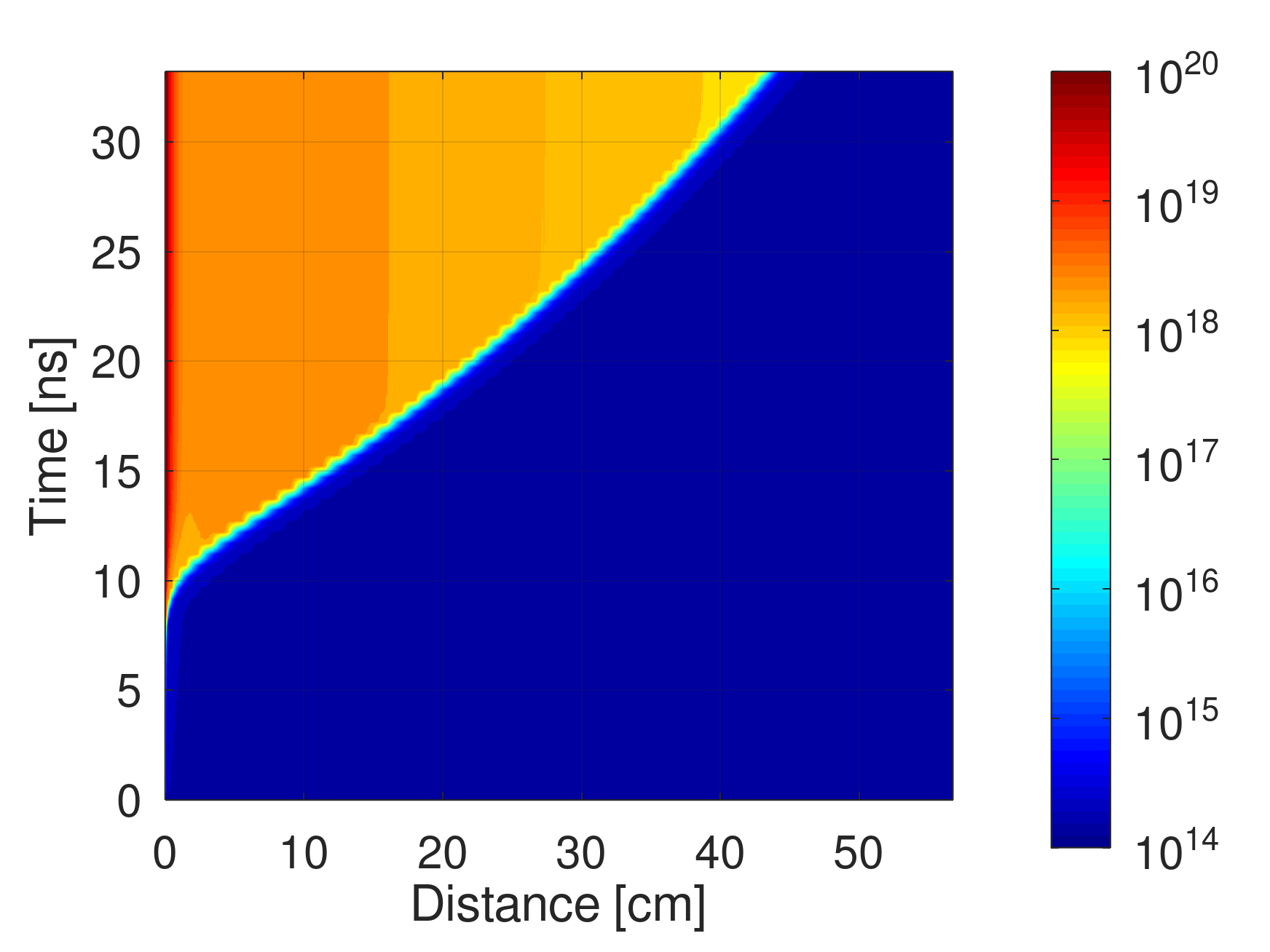}
\caption{ Electron density[$\mathrm{m^{-3}}$]   }
 \end{subfigure}
  \begin{subfigure}{.5\textwidth}
  \centering
\includegraphics[width=1\linewidth]{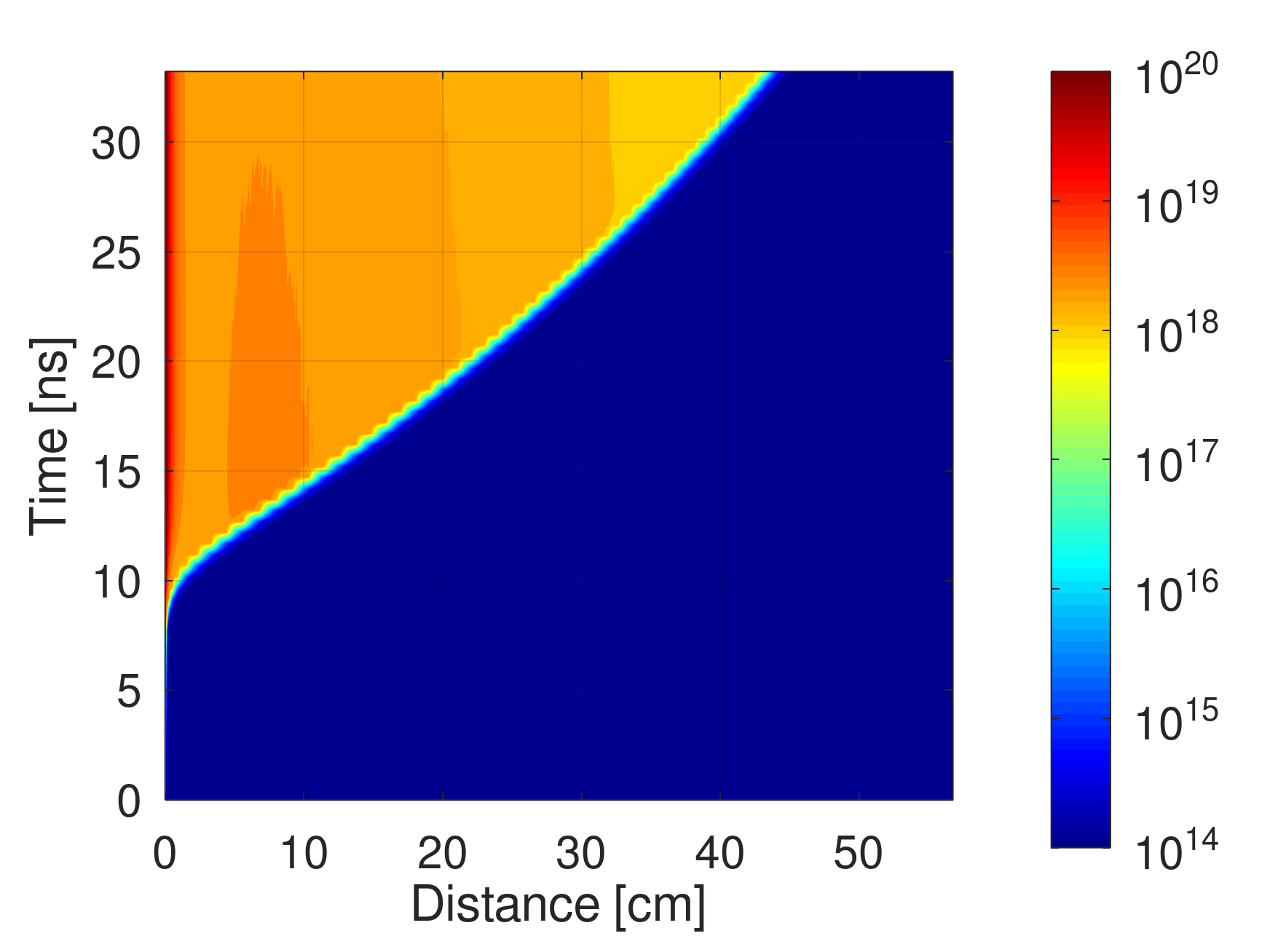}
\caption{ N2+ density[$\mathrm{m^{-3}}$] }
 \end{subfigure}
   \begin{subfigure}{.5\textwidth}
  \centering
\includegraphics[width=1\linewidth]{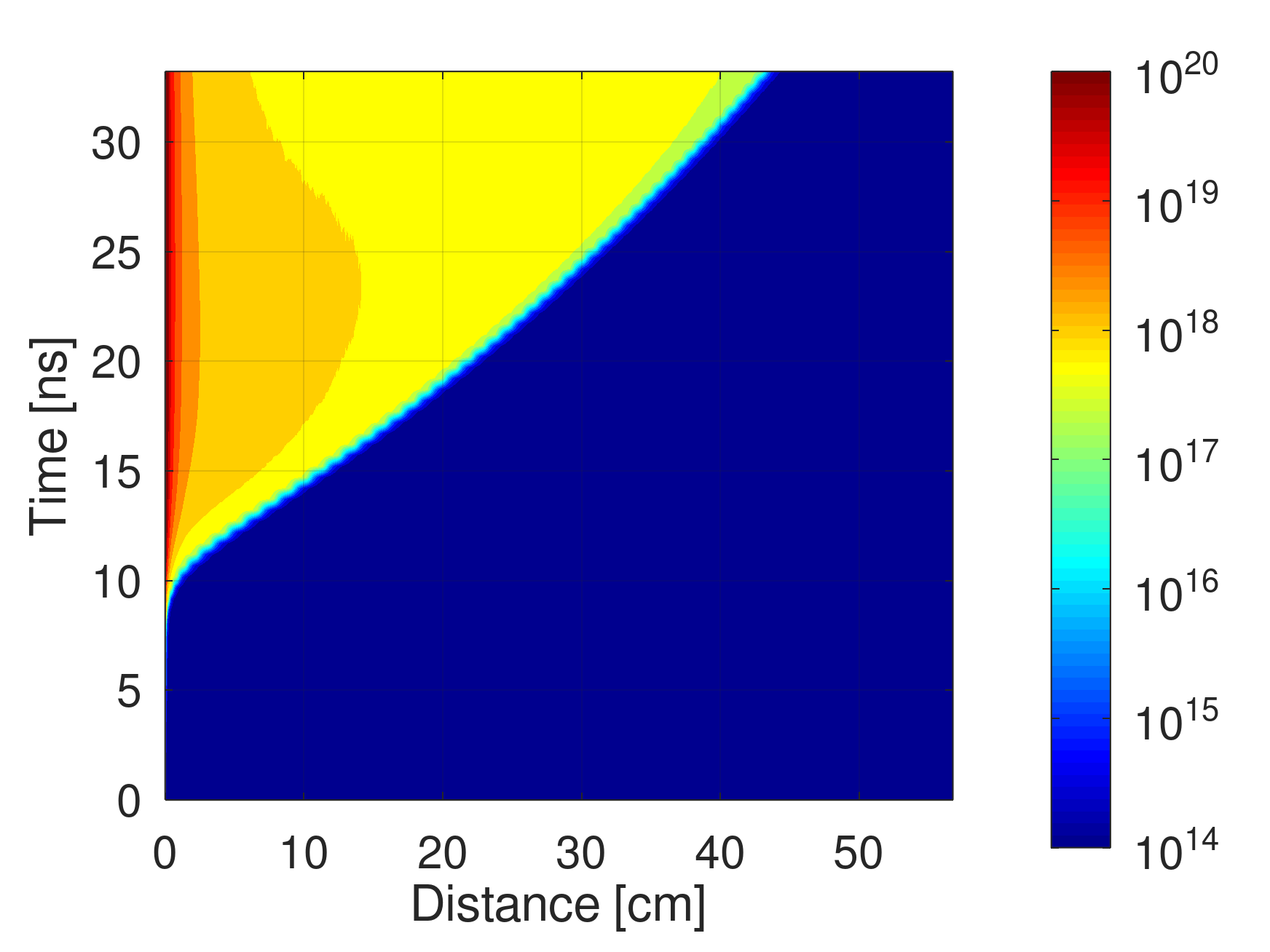}
\caption{ N2C density[$\mathrm{m^{-3}}$]  }
 \end{subfigure}%
  \begin{subfigure}{.5\textwidth}
  \centering
\includegraphics[width=1\linewidth]{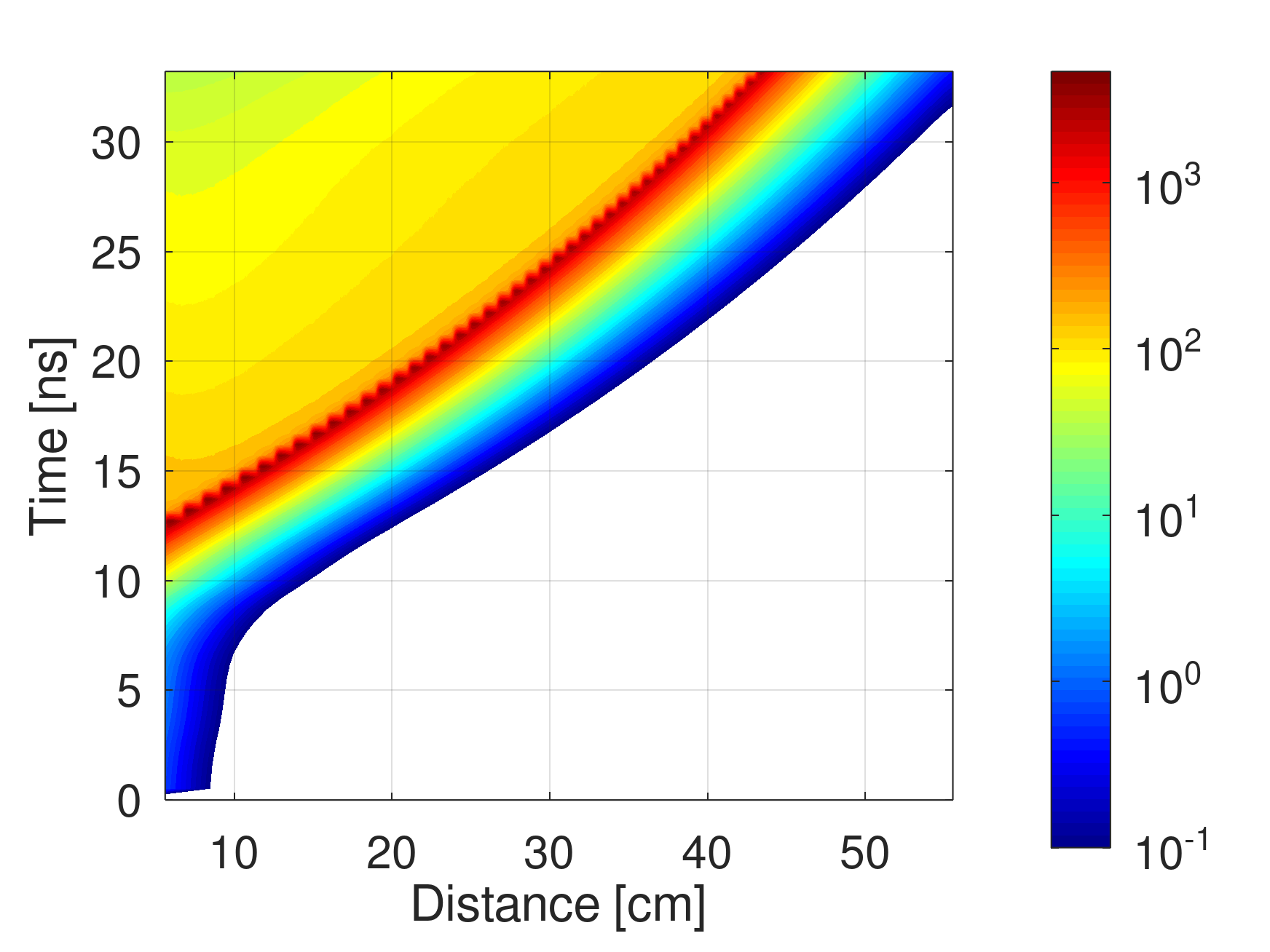}
\caption{ $\mathrm{|E_z|/N_{g}}$ [Td]  }
 \end{subfigure}
    \begin{subfigure}{.5\textwidth}
  \centering
\includegraphics[width=1\linewidth]{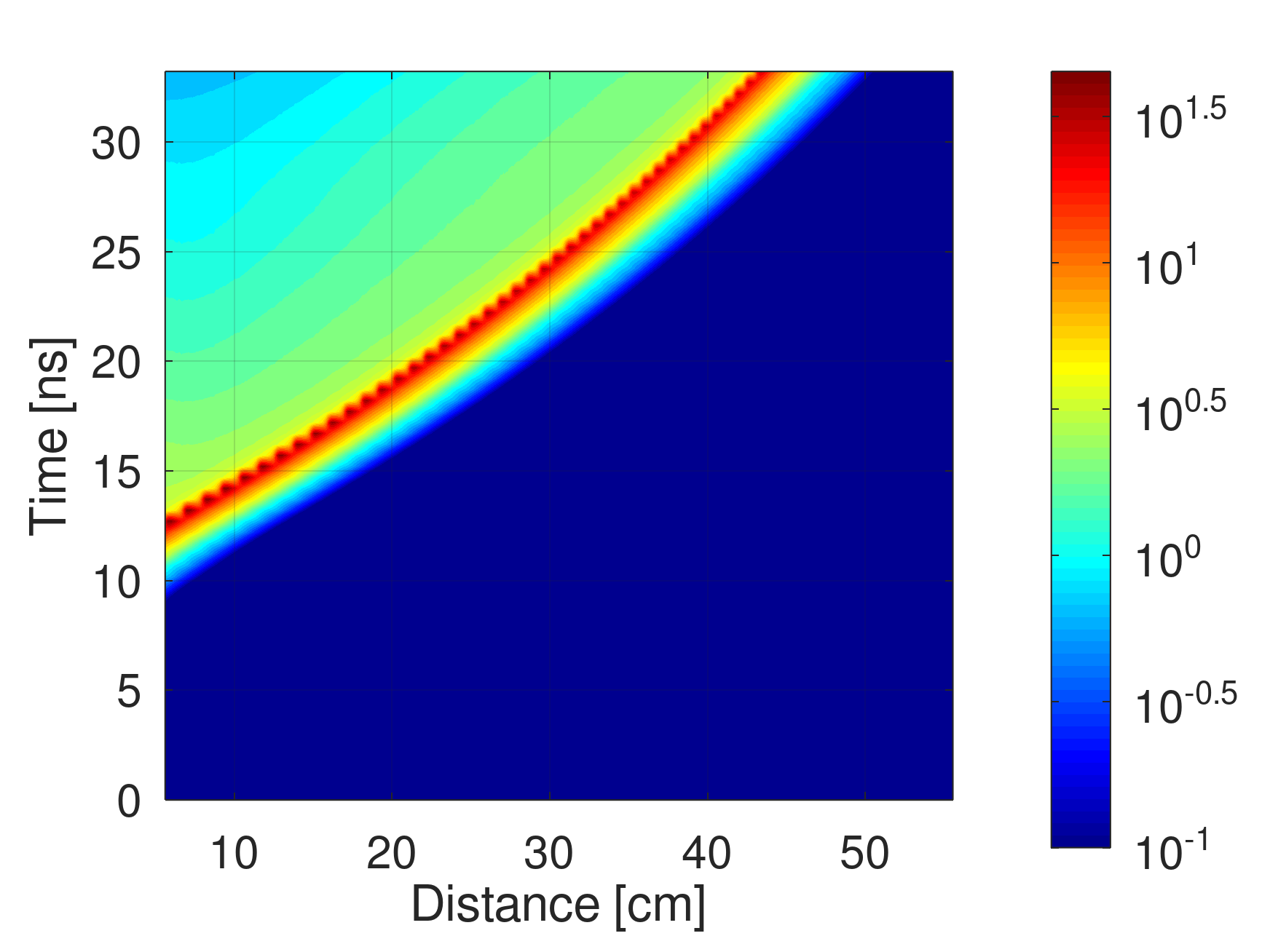}
\caption{ $\mathrm{T_{e}}$ [eV]  }
 \end{subfigure}
  \caption{Positive polarity at 10 Torr: Axial distance vs time contour plot of electron,  N$_2^+$,  N$_2$(C$^3\Pi_u$) species density [$\mathrm{m^{-3}}$], axial component of reduced electric field magnitude [Td] and electron temperature [eV] at r$=$0. Positive polarity at 10 Torr. All plots in log-scale. }
    \label{fig:10torr_+_xt}
 \end{figure}


\subsubsection{Positive polarity}

Figures~\ref{fig:1torr_+_xt},\ref{fig:5torr_+_xt},\ref{fig:10torr_+_xt} show $x-t$ diagrams of the electron and ion densities,  density of N$_2$(C$^3\Pi_u$), the longitudinal component of the reduced electric field $|E_z|/N_g$ and the electron temperature $T_e$ for positive polarity of the high-voltage pulse. As everywhere in this work, the discharge is in nitrogen. 

Similar to the negative polarity discharge, the discharge starts from the high-voltage electrode at approximately 8~ns. Differently from the negative polarity, the FIW propagates at an almost constant velocity, especially for lower pressures. Similar to the negative polarity, the FIW front presents a gradient of both electron and $\mathrm{N_{2}^{+}}$ ions, with a length of approximately 4~cm. Behind the FIW front, the quasi-neutral plasma body sustains electron and $\mathrm{N_{2}^{+}}$ densities in the order of $10^{12}$ - $10^{13}$ $cm^{-3}$, with higher values observed at lower pressure. N$_2$(C$^3\Pi_u$) distribution follows the FIW development, but densities inside the plasma bulk are higher and in the order of $10^{12}$ $cm^{-3}$. 

At moderate and higher pressure, and similar to the negative polarity discharge, N$_2$(C$^3\Pi_u$) front presents a weaker gradient with a length of more than 10~cm, owed to a convolution of increasing electron density and decreasing electric field. Even if, similar to the negative polarity discharge, the electric field in the FIW front is much higher than behind the front, the difference between the field in the front and after the front is much higher for positive polarity. The calculated reduced electric field and electron temperature inside the plasma bulk are similar to negative polarity (in the order of 10 - 100 Td and 1 - 5 eV respectively) but the FIW front sustains substantial higher fields in the order of $10000$~Td , resulting to electron temperatures around and even higher than 100~eV. As mentioned also before, such high values of $E/N$ are consistent with several modeling results~\cite{klochko2014investigation, zhu2020scaling,chen2021modeling}. Without going into details, will note that so high values of electron energy observed in numerical simulation in the FIW front, may indicate the necessity to take into account non-local effects in the front. Non-locality is beyond the scope of this paper, will just mention that the over-population of the EEDF of nanosecond discharge of positive polarity in N$_2$ at 5~Torr by high energy electrons was indirectly observed in experiments \cite{Anikin2002}. 
\color{black}

At $P=5$~Torr, the calculated FIW propagation speed is substantially higher and equals approximately 3.17 cm/ns. Owed to this high velocity, the FIW is reaching the grounded electrode in less than 30~ns.  At $P=10$~Torr, the FIW speed slightly decreases, reproducing the experimental observations. 

\begin{figure}[t!]
\centering
\includegraphics[width=\linewidth]{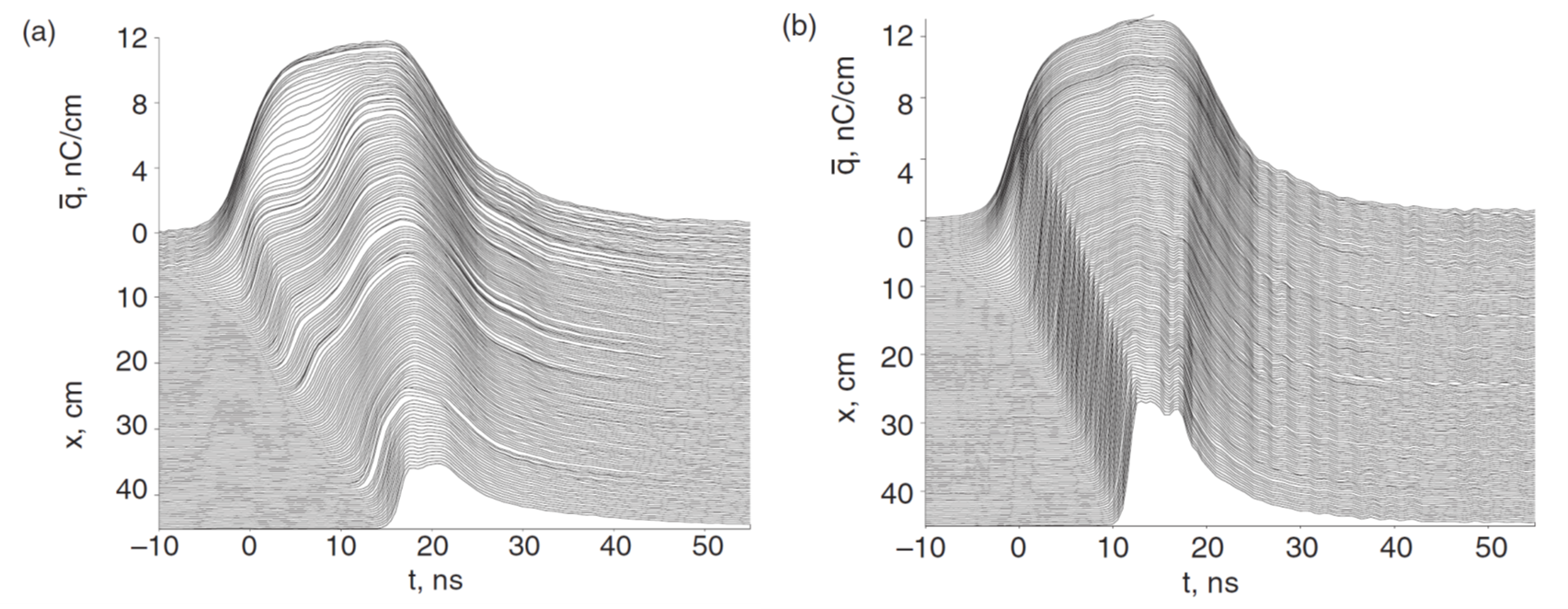}
  \caption{Dynamics of charge measured by capacitive probe \protect{\cite{Anikin2002}} for FIW developed in 5 Torr N$_2$:(a) negative polarity, (b) positive polarity. The high voltage electrode tip corresponds to coordinate $x = 0$~cm.}
 \label{xt_experiment}
\end{figure}

The experimental results on charge behaviour in space and time for $P=5$Torr nitrogen discharge for both polarities \cite{Anikin2002} are presented in Fig.~\ref{xt_experiment}. Two major differences are observed in the experiment for negative and positive polarities: (i) the presence, at negative polarity, of the developed precursor — a primary wave of low amplitude which the main wave catches up with only at a distance of $\sim 30$~cm from the high-voltage electrode; (ii) slowing down of the velocity of this precursor with increasing the distance from the high voltage electrode, while the velocity of the ionization wave initiated by a positive polarity pulse on the high-voltage electrode, is constant. The latter effect is well reproduced by numerical simulations of the present work (compare to Figs~\ref{fig:5torr_-_xt} and \ref{fig:5torr_+_xt} respectively). 
\color{black}

\subsubsection{Temporal profiles of $E/N$ and electron density}

It is interesting to compare calculated absolute values of the reduced electric field and the electron density as a function of time with the experimental data available in the literature. As far as capacitive probe allows measuring only longitudinal component of the electric field, $E_Z$, all the discussion below is about the longitudinal component. It is also important to mention that the capacitive probe has a space sensitivity function, and experimentally observed peak of the field should be smoother than in reality. 
\begin{figure}[!t]
\centering
\includegraphics[width=\linewidth]{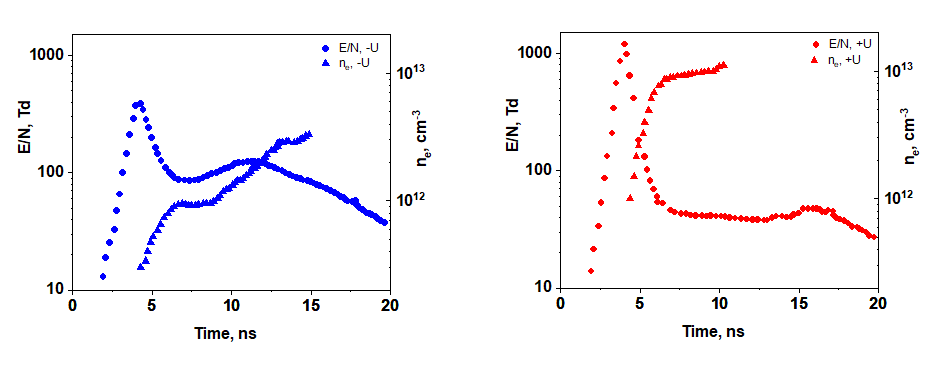}
  \caption{Experimentally measured longitudinal component of the reduced electric field and averaged over the cross-section of the tube electron density at the distance 13~cm from the high-voltage electrode \protect{\cite{Anikin2002}}: (a) negative polarity; (b) positive polarity. Nitrogen, pressure 5 torr.}
    \label{Minus_plus_field_ne}
 \end{figure}

 \begin{figure}[!b]
\centering
 \begin{subfigure}{.5\textwidth}
 \centering
\includegraphics[width=1\linewidth]{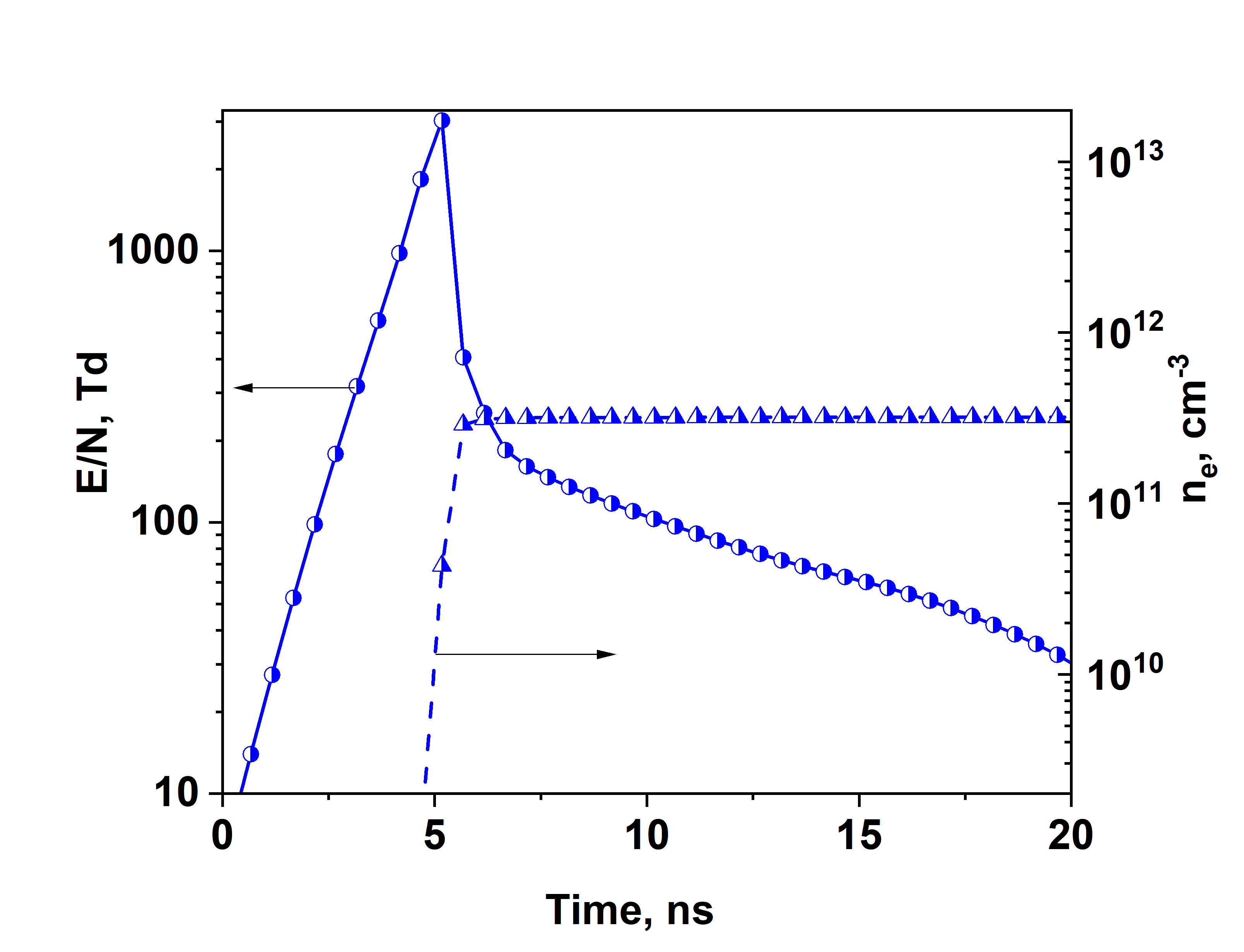}
\caption{ Negative Polarity  }
 \end{subfigure}
  \begin{subfigure}{.5\textwidth}
  \centering
\includegraphics[width=1\linewidth]{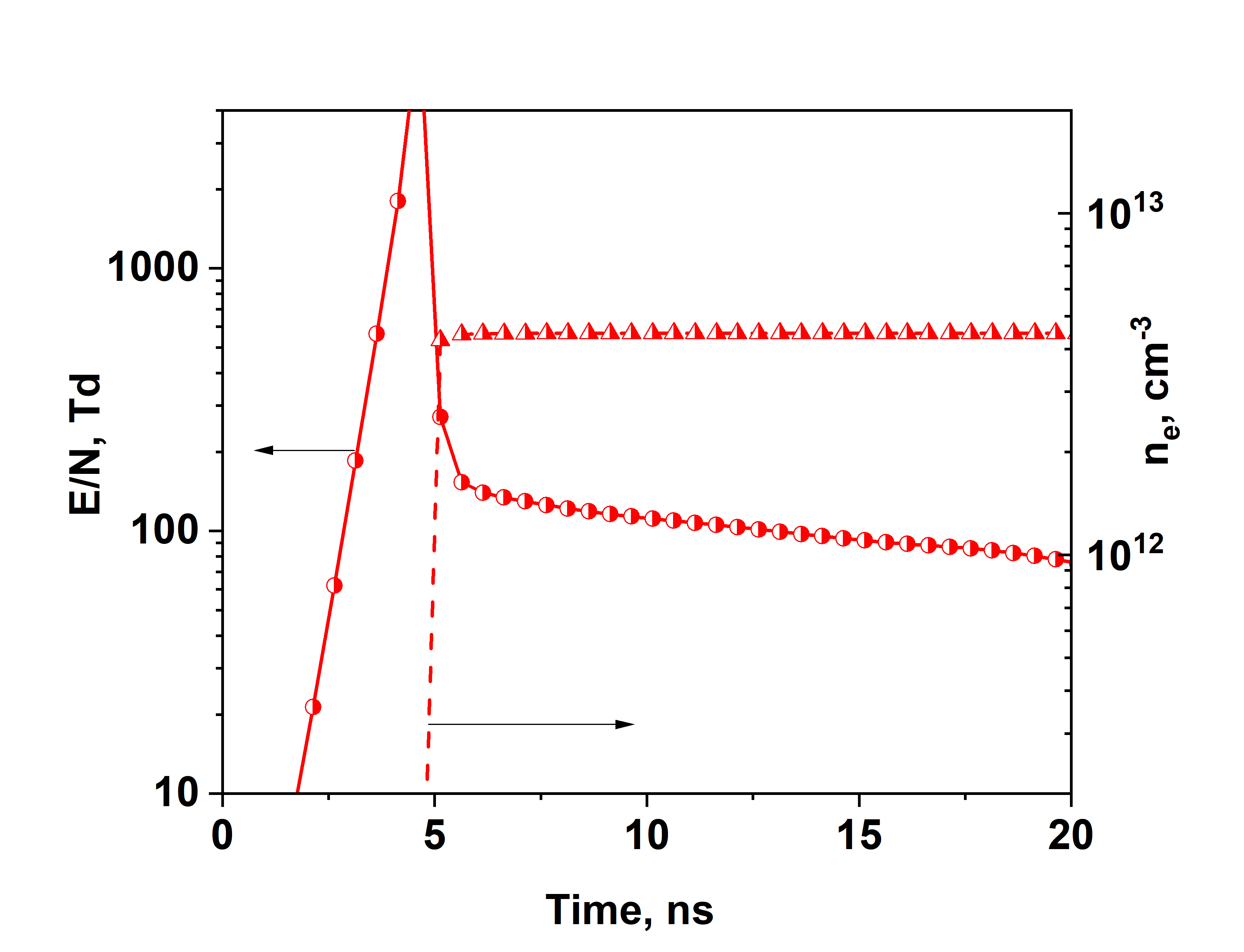}
\caption{ Positive Polarity }
 \end{subfigure}
  \caption{Calculated longitudinal reduced electric field and electron density at the distance 13~cm from the high-voltage electrode. Nitrogen, pressure 5 Torr.}
    \label{fig:timePlot_-+}
 \end{figure}

The experimentally measured waveforms of the longitudinal electric field at 13~cm from the high-voltage electrode and of the averaged over the cross-section electron density \cite{Anikin2002} are presented in Fig.~\ref{Minus_plus_field_ne}. The duration of the high fields in the front of the fast ionization wave is 2-3~ns for any of the polarities. For negative polarity, the peak values of the reduced electric field do not exceed 500~Td, decreasing after the front to approximately 100~Td. For positive polarity, the fields are higher than 1000~Td in the front, decreasing to 40~Td after the front. The electron density is lower at negative polarity, $n_e^{exp}\sim 4 \cdot 10^{12}$~cm$^{-3}$, while for positive polarity $n_e^{exp}\sim 10^{13}$~cm$^{-3}$.

Fig.~\ref{fig:timePlot_-+} shows the same temporal profiles extracted from the numerical simulations. The calculated $E/N$ is noticeably higher at the plasma front for positive polarity compared to negative polarity, which agrees with the experimental data. Similarly, good agreement is observed on the electron density values, which are lower for negative polarity. After the front, the reduced field is decreasing to values close to the ones observed in the experiments and around 100~Td for both polarities. The build-up of the electron density in numerical modeling is more sharp than in the experiment, while the delay between the field onset and the increase of the electron density is longer than in the experiment. The higher values of both electron density inside the plasma bulk and electron temperature in the FIW front at positive polarity, lead also to higher values of the photoionization source term which is concentrated at the FIW front (i.e. an order of magnitude difference between positive and negative polarities at 5 torr and at the same time-instant). These aspects explain the higher velocity observed under positive polarity, as volume ionization (both electron-impact and photo-related) is stronger in this case. 
 
Will now consider more in details the radial distribution of different components behind the fast ionization wave front, and how this distribution influences N$_2$(C$^3\Pi_u$) density, and so the optical emission from the discharge). 

  \begin{figure}[!t]
\centering
\includegraphics[width=0.75\linewidth]{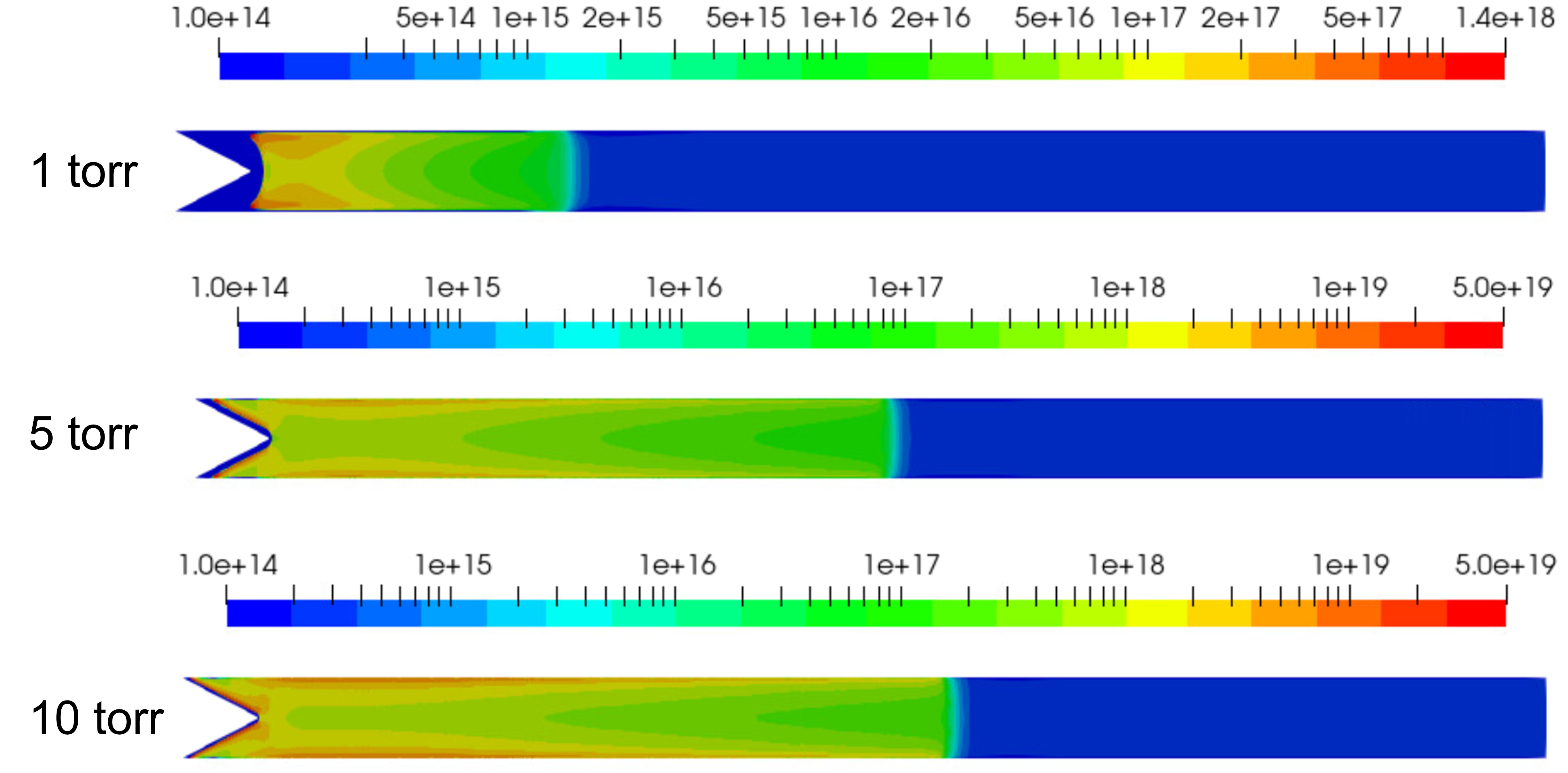}
  \caption{Electron density contours [$\mathrm{m^{-3}}$] for different gas pressure at t$=$32.65 ns. Negative polarity. Min/max values are limited to $10^{14}$ $m^{-3}$ and $5\times10^{19}$ $m^{-3}$ respectively for visualization purposes.}
    \label{fig:ne_comp_-}
 \end{figure}


\color{black}
 \begin{figure}[!b]
\centering
 \begin{subfigure}{0.35\textwidth}
 \centering
\includegraphics[width=1\linewidth]{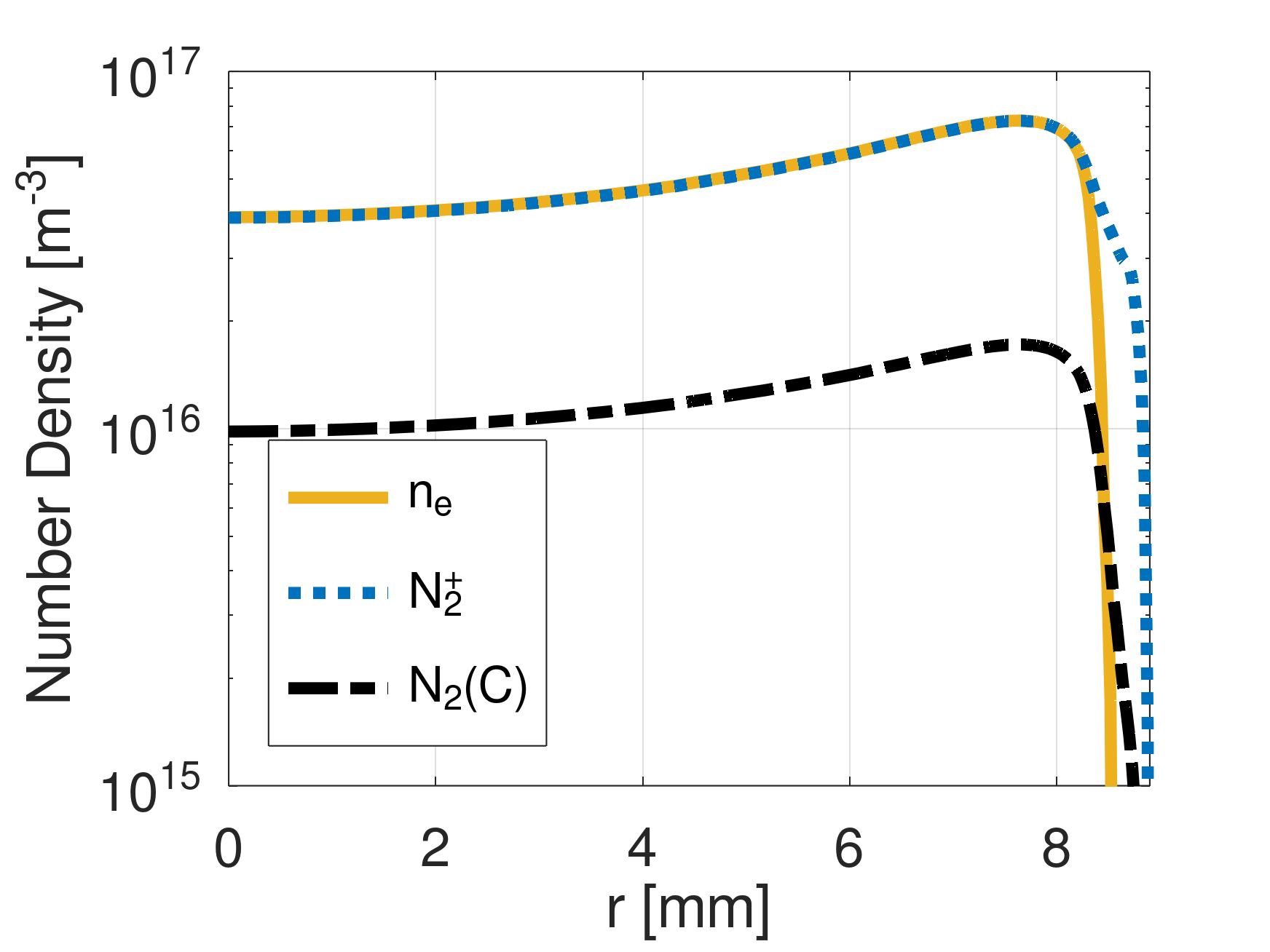}
\caption{ p=1 Torr   }
 \end{subfigure}
  \begin{subfigure}{.35\textwidth}
  \centering
\includegraphics[width=1\linewidth]{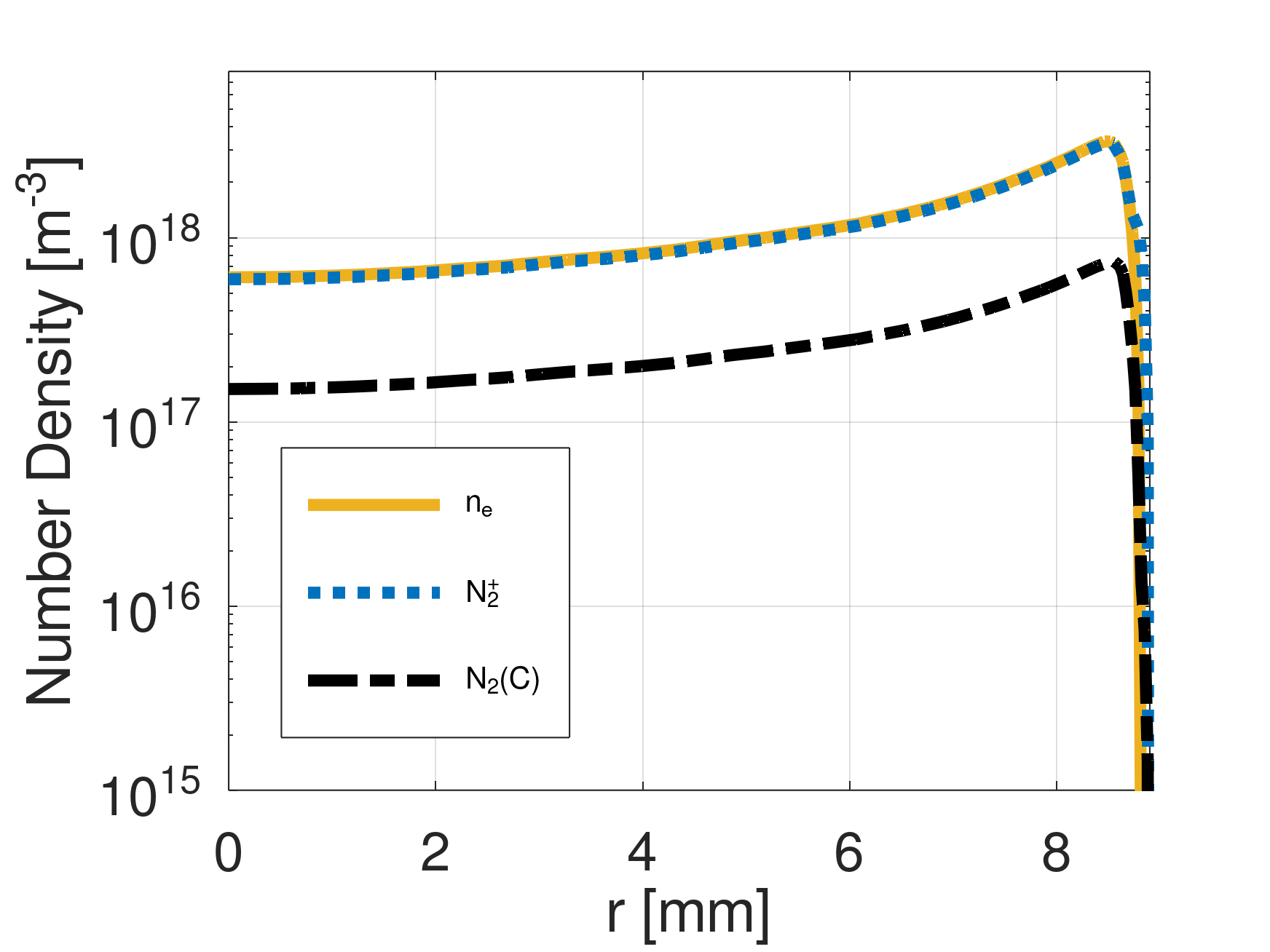}
\caption{ p= 5 Torr  }
 \end{subfigure}
    \begin{subfigure}{.35\textwidth}
  \centering
\includegraphics[width=1\linewidth]{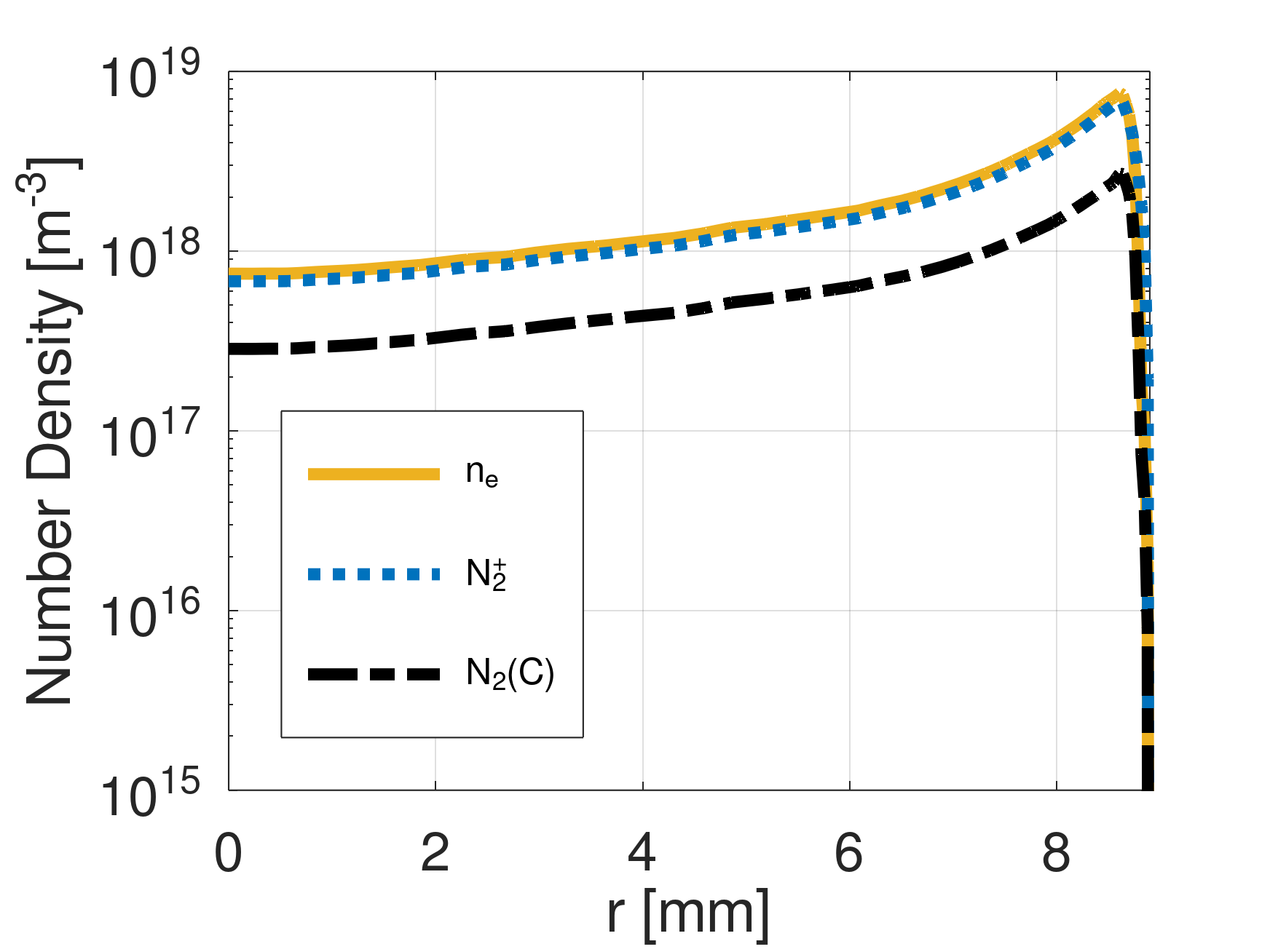}
\caption{ p= 10 Torr  }
 \end{subfigure}
  \caption{ Negative polarity: radial profiles of electron, N$_2^+$ and N$_2$(C$^3\Pi_u$) densities at t$=$32.68 ns and at a distance of 6.75 cm from the HV electrode for different gas pressure.}
    \label{fig:xline_-}
 \end{figure}

\subsection{Radial distribution of electron density in the nanosecond discharge. Comparison to the experimental discharge appearance measured by optical imaging.}

    \begin{figure}[!t]
\centering
\includegraphics[width=0.75\linewidth]{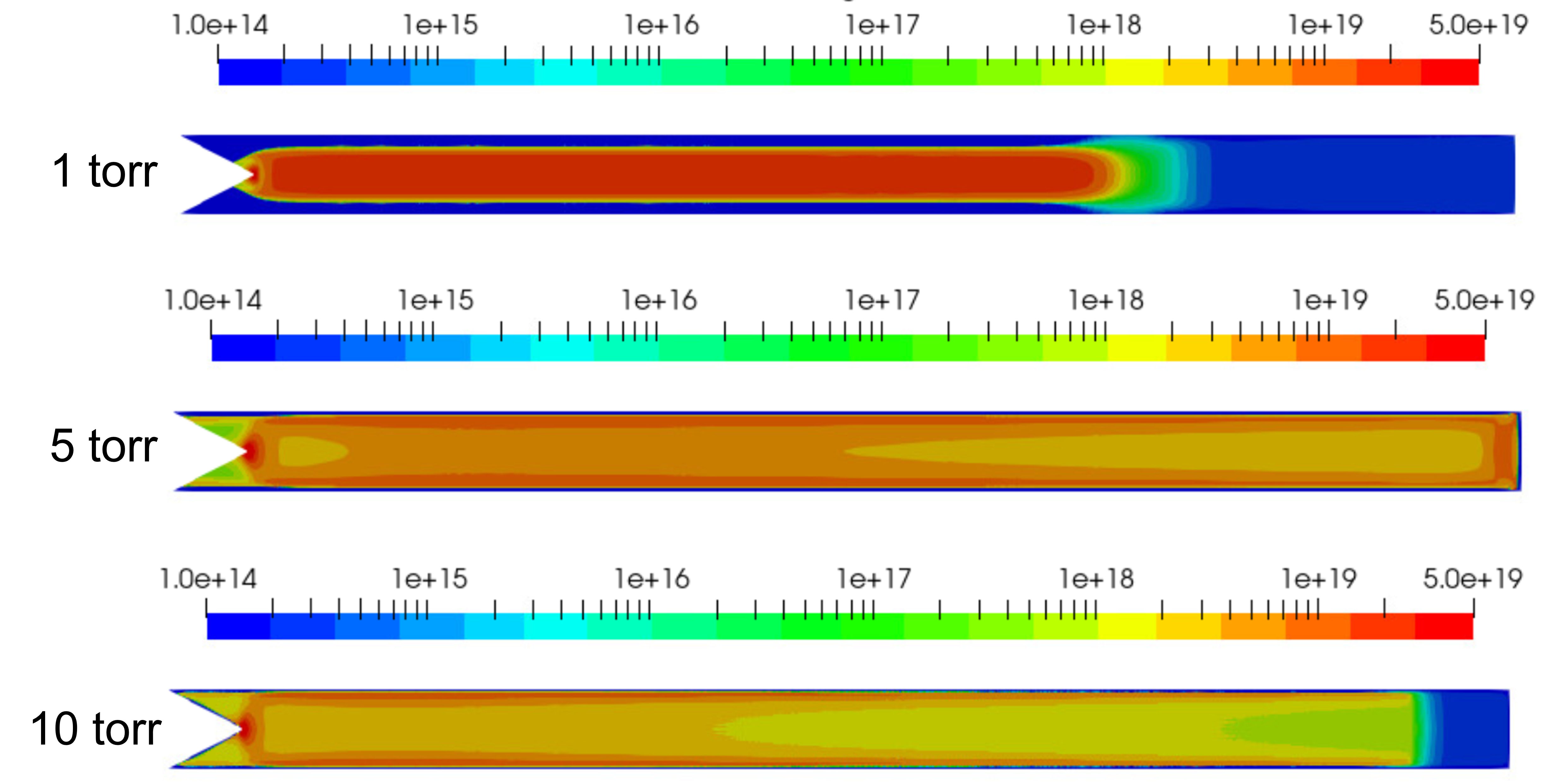}
  \caption{Electron density contours [$\mathrm{m^{-3}}$] for different gas pressure at t$=$32.65 ns. Positive polarity. Min/max values are limited to $10^{14}$ $m^{-3}$ and $5\times10^{19}$ $m^{-3}$ respectively for visualization purposes.}
    \label{fig:ne_comp_+}
 \end{figure}

 \begin{figure}[!b]
\centering
 \begin{subfigure}{.35\textwidth}
 \centering
\includegraphics[width=1\linewidth]{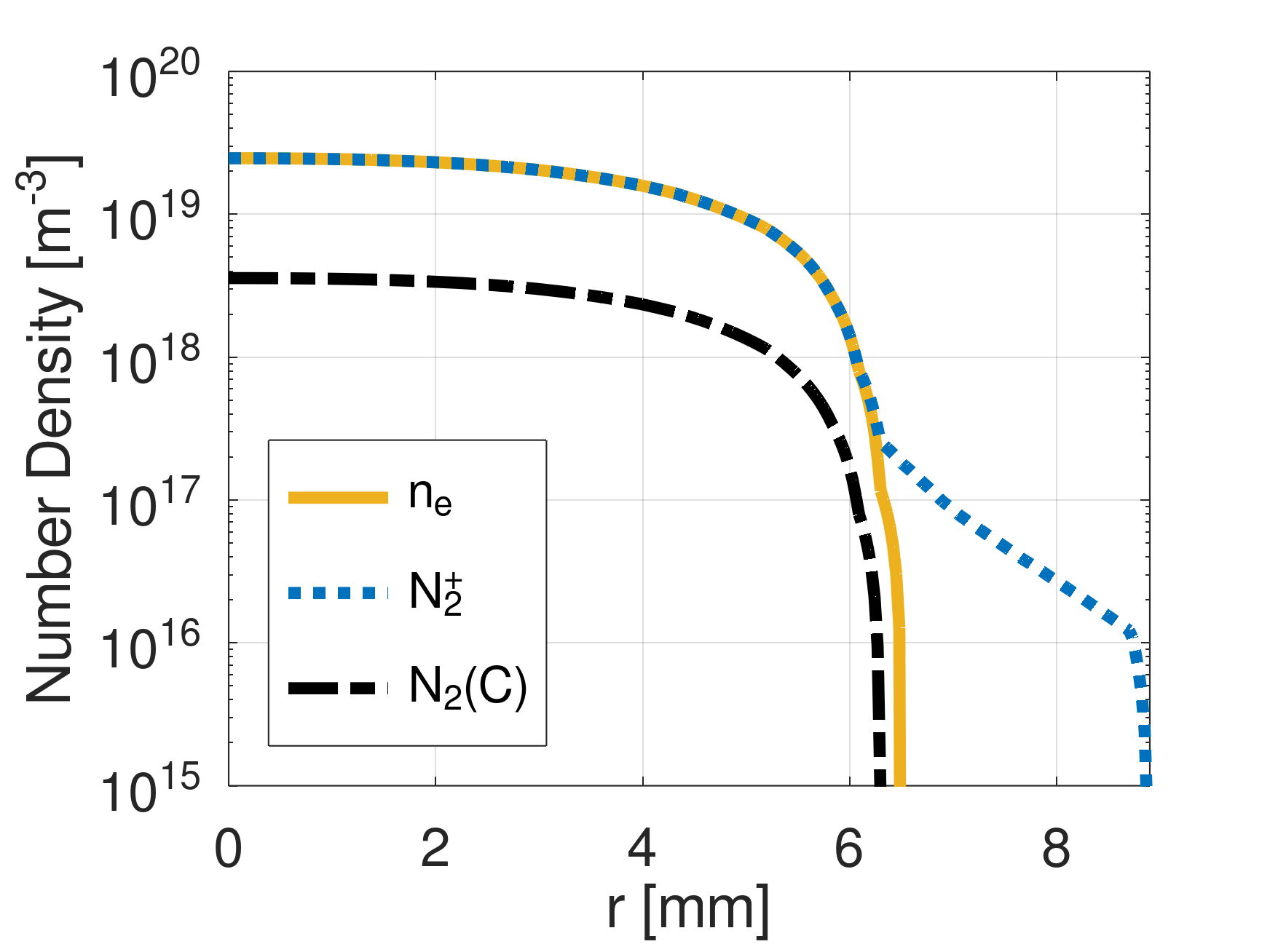}
\caption{ p=1 Torr   }
 \end{subfigure}
   \begin{subfigure}{.35\textwidth}
  \centering
\includegraphics[width=1\linewidth]{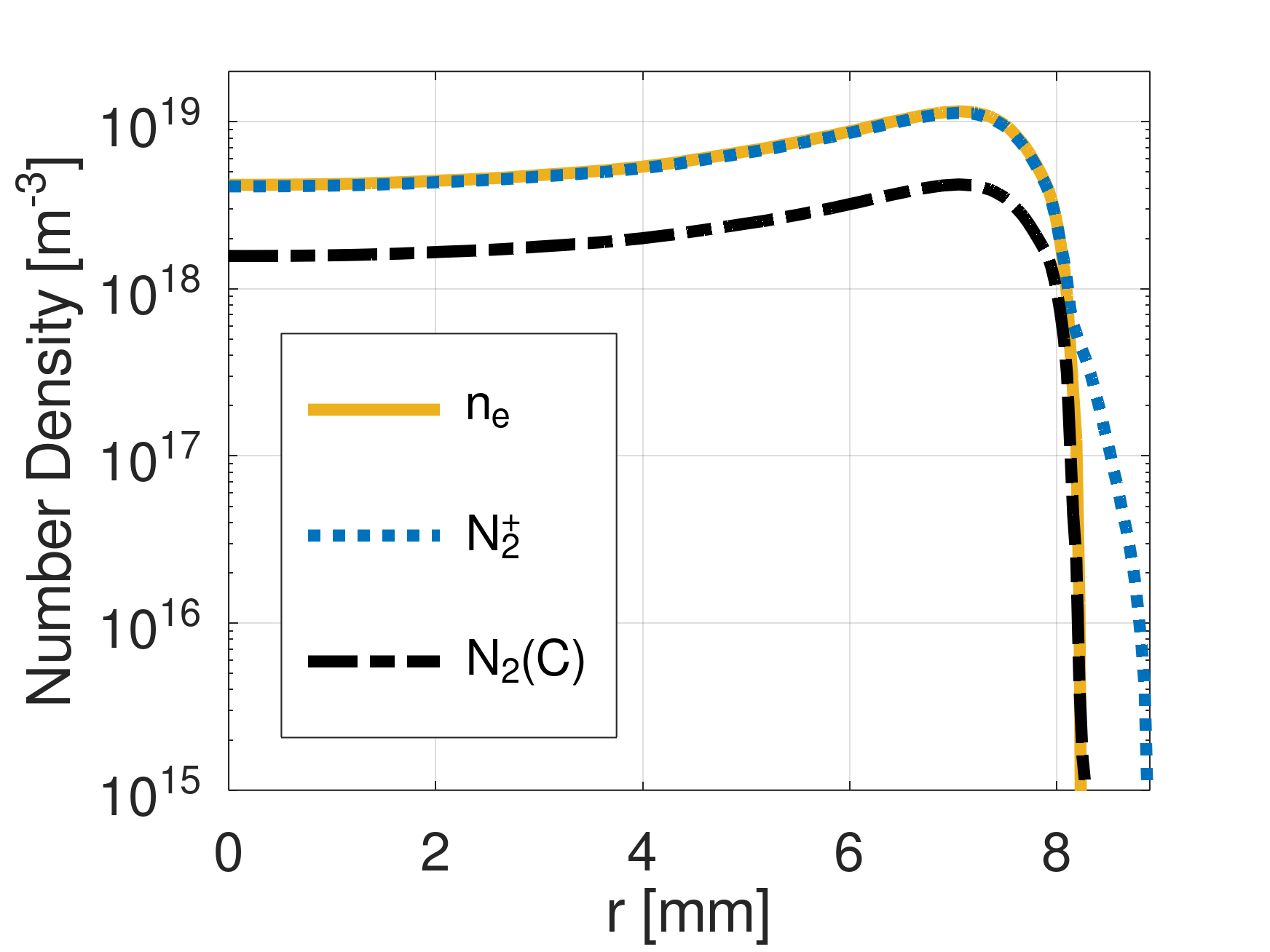}
\caption{ p= 5 Torr  }
 \end{subfigure}
   \begin{subfigure}{.35\textwidth}
  \centering
\includegraphics[width=1\linewidth]{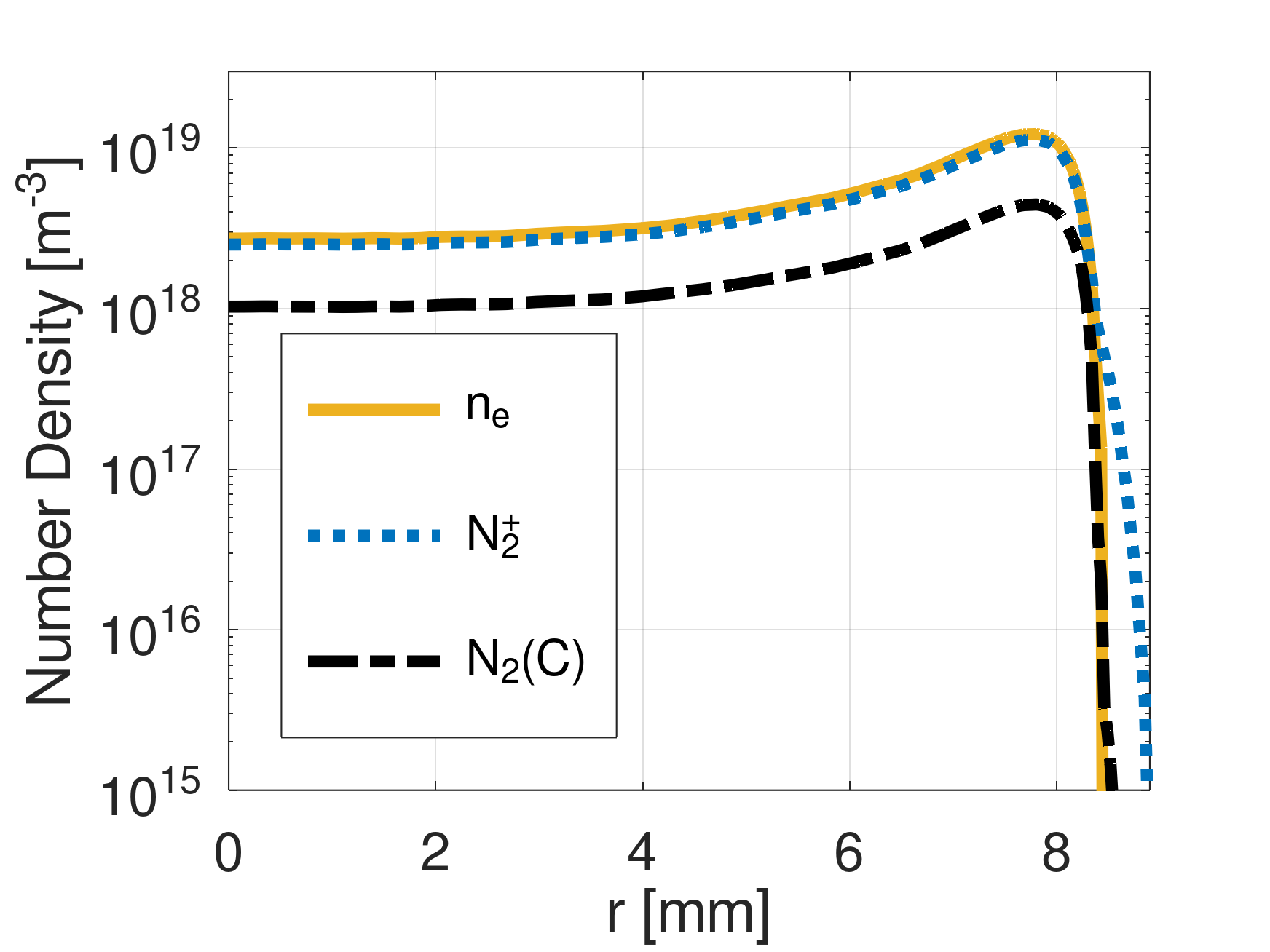}
\caption{ p= 10 Torr  }
 \end{subfigure}
  \caption{Positive polarity: radial profiles of electron, N$_2^+$ and N$_2$(C$^3\Pi_u$) densities at t$=$32.68 ns and at a distance of 6.75 cm from the HV electrode for different gas pressure.}
    \label{fig:xline_+}
 \end{figure}

\subsubsection{Negative polarity}

  Comparative studies are presented herein for the case of negative polarity.  In Fig.~\ref{fig:ne_comp_-}, we plot the electron density contours in the whole discharge tube, at t$=$32.68 ns for different gas pressure.

It is clearly seen that the discharge propagates further with increasing pressure, featuring a less wide sheath near the HV electrode. The radial distribution of the electron density is non-uniform in all pressures, with higher densities observed near the dielectric walls. To shed more light on this, in Fig.~\ref{fig:xline_-}, we plot the radial profiles of electron, N$_2^+$ and N$_2$(C$^3\Pi_u$) densities at t$=$32.68 ns and at a distance of 6.75 cm from the HV electrode's tip for different gas pressures.

  All species follow the same radial distribution, showing maximum near the dielectric walls. The density gradient is higher at high pressure than at lower ones e.g. at 10~Torr, the plasma density near the dielectric reaches values of almost $10^{19}$ $m^{-3}$ while at the axis, is one order of magnitude lower. At low pressure (1~Torr) the sheath formed near the dielectric is more pronounced as expected by plasma theory. As the emission relates mainly to the density of N$_2$(C$^3\Pi_u$), the numerical results confirm the experimental observations of higher emission near the walls for negative polarity. 
  
  \subsubsection{Positive polarity}

  Comparative studies are presented herein for the case of positive polarity.  
    In Fig.~\ref{fig:ne_comp_+}, we plot the electron density contours in the whole discharge tube, at t$=$32.68 ns for different gas pressure.

Here, the discharge propagates further with increasing pressure until it reaches a maximum speed, which is then reduced - at 5~Torr the discharge has already filled in the gap at this time-instant, while at 1 and 10~Torr it stops propagating before bridging the gap. The radial distribution of the electron density is rather uniform at 1~Torr while it begins to take a non-uniform profile with increasing pressure. Near the HV electrode, an intense discharge is produced which is in agreement with the experimental emission profiles. As in the negative polarity case above, and to shed more light on this, in Fig.~\ref{fig:xline_+}, we plot the radial profiles of electron, N$_2^+$ and N$_2$(C$^3\Pi_u$) densities at t$=$32.68 ns and at a distance of 6.75~cm from the HV electrode's tip for different gas pressures.

Similar to the negative polarity case, all species follow the same radial distribution. In contrast to the negative polarity though, the discharge is more uniform at low pressure, showing maximum near the axis or presenting a smoother density gradient. As the pressure increases, the discharge seems to obtain a non-uniform radial profile and approach the distribution of the negative polarity case. The sheaths formed near the dielectric are thicker at all pressure levels in this case. 

The change of radial profiles observed between the negative and positive polarities is related to the presence of the dielectric and associated surface charging effects. Similar to nanosecond surface Dielectric Barrier Discharges (DBDs) \cite{Soloviev2008}, the effects of surface charge accumulation as well as secondary electron emission on the dielectric surfaces, which are both dependent on the voltage polarity and thus the orientation of the electric field vectors which dictate charged species transport towards the dielectric tube, seem to play an important role on the discharge dynamics.

    \begin{figure}[!t]
\centering
\includegraphics[width=1\linewidth]{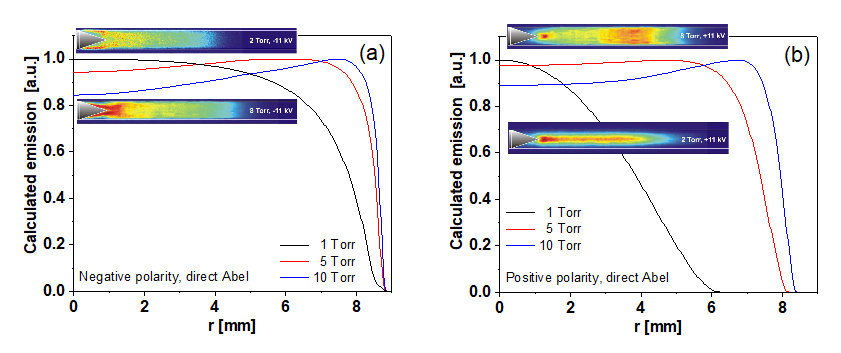}
  \caption{Direct Abel inversion of calculated N2(C) density profiles in pure N2 for (a) the negative polarity discharge, the calculated profiles are taken from Fig.~\ref{fig:xline_-}; (b) the positive polarity discharge, the calculated profiles are taken from Fig.~\ref{fig:xline_+}. ICCD images for corresponding polarity and pressure experiments in air taken from Fig.~\ref{2D_ICCD} are presented in the same figure (see text for details).}
    \label{fig:Emission_by_Abel}
 \end{figure}

The calculated distribution of discharge emission over the cross-section reproduces the main features observed in the experiment. To provide better comparison, Fig.~\ref{fig:Emission_by_Abel} presents a direct Abel transform of the calculated radial distribution of N$_2$(C$^3\Pi_u$) density taken from Figs~\ref{fig:xline_-} and~\ref{fig:xline_+}. The forward Abel transform is done using the online PlasmAbel software \cite{PlasmAbel}. The ICCD images of  N$_2$(C$^3\Pi_u$) $\rightarrow$ N$_2$(B$^3\Pi_g$) emission in air are shown in the same figure (see Sec.~\ref{sec:experimental} for more details). Excellent qualitative correlation between ICCD imaging and [N$_2$(C$^3\Pi_u$)] calculations is observed. Will underline that, in spite of visible ``homogeneity'' with a weakly expressed minimum, at negative polarity on the high-voltage electrode, the discharge develops near walls, leaving the near-axis part "empty". The difference between the density of particles produced near the walls and on the axis can reach an order of magnitude or more (see Fig.~\ref{fig:xline_-}), and this should be taken into account in any measurements of the density of active species. At positive polarity of the high-voltage electrode, especially at low pressures, the discharge develops along the axis, the distance between a quasi-neural plasma and the walls (sheath thickness) can reach 2~mm for a tube diameter of 20~mm.

\section{Conclusions}\label{sec:conclusion}

\color{black}

In this study, fast ionization waves (FIWs) propagating along a cylindrical coaxial discharge setup are investigated using 2D-axisymmetric numerical modeling. FIWs are initiated by a single nanosecond pulse with a 27 kV amplitude applied to the high-voltage electrode. The pulse has a FWHM of 25~ns and a rise time of 4~ns, corresponding to a voltage increase rate of $\partial U / \partial t \approx 6$~kV/ns. Both negative and positive polarity pulses are considered. The study examines pressures ranging from 1 to 10 Torr in molecular nitrogen. The numerical results are validated against available experimental data.

Consistent with experimental observations, the numerical modeling reveals a characteristic temporal behavior of the FIW front: the electron density builds up following a sharp and brief (a few nanoseconds) peak in the electric field. The peak electric field reaches approximately $10^3$~Td, with values 20–30\% higher for positive polarity pulses. After the front, the field drops to around $\sim 100$~Td. The electron density behind the FIW front is in the range of $10^{11}-10^{12}$~cm$^3$/s for FIWs initiated by negative polarity pulses and $10^{12}-10^{13}$~cm$^3$/s for FIWs initiated by positive polarity pulses. The synchronized waveforms of $E/N(t)$ and $n_e(t)$ observed in experiments are reproduced by the numerical modeling.

For both polarities, FIWs originate at the high-voltage electrode and propagate along the tube. The numerically obtained FIW velocities are in good agreement with experimental results, exhibiting a clear maximum as a function of pressure. For a given voltage amplitude and discharge tube diameter, this maximum occurs at $5–6$~Torr, with FIW velocities of $2–3$~cm/ns for negative polarity pulses and $3–4$~cm/ns for positive polarity pulses.

Furthermore, the radial distributions of densities of electrons, ions and N$_2$(C$^3\Pi_u$) molecules are analyzed. The numerical modeling indicates similar radial distributions for  $n_e(t)$ and  [N$_2$(C$^3\Pi_u$)] for all considered cases, while a well-pronounced sheath with a non-compensated charge of positive ions is observed near the tube wall. The sheath thickness reaches about 2~mm at $P=1$~Torr for a positive polarity pulse. The density of N$_2$(C$^3\Pi_u$) reaches $10^{11}-10^{11}$~cm$^{-3}$, with higher values for positive polarity pulses. The pressure-dependent radial distribution of optical emission is well explained by the modeling and shows a good agreement with experimental observations. At negative polarity, the FIW in nitrogen develops near the tube walls, resulting in electron and N$_2$(C$^3\Pi_u$) densities in the tube center that can be an order of magnitude lower. At positive polarity and low pressures, the significant sheath thickness creates the impression that the FIW propagates in a beam-like mode along the tube axis.

Thus, the proposed modeling approach successfully reproduces the key features of FIW development in nitrogen at moderate pressures and can be used as a valuable tool for optimizing nanosecond discharge experiments. Further parametrization for a broader pressure range, different gas mixtures (e.g. air), tube geometries (e.g. diameter to length ratio) and pulse characteristics as well as investigation of the sensitivity of the results to chosen parameters (namely the chosen plasma-chemistry set) are left for future studies. 

\color{black}

\section*{Acknowledgments}

 This work was supported by the European Union Project CAIPIRINH3A, under the GA number 101191768. Views and opinions expressed are however those of the author(s) only and do not necessarily reflect those of the European Union or CINEA. Neither the European Union nor the granting authority can be held responsible for them.
 The work of S.M.Starikovskaia was also partially supported by the Energy4Climate Interdisciplinary Center (E4C) of IP Paris and Ecole des Ponts ParisTech in the framework of the 3rd Programme d’Investissements d’Avenir [ANR-18-EUR-0006-02] and by Cellule Énergie du CNRS (PEPS 2024 Project ZEPhiR).

\section*{References}

\bibliography{mybib}
\bibliographystyle{unsrt}

\end{document}